\documentclass[a4paper,fleqn,usenatbib]{mnras}
\usepackage{newtxtext,newtxmath}
\usepackage[T1]{fontenc}
\usepackage{ae,aecompl}
\usepackage{multicol}
\usepackage{graphicx}	% Including figure files
\usepackage{amsmath}	% Advanced maths commandsfapp
\usepackage{multirow}
\usepackage{longtable}
\usepackage{morefloats}
\usepackage{float}
\usepackage{cleveref}
\usepackage{pdflscape}
\usepackage{placeins}
\usepackage{capt-of}
\usepackage{subfigure}
\usepackage[dvipsnames]{xcolor}

\newcommand{\kms}{{\mathrm{km~s^{-1}}}}

\title[Asteroseismology of blue stragglers in $\omega$ Cen]
{Asteroseismology of blue stragglers in $\omega$ Centauri: 
the case of five double-mode radial SX Phoenicis pulsators}
\author[Daszy\'nska-Daszkiewicz et al.]{ J. Daszy\'nska-Daszkiewicz\thanks{E-mail:daszynska@astro.uni.wroc.pl},
P. Walczak,  P. G\'ora, W. Szewczuk \\
Instytut Astronomiczny, Uniwersytet Wroc{\l}awski, Kopernika 11, 51-622 Wroc{\l}aw, Poland\\
}

\date{Accepted XXX. Received YYY; in original form ZZZ}

\pubyear{2025}

\hypersetup{draft}
\begin{document}
\label{firstpage}
\pagerange{\pageref{firstpage}--\pageref{lastpage}}
\maketitle

\begin{abstract}
We present the first extensive seismic modelling of SX Phe stars in the stellar system $\omega$ Cen. 
First, using the new values of reddening $E(B-V)$ and distance modulus $(m-M)_V$, and bolometric corrections from Kurucz model atmospheres,
we determine the effective temperatures and luminosities of all SX Phe variables in $\omega$ Cen with available $(B-V)$ colours.
Next, we carefully select SX Phe stars that have a frequency ratio strongly suggesting excitation of two radial modes,
and, in addition, their preliminary pulsational models have the values of $(T_{\rm eff},~ L)$ consistent with observational determinations.
For five double-mode radial pulsators, we perform an extensive seismic modeling using the Bayesian analysis based 
on Monte Carlo simulations. We study the effect of opacity tables and helium abundance.
%with a high probability are double-mode radial pulsators, %and have based on preliminary pulsational computations. 
With the OPAL data and $Y=0.30$, we obtained masses in the range (1.0,~1.2)\,M$_{\odot}$, metallicity $Z\in(0.0007,~0.0029)$
and the age of about (1.9,~3.8)\,Gyr. The OP and OPLIB seismic models have always higher metallicites, sometimes outside 
the allowed range for $\omega$ Cen. In the case of three stars, we find seismic models within the observed range of $(T_{\rm eff},~L)$ 
with all three opacity tables. In the case of two stars, with the highest metallicity, seismic models computed with the OP and OPLIB tables
are located far outside the observed error box. 
The OPAL seismic models follow the age$-$metallicity relation known for $\omega$ Cen from the literature.
%Our modelling is based on the assumptions that after formation via collision or binary merge, these pulsating blue straggles follow the single-star evolution. 
%Our modeling is based on the assumption that the evolution of pulsating blue stragglers, after their formation via the collision or merger of binary stars, can be approximated by the single-star evolution. This first attempt of seismic modelling of pulsating blue straggles is a good starting point for more advanced studies also considering formation 
%by collisions or in binary systems. This is our task for the near future.  
\end{abstract}


%%%%%%%%%%%%%%%%%%%%%%%%%%%%%%%%%%%%%%%%%%%%%%%%%%
%%%%%%%%%%%%%%%%% BODY OF PAPER %%%%%%%%%%%%%%%%%%

\section{Introduction}

SX Pheonicis stars are Population II pulsators exhibiting similar spectral types, effective temperatures and pulsational periods
to $\delta$ Scuti stars, which belong  to Population I  \citep{Breger1980}. Most SX Phe stars reside in globular clusters and 
they occupy the region of blue straggler stars (BSSs) on the colour-magnitude diagram. 
BSSs are very intriguing objects which are bluer and brighter than the cluster stars at the main sequence (MS) turn-off \citep{Sandage1953}.
Thus, they appear much younger than the population they belong to and in which they formed. 
Three main mechanisms explaining BSS origin were proposed:  1) mass transfer during a binary evolution \citep[e.g.,][]{McCrea1964,Ivanova2015}, 
2) coalescence in a binary system \citep[e.g.,][]{Nelson2001,Chen2008} or
3) direct collision of two or more stars in high-density regions \citep[e.g.,][]{Benz1987,Sills2002}. 
These mechanism can act simultaneously within the same cluster and the discovery of a double BSS sequence in M30 by \citet{Ferraro2009} could be an example of this.
Each of the BSS formation channel mentioned above is intrinsically a three-dimensional (3D) problem. In the first approximations, binary evolution 
with mass transfer can be computed with one-dimensional (1D) evolutionary codes. 
However, even with this approximation, the evolution of binary stars is very complicated.
In addition to the fact that each star is characterized by a set of its own parameters (e.g., mass, chemical composition, rotation), 
we also have many other parameters like mass-transfer rate, its stability and conservativity, orbital elements \citep[e.g.,][]{Langer2012}. 
Mergers (from binaries or collisions) are dynamical processes and demand time-consuming 3D modelling \citep[e.g.,][]{Lombardi1996,Sills1997,Glebbeek2008}, 
which is not feasible for long-term evolution occurring on a nuclear timescale.
Yet another scenario for the formation of blue stragglers is the triple evolution which adds another layer of complexity.
In the hierarchical triple system,
the merger rate of inner primordial binary can be increased through a process involving Kozai cycles and tidal friction \citep[e.g.,][]{Perets2009}.
The most recent overview of the physical properties and current understanding of BSSs was provided by \citet{WangRyu2025}.

Detecting pulsations in blue stragglers enables to constrain their parameters and interior properties via asteroseismology. 
Despite the importance of blue stragglers in terms of evolutionary theory and cluster dynamics, few attempts have been made 
to seismically model these puzzling objects. Asteroseismology of pulsating BSS, including SX Phe stars,
is still in its infancy and waits to realize its enormous potential.

Linear nonadiabatic pulsations for SX Phe star models were computed  by \citet{Gilliland1998} and \citet{Santolamazza2001}.
They showed that pulsations in these variables are excited by the classical $\kappa$ mechanism operating in the second helium ionization zone. 
Both radial and non-radial modes are excited. \cite{Santolamazza2001} derived a theoretical relation between 
pulsational period of radial modes, mass, effective temperature, luminosity and metallicity. 
\citet{Gilliland1998} and \citet{Bruntt2001} studied SX Phe stars in the globular cluster 47\,Tuc and estimated masses of a few variables
for a fixed value of metallicity. To this end, these authors combined the location of SX Phe stars on the colour-magnitude diagrams 
and on the Petersen diagram for double-mode radial variables.
\citet{Petersen1996} estimated a mass of the prototype SX Phe at $M=1.0 {\rm M}_{\odot}$ and the age at 4.07 Gyr assuming the metallicity $Z=0.001$.
\citet{Templeton2002} studied the effect of chemical composition on pulsational properties and the period-luminosity relation.
They globally enriched helium content to simulate the effects of stellar collisions and global mixing possible in blue stragglers.
All above mentioned  authors adopted the OPAL opacity tables and assumed fully radiative envelopes. 
\citet{Olech2005} computed non-adiabatic pulsational models appropriate for SX Phe stars in $\omega$ Cen and found that almost 
all observed frequencies fall in the ranges predicted for unstable modes.
Nonlinear convective radial pulsations were computed by \citet{Fiorentino2015} for three values of metallicity. They determine
the instability strip for the first four radial modes. Using the theoretical dependence between mass, period and luminosity,
\citet{Fiorentino2015} constrained masses for SX Phe stars in $\omega$ Cen.

However, an extensive seismic modelling of oscillating blue stragglers is rather lacking. The only detailed analysis has been performed 
for the prototype SX Phe by \cite{JDD2020,JDD2023}. From the fitting of two radial modes,
fundamental and first overtone, as well as the photometric amplitudes and phases for the dominant mode, strong constraints on the 
parameters and convective efficiency in the envelope were obtained.
The best seismic OPAL models had the increased helium abundance $Y\approx 0.32$, metallicity $Z\approx 0.002$, 
the mass $M\approx 1.08 {\rm M}_{\odot}$ and the age of about 2.90 Gyr.
%Seismic modelling indicated a slight increase in helium abundance in the SX Phe prototype. 
Moreover, seismic models turned out to be very sensitive to the adopted opacity data
and only the OPAL seismic models had effective temperatures and luminosities consistent with the observational determinations.
The same "opacity" problem was obtained for several high-amplitude $\delta$ Sct stars \citep{JDD2022,JDD2023}.

There are also a few examples of the studies of BSSs exhibiting $\delta$ Sct pulsations in old open clusters.
\cite{Arentoft2007} tried to identify overtones of four blue stragglers in NGC\,2506 assuming they pulsate radially.
A large number of frequencies (41 and 26) in two oscillating BSSs in M\,67 was detected by \citet{Bruntt2007}. 
They also computed a corresponding grid of pulsational models and showed that radial and nonradial modes can be excited in a wide range of frequencies. 
\citet{Guzik2023} identified four oscillating BSSs in the open cluster NGC\,6819 based on the analysis of the Kepler light curves.
These BSSs exhibit pulsation both in p- and g-modes. The authors identified frequency separations in the two $\delta$ Sct pulsators
and estimated their masses and radii.
All papers dealing with pulsational computations are based on 1D single-star evolutionary models or even envelope models.

Here, we present an extensive seismic modelling of five double-mode radially pulsating SX Phe stars located in the central region of the globular cluster $\omega$ Cen.
%In such dense environments, the collisional formation of BSSs is very probable. 
The predictions of 3D simulations showed that mergers have quite similar properties
to normal (single) stars, i.e., they have similar surface chemical composition and, in general, BSSs follow evolution close to single stars of the same mass
\citep[e.g.,][]{Lombardi1996,Sills2001,Glebbeek2013}. The formation of BSSs by collisions in globular clusters, in particular in their central regions, can be very frequent. It is therefore quite likely that the SX Phe stars 
in the central regions of $\omega$ Cen were formed by collisions.
%and even plays a dominant role.  %and this is the main assumption in our paper. 
Moreover, given the large uncertainties and many free parameters in the calculations of binary evolution (e.g. mass transfer) and simulations of merger processes, seismic modelling of BSSs using 1D evolutionary codes is 
a reasonable approach as a first approximation. Such seismic models provide a solid starting point for more advanced modelling.

In Sect.\,2, we determined the values of effective temperature and luminosity for all known SX Phe stars in $\omega$ Cen with
the available colours $(B-V)$ and visual magnitude $V$. We also select the most suitable SX Phe stars for seismic modelling
and  re-determine the frequencies from the CASE light curves of these stars.
In Sect.\,3, we present the results of our extensive seismic modelling which involves the use of Bayesian analysis based on Monte Carlo simulations.
Sect.\,4 contains a discussion of the obtained results and Sect.\,5 summarizes the paper. Additional materials are given in the Appendix.

%Till now the only extensive seismic, which scanned a wide range of mass, metallicity, hydrogen abundance, efficiency of convection i th eouter layers  SX ...

\section{SX Phe stars in $\omega$ Centauri}

$\omega$ Centauri (NGC\,5139) is a very complex stellar system, most probably the remnant of a dwarf galaxy captured by the Milky Way (MW) billions of years ago
\citep[e.g.,][]{Hilker2000}. The recent detection of fast-moving stars in the centre of $\omega$ Cen strongly suggest the existence of an
intermediate-mass black hole with a lower-mass limit of about 8200\,M$_{\odot}$ \citep{Haberle2024}.
One of the characteristic features of this most massive MW globular cluster is a large spread in metallicity  \citep[e.g.,][]{Villanova2014,Nitschai2023}, which is a consequence of its complex formation and evolution.
At least six sub-populations with different metallicities can be singled out  \citep{Villanova2014}.
Recently, \citet{Nitschai2023} derived the distribution of [m/H] (see their Fig.\,11). 
To this end, these authors used massive MUSE spectroscopy for main-sequence and red giant stars. 
They obtained the values of [m/H] in the range of about $(-2.2, -0.6)$ with the maximum at [m/H]$\approx -1.55$.
This corresponds to the metal abundance by mass $Z$ of about (0.0001,~0.004).
Moreover, a correlation between age and metallicity has also been obtained \citep{Villanova2014,Clontz2024}.
The majority of stars in $\omega$ Cen have ages between 10 and 13 Gyr \citep{Clontz2024}.

$\omega$ Cen also contains many variable stars of various types. In particular,  it hosts a large population of RR\,Lyr stars \citep{Navarrete2015}
and the largest number of SX Phe stars of any globular cluster \citep{Kaluzny2004,Olech2005}.
SX\,Phe stars were analysed by \citet{Olech2005} in the framework of {\it Cluster AgeS Experiment} (CASE). The authors identified 69 SX Phe stars in the field 
of $\omega$ Cen with frequencies in the range of about (11,~56)\,d$^{-1}$. One variable V65 appeared to be a foreground star.
Many of these SX Phe stars exhibit multi-periodic variations and some of them have frequency ratios
indicating pulsations in radial modes. 

\subsection{Determination of the effective temperatures and luminosities}

\begin{figure*}
	\includegraphics[width=88mm,height=7.05cm,clip]{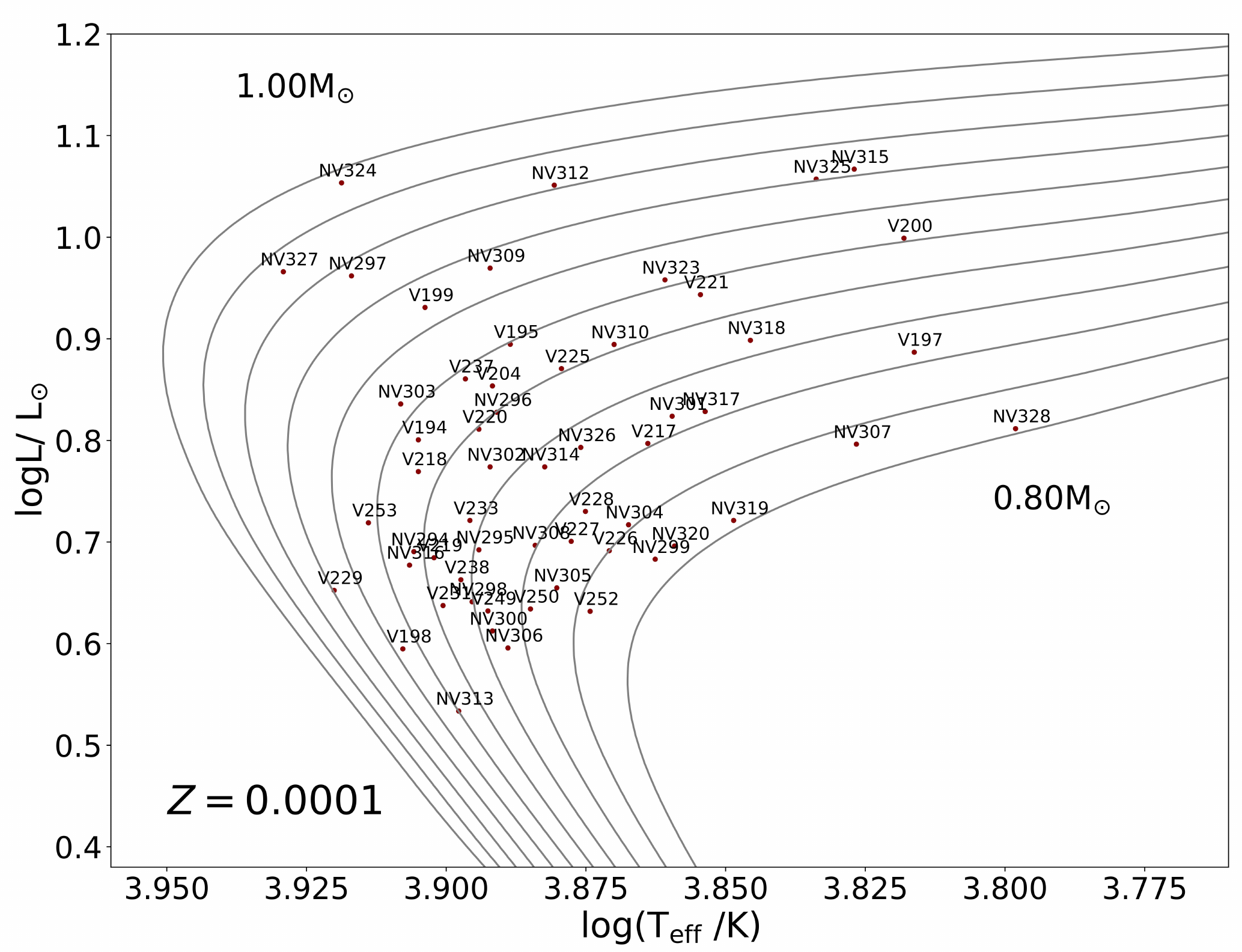}
	\includegraphics[width=88mm,height=7cm,clip]{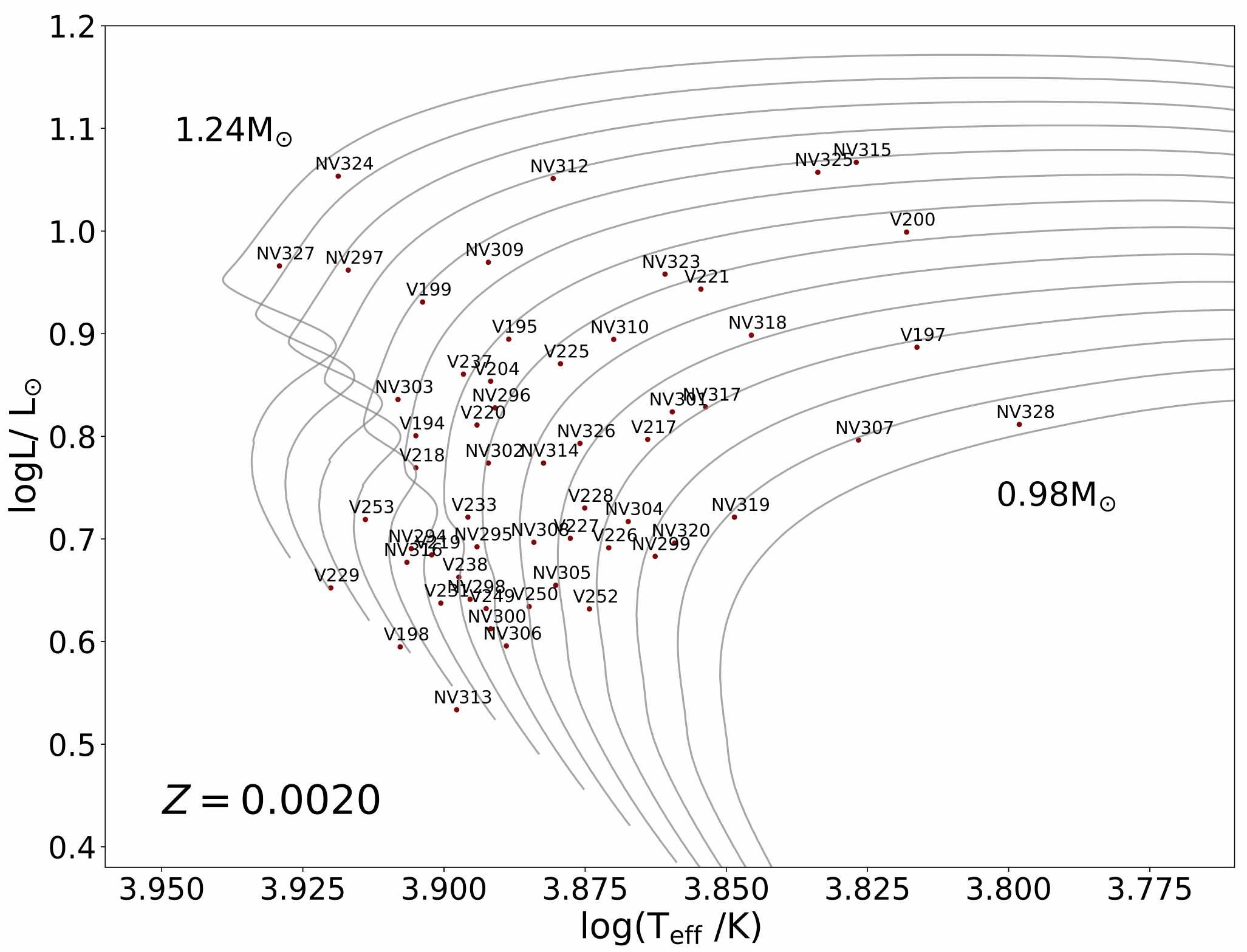}
	\caption{The position of the central values of $(\log T_{\rm eff},~\log L/{\rm L}_{\odot})$ of 58 SX Phe stars
		in $\omega$ Cen on the HR diagram.
		For a comparison, we depicted the evolutionary tracks computed for the metallicity $Z=0.0001$ (left panel) and $Z=0.002$ (right panel).
		The maximum and minimum values of mass are indicated. The track were computed with the OPAL opacities, AGSS09 mixture, initial hydrogen abundance $X_0=0.70$, 
		mixing length parameter $\alpha_{\rm MLT}=1.0$ and zero-rotation velocity.}
	\label{fig:HR_all_stars}
\end{figure*}

Here, we focus on 58 SX Phe pulsators which have determinations of, both, colours $(B-V)$ 
and visual magnitude $V$.
We determined the effective temperature and luminosity using data collected by
\citet{Olech2005} and Kurucz models of stellar atmospheres.
To this aim, we used the newest value of a reddening $E(B-V)=0.185\pm 0.040$ derived 
by \citet{Clontz2024} for the central region of $\omega$ Cen.
An increase in extinction toward the cluster centre has been already 
indicated by \citet{Schlegel1998} and \citet{McNamara2011}, who
gave $E(B-V)=0.139$\,mag and  $E(B-V)=0.16\pm 0.01$\,mag, respectively.
Moreover, we adopted the true distance modulus $(V-M_V)_0=13.672\pm 0.019$\,mag, reported 
by  \citet{Clontz2024}, which corresponds to the mean value of the literature distances
 $d=5.426 \pm 0.047$\,kpc, as derived by \citet{Baumgardt2021}.
With a standard extinction law of $R_V = 3.1$, the apparent distance modulus amounts to $(m-M)_V=14.246\pm 0.059$. 
This value is greater than the one used by \citet{Olech2005}. i.e.,  $(m-M)_V=14.09\pm 0.04$, 
which was derived from the analysis of an eclipsing binary OGLE GC17 by \citet{Kaluzny2002}.
The new values of dereddened colours $(B-V)_0$ and visual absolute magnitude $M_V$ of 58 SX Phe stars
are given in Table\,A1 in the Appendix A. 
%V65 was omitted because this is a field star. 
 
Having the new values of $(B-V)_0$ and $M_V$, we derived the effective temperature and luminosities
using the Kurucz atmosphere models. Firstly, we determined the effective temperature $T_{\rm eff}$  by interpolating the Kurucz values of  $(B-V)_0$ to the observed ones. We considered three values 
of the surface gravity $\log g=3.5, 4.0,4.5$  and four values of the metallicity  [m/H]$=-2.0, -1.5, -1.0$ and $-0.5$. 
In the case of SX Phe stars, the value of $\log g$ is around 4.0 dex,  so it is safe to adopt the range [3.5, 4.5]. 
In the case of metallicity values, we were guided by the latest determinations of \citet{Nitschai2023}, 
who derived the distribution of [m/H] in $\omega$ Cen. In that way we derive the effective temperatures taking into account uncertainties in $\log g$ and [m/H].

Adopting the Kurucz bolometric corrections $BC$ for the determined range of $T_{\rm eff}$ and including the above-mentioned ranges in $\log g$ and [m/H],  we calculated luminosities using the elementary formula:
\begin{equation}
\log L/{\rm L}_{\odot}=\frac{4.74-(M_V+BC)}{2.5},
\end{equation}
where $M_{\rm bol,\odot}=4.74$ is the absolute solar bolometric magnitude according to the IAU\,2015 resolution \citep{Mamajek2015}.
% for a system where BC'_sun=-0.07, one has BC'=BC+0.124 Kurucz (1979)

In that way, we determined the values of $(T_{\rm eff},~L)$ for all SX Phe stars in $\omega$ Cen which have the observed colours $(B-V)$, that is for 58 in total. 
All of them are given in Table\,A1 in the Appendix A, together with the new values of $(B-V)_0$ and $M_V$. 
In Fig.\,1, we plotted the central values of  $(\log T_{\rm eff},~\log L/{\rm L}_{\odot})$  on the Hertzsprung-Russell diagram together with the evolutionary tracks calculated for a metal abundance by mass of $Z=0.0001$ (left panel) and $Z=0.002$ (right panel). Evolutionary tracks were computed with the Warsaw-New Jersey code
\citep[e.g.,][]{Pamyatnykh1999} adopting the OPAL opacity tables  \citep{Iglesias1996} and chemical mixture of \citet{Asplund2009}, hereafter AGSS09. 
At the lower temperature range ($\log T<3.95$),  opacity data from \citet{Ferguson2005} were used.
%$BC$ vs $T_{\rm eff}$  the effect of [m/H]$=-1.0,~ -2.0$ and $\log g=4.0,~4.5$     $(B-V)_0$ vs $T_{\rm eff}$
As one can see, the evolutionary stage of a given star depends on the assumed metallicity. For lower metallicities, the star will be in a more advanced phase of evolution.

\subsection{Selection of SX Phe stars for seismic analysis}

\begin{figure*}
	\includegraphics[width=88mm,height=7.35cm,clip]{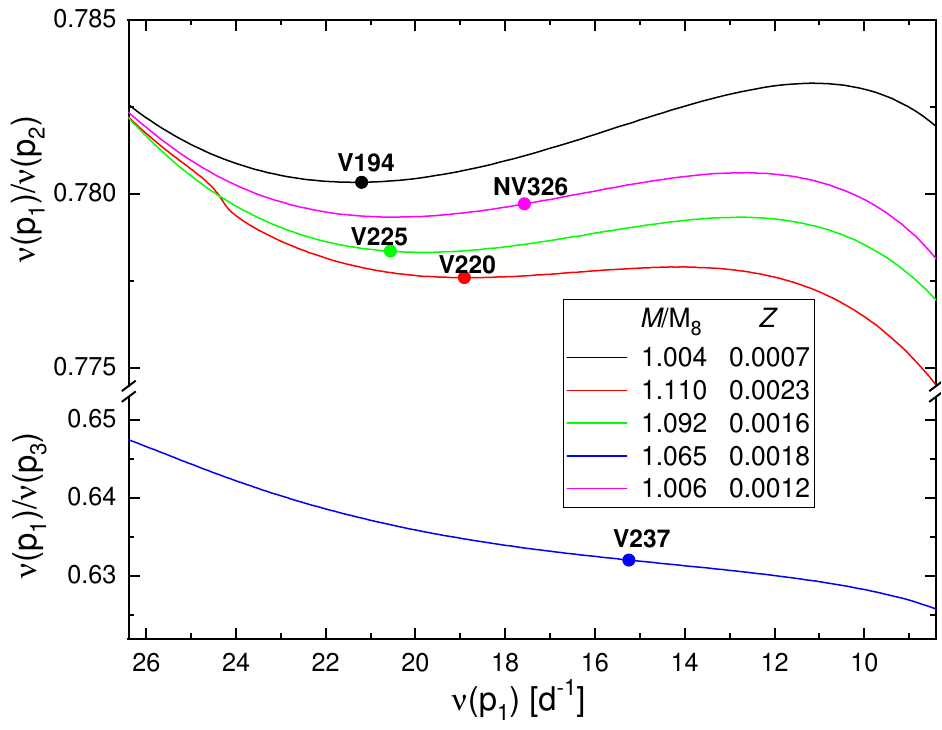}
	\includegraphics[width=88mm,clip]{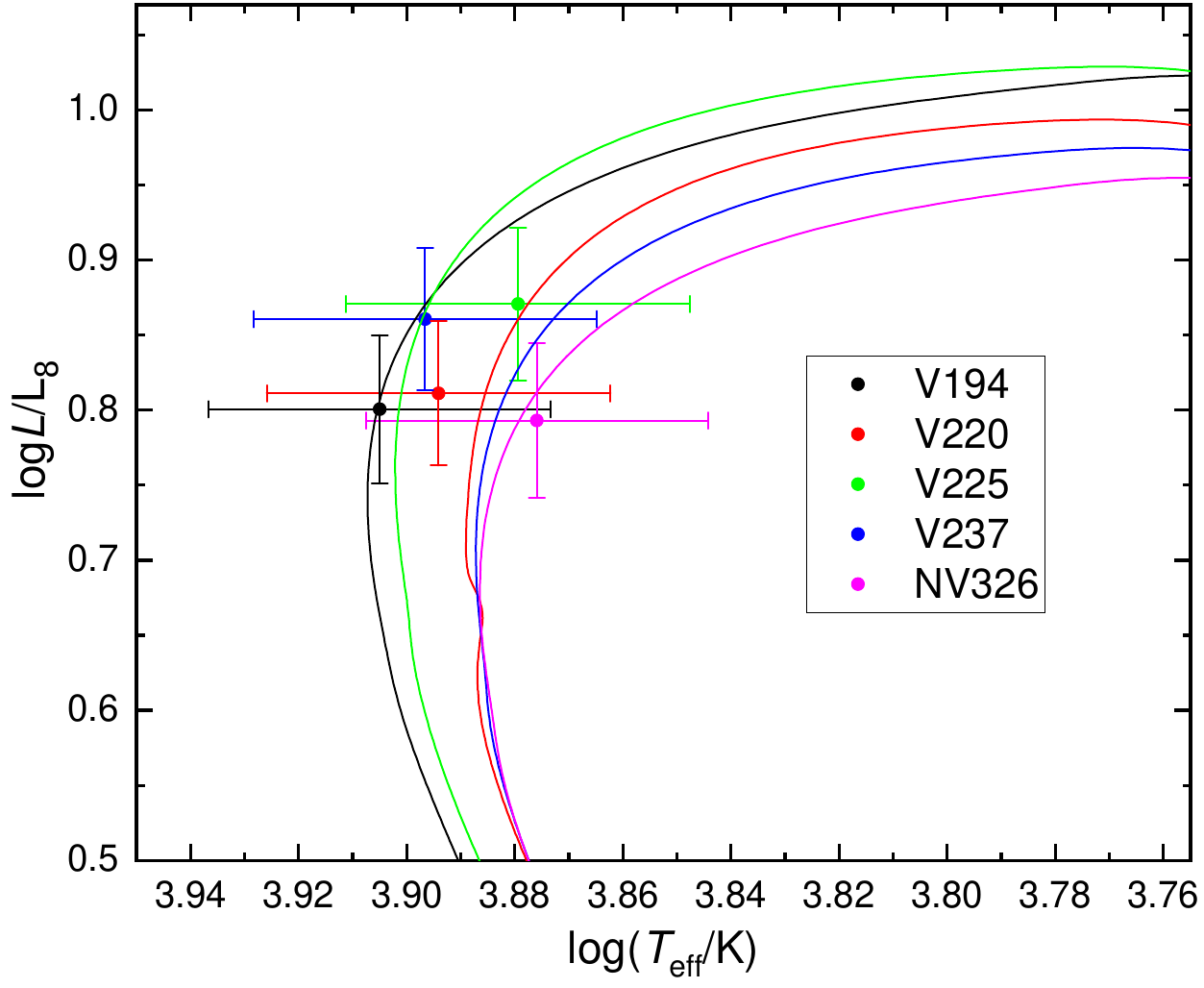}
	\caption{Left panel: the frequency ratio of the fundamental and first overtone radial modes (the OY axis above the break) and 
		of the fundamental and second overtone radial modes (the OY axis below the break) as a function of the fundamental-mode frequency.
		There are shown models which reproduce the two frequencies of the five selected  SX Phe stars marked as dots.
		The values of mass and metallicity are given in the legend. 
		Right panel: a comparison of  the position  of the 5 SX Phe stars on the HR diagrams with evolutionary tracks of the models considered on the left panel. }
	\label{fig:Petersen_HR_5stars}
\end{figure*}
In the framework of the CASE project, \citet{Olech2005} analysed the light curves of  69 variables  in the field of $\omega$ Cen 
and located in the BSS region on the colour-magnitude diagram. Most of the stars appeared to pulsate in more than one frequency.
Moreover, in the case of several stars, the frequency ratios indicate the possibility of pulsations in two radial modes.
This fact  enable seismic modelling of these pulsators by fitting the radial mode frequencies.
The period ratios of radial modes can be determined very precisely and take values in a very limited range.  In the case of SX Phe stars, the frequency ratio of radial fundamental 
to first overtone modes is about 0.78. For the fundamental to second overtone modes -- about 0.63, 
while for the second to third overtones -- about 0.81. In Table\,1, we collected all stars (18 in total) that have determined parameters 
$(T_{\rm eff},~L/{\rm L}_{\odot})$ and frequency ratios indicating pulsations in radial modes.
$\nu_1$ always represents the value of dominant frequency.

As mentioned in the introduction, in this paper we assumed that the evolution of BSSs after their formation can be described, 
to a first approximation, by single-star evolution.
We computed evolutionary models by means of the 1D Warsaw-New Jersey evolutionary code \citep[e.g.,][]{Pamyatnykh1999}.
This code includes the mean effects of the centrifugal force, assuming solid-body rotation
and constant global angular momentum during the evolution.
The rotational mixing and transport of angular momentum are not included.
The treatment of convection in the stellar envelope relies on the standard mixing-length theory (MLT) 
and the efficiency of convection is measured by the mixing length parameter $\alpha_{\rm MLT}$.
We adopted the AGSS09 chemical mixture and the OPAL2005 equation of state \citep{Rogers2002}.
To compute stellar pulsations, we used linear non-adiabatic code of \citet{Dziembowski1977} which takes into account
the effects of rotation on pulsational frequencies up to the second order in the framework of perturbation theory.
Moreover, the frozen convection approximation is adopted, i.e., the convective flux does not change during the pulsations,
which is a reasonable approach if convection is not very efficient in the envelope.

After calculating a large grid of pulsational models and trying to fit two frequencies as radial modes as well as the effective temperature and luminosity 
within determined ranges (given in Table\,A1), we selected five stars for detailed seismic analysis, i.e.,  V194, V220, V225, V237 and NV326.
Four stars (V194, V220, V225, NV326) have the values of $\nu_1/\nu_n$ corresponding to the radial fundamental 
$(\ell=0,~p_1)$ and first overtone $(\ell=0,~p_2)$ modes,
whereas in the case of V237 this ratio indicates pulsations in the radial fundamental and second overtone $(\ell=0,~p_3)$ modes.  The selected stars are marked in bold in Table\,1.
The parameters $(T_{\rm eff},L)$ of these stars determined in the previous subsection are given in Table\,2. 
For three stars, determinations of the projected rotational velocity  are also available in the literature.
Using high-resolution spectra, \citet{Mucciarelli2014} determined $V_{\rm rot}\sin i$ for V220 and V225 to be 22\,$\kms$ and 20\,$\kms$, respectively.  In addition, they estimated the values of effective temperature,
$\log T_{\rm eff}\approx 3.87 $ for V220 and $\log T_{\rm eff}\approx 3.91$   for V225, 
which agree with our determinations.
\citet{Simunovic2014} derived $V_{\rm rot}\sin i =55.9\pm 24.7\,\kms$ for NV326 on the basis of medium-resolution spectra. 

In the left panel of Fig.\,2, we plotted the Petersen diagram showing evolution of the frequency ratio of the radial fundamental to first overtone
modes $\nu(p1)/\nu(p_2)$ and the radial fundamental to second overtone modes $\nu(p1)/\nu(p_3)$, as a function of 
the fundamental-mode frequency. The observed values $(\nu_1,~\nu_1/\nu_n)$ of the five SX Phe stars are marked with dots. The right panel of Fig.\,2 shows positions of these stars on the HR diagram together with corresponding evolutionary tracks.
For each star, the frequencies of two radial modes have been reproduced, and the models 
have effective temperatures and luminosities that are within the error box. Moreover, their metallicities are
in the allowed range for $\omega$ Cen.  All models were computed assuming the helium abundance by mass $Y=0.30$, OPAL opacities, the initial rotational velocity $V_{\rm rot,0}=10\,\kms$ and $\alpha_{\rm MLT}=1.0$. 
\begin{table}
	\centering
	\caption{The list of stars with the frequency ratio indicating the two radial modes. The values of frequencies were taken from \citet{Olech2005}. There are given the dominant  frequency $\nu_1$ and the frequency $\nu_n$ giving the ratio $\nu_1/\nu_n$ which may indicate radial modes. In the case of stars  NV305  and NV323, the inverse quotient was given, i.e., $\nu_n/\nu_1$. Stars selected for further seismic analysis are marked in bold.}
	\label{tab:freq_ratio}
	\begin{tabular}{r ccc c r} % four columns, alignment for each
		\hline
		star       &   $\nu_1$ &   $\nu_n$ & $\nu_1/\nu_n$&    preliminary \\
	      & $[\mathrm d^{-1}]$ & $[\mathrm d^{-1}]$  &     & identification \\
		\hline
    	 {\bf V194} & {\bf 21.1964(1)}  & {\bf 27.1633(5)} & {\bf 0.78033(2)} & \bf{F+1O}\\
 V197 &  21.2219(2)  &  27.1838(8) & 0.78068(2) & F+1O \\
            V204 & 20.2529(1) & 26.1552(5) & 0.77433(2) & F+1O \\        
            V218 & 22.8627(5) & 27.0008(6) & 0.84674(3) & 2O+3O \\
 {\bf V220} & {\bf 18.9083(2)} & {\bf 24.3161(5)} & {\bf 0.77760(2)} & \bf{F+1O} \\           
       V221 & 27.6749(4) & 32.8848(5) & 0.84157(2) & 2O+3O \\
 {\bf V225} & {\bf 20.5600(2)} & {\bf  26.4145(5)} & {\bf 0.77836(2)} & \bf{F+1O} \\
 {\bf V237} & {\bf 15.2433(2)} & {\bf  24.1162(5)} & {\bf 0.63208(2)} & \bf{F+2O} \\
 V249 & 28.6149(4) & 35.0000(9) & 0.81757(2) & 1O+2O \\
 		    V250 & 24.6142(2) & 37.6484(8) & 0.65379(1) & F+2O \\
		    V253 & 25.0196(5) & 31.1610(7) & 0.80291(2) & 1O+2O or F+1O \\	    
	      NV301 & 28.2141(5) & 35.3744(7) & 0.79759(2) & F+1O \\	      
	      NV305 & 27.3469(5) & 22.6529(6) & 0.82835(2) & 2O+1O \\
	      NV306 & 26.0387(5) & 40.7315(6) & 0.63928(2) & F+2O \\      
%		  NV309 & 25.1601(6) & 38.8636(7) & 0.64739(2) & F+2O \\		  
	 	  NV312 & 23.0802(5) & 36.7987(8) & 0.62720(2) & F+2O \\
%	 	  NV322 & 20.8524(6) & 26.8419(6) & 0.77686(3) & F+1O \\
	 	  NV323 & 20.2615(6) & 15.7478(6) & 0.77723(3) & 1O+F \\
		  NV324 & 19.4953(4) & 30.3112(9) & 0.64317(2) & F+2O \\
  {\bf NV326} & {\bf 17.5728(6)} & {\bf 22.5375(6)} & {\bf 0.77971(3)} & \bf{F+1O}\\
		\hline
	\end{tabular}
\end{table}        
%	V194  &        &       &       \\ V220  &        &       &       \\ V225  &        &       &       \\ V237  &        &       &       \\ NV326  &        &       &       \\		
%
\begin{table}
	%	\small
	\centering
	\caption{The values of effective temperature, luminosity and available projected rotation velocity for SX Phe stars selected for detailed seismic analysis.}
	\label{tab:TL_5stars}
	\begin{tabular}{|c|c|c|c|c|c|}
		\hline
		 Star & $\log(T_{\rm eff}/\rm K)$ & $\log(L/\rm L_{\odot})$ & $V_{\rm rot}\sin i$ \\
		      &                           &                          &      $[\kms]$      \\ 
		\hline
		 V194 &  3.905(32) & 0.801(49) & -- \\		
		 V220 &  3.894(32) & 0.811(48) & 22$^1$ \\
		 V225 &  3.879(32) & 0.871(51) & 20$^1$\\
		 V237 &  3.897(32) & 0.861(48) & -- \\
		NV326 &  3.876(32) & 0.793(51) & $55.9\pm24.7^2$\\
        
\hline
\end{tabular}\\
\footnotesize $1$ -- \citet{Mucciarelli2014} ~~~
\footnotesize $2$ -- \citet{Simunovic2014}
\end{table}

\subsection{Frequency analysis of the {\it CASE} light curve of the five SX Phe stars}

Before detailed seismic modelling of the selected SX Phe stars, we re-analysed the $BV$ light curves obtained 
in the framework of the CASE project \citep{Kaluzny2004}. Our program stars were observed in the interval 
from 1999 February 6/7 to 2000 August 9/10. CCD photometry was performed using the 1.0-m Swope telescope at Las 
Campanas Observatory over 59 nights. For our target stars, the number of observations ranges from 590 to 739 
in the $V$ filter for NV326 and V194, respectively, and from 145 to 190 in the $B$ filter for V237 and V220, respectively.
The total time span of observations is 550 days, corresponding to a Rayleigh resolution of $0.0018\,d^{-1}$,
except for observations in the $B$ filter for NV326 and V237, where the time span is 479 days.
An example of the light curve for NV326 is shown in the top panel of Fig.\,B1 in the Appendix\,B, 
The data were phased with the dominant frequency $\nu_1 = 17.572793$\,d$^{-1}$.
	
We performed the frequency analysis for a combined data set in the $B$  and $V$ filters.
Such approach allowed to increase the signal-to-noise ratio $S/N$. 
We used an algorithm based on the fast Fourier transforms for non-equally spaced data \citep[e.g.,][]{Leroy2012}
and proceeded the standard pre-whitening procedure. As a significance criterion of a given frequency peak, 
we chose $S/N = 4$. The noise $N$ was calculated as a mean amplitude in a 20 day window 
centered at the significant frequency before its extraction.
An example periodograms along with the spectral window for NV326 are shown in the panels from the second one
downward in Fig.\,B1 in Appendix B. The Rayleigh resolution for the CASE light curve is about 
0.0018\,d$^{-1}$ and with such accuracy the combinations and harmonics were determined.
With a few exceptions all frequencies found by \citet{Olech2005} (Olech2005) were confirmed.
In the Appendix\,B , we give  all significant frequencies determined from combined $B$ and $V$ data. 
The amplitudes and phases were fitted separately for each filter. We give also the value of $S/N$, possible 
harmonic/combination and the corresponding Olech2005's frequency. For V194 we did not find the two frequencies 
of Olech2005, i.e., their $f_2=43.7932$\,d$^{-1}$ and $f_4=22.1462$\,d$^{-1}$. The latter one was 
close to the detection limit in Olech2005. Besides, we found  $\nu_4=21.7557$\,d$^{-1}$ instead of Olech2005's 
$f_3=20.7471$\,d$^{-1}$. The relation between them is given in the last column of Table\,B1.
In the case of the other four stars, all Olech2005's frequencies were detected.

\section{Seismic modelling  with the Monte Carlo-based Bayesian analysis}

All evolutionary computations were performed with the 1D Warsaw-New Jersay code and nonadiabatic linear pulsation with Dziembowski code.
Both codes have already been described in Sect.\,2.2. We assume that the star is chemically homogeneous on the Zero Age Main Sequence,
i.e. the merger event has erased the chemical composition gradients.

We computed extensive grids of evolutionary and pulsational models for the five SX Phe stars in $\omega$ Cen, i.e., 
V194, V220, V225, V237 and NV326. To find their seismic models, we employed the Bayesian analysis based on Monte Carlo simulations.
The analysis is based on the Gaussian likelihood function as describe in, e.g., \citet{2005A&A...436..127J, 2006A&A...458..609D, 2017MNRAS.467.1433R,Jiang2021}
\begin{equation}
	{\cal L}(E|{\mathbf H})=\prod_{i=1}^n \frac1{\sqrt{2\pi\sigma_i^2}} \cdot
	{\rm exp} \left( - \frac{  ({\cal O}_i-{\cal M}_i)^2}{2\sigma_i^2}  \right),
\end{equation}
where ${\mathbf H}$ is the hypothesis that represents adjustable model and theory parameters. e.g., 
mass $M$, initial hydrogen abundance $X_0$ (or helium abundance $Y$), metallicity $Z$, initial rotational velocity $V_{\rm rot,0}$,
convective overshooting parameter $\alpha_{\rm ov}$ and the mixing length parameter $\alpha_{\rm MLT}$.
The evidence $E$ represents the calculated observables ${\cal M}_i$, %e.g.,  the effective temperature $T_{\rm eff}$, luminosity $L/L_{\odot}$, pulsational frequencies,
that can be directly compared with the observed parameters ${\cal O}_i$ determined with the errors $\sigma_i$.

\subsection{Fitting the two radial modes}

To construct seismic models of the selected SX Phe stars,
we used the following observations: the frequencies of two radial modes $\nu_1$ and $\nu_n$,
as well as the effective temperature $T_{\rm eff}$ and luminosity $L/L_{\odot}$.  In the case of four SX Phe stars,
V194, V220, V225 and NV326,  we fitted the two frequencies as the fundamental and first overtone modes,
whereas in the case of V237  -- the fundamental and second overtone modes.
Since we only have four observables, it is safe to determine a maximum of three parameters.
In the considered mass range, the models have no convective core at all or it is very small and only in models with masses from about 1.15. Therefore,  we set  $\alpha_{\rm ov}=0.0$. As for the mixing length parameter  $\alpha_{\rm MLT}$,  it was demonstrated by \citet{JDD2020,JDD2022,JDD2023}, that  constraints
on $\alpha_{\rm MLT}$ can be derived only if the photometric amplitude and phases are available in at least three
passbands. Because no such data are available for these SX Phe stars, we fixed the  $\alpha_{\rm MLT}$ parameter to be 1.0 in all computations. 
We were guided by the results obtained from extensive seismic modelling of the prototype SX Phe \citep{JDD2023}.
Besides, in this range of $T_{\rm eff},~L$ and low metallicites, the frequencies of low order radial modes 
are weakly dependent on the envelope convection. We demonstrated this in Fig.\,3 which shows the Petersen diagram 
computed for models with $M=1.05$\,M$_{\odot}$, three values of $\alpha_{\rm MLT}=0.5, ~1.0, ~1.8$ 
and two values of metallicity, $Z=0.002$ and $Z=0.0002$. As can be seen, the effect of $\alpha_{\rm MLT}$ becomes important only for low frequencies
which correspond to $\log (T_{\rm eff}/K)<3.83$. None of the stars we analyzed has such a low effective temperature.
\begin{figure}
	\includegraphics[width=88mm,height=10.5cm,clip]{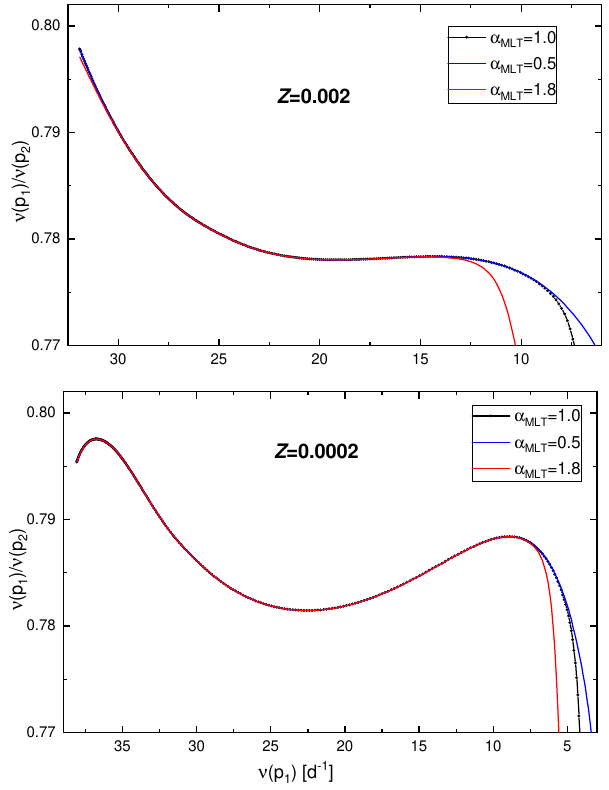}
	\caption{The effect of the mixing length parameter $\alpha_{\rm MLT}$ on the Petersen diagram showing the frequency ratio of radial fundamental to first overtone modes as a function of the fundamental-mode frequency. The models were computed for a mass $M=1.05$\,M$_{\odot}$, OPAL tables and AGSS09 mixture assuming the zero-rotation. Two values of metallicity were considered:
		$Z=0.002$ (top panel) and $Z=0.0002$ (bottom panel). The scale on OY axes is the same on both plots whereas the scale on OX axes is adapted to the frequency range.}
	\label{Petersen_aMLT}
\end{figure}
\begin{figure*}
	\includegraphics[width=88mm,height=6.7cm,clip]{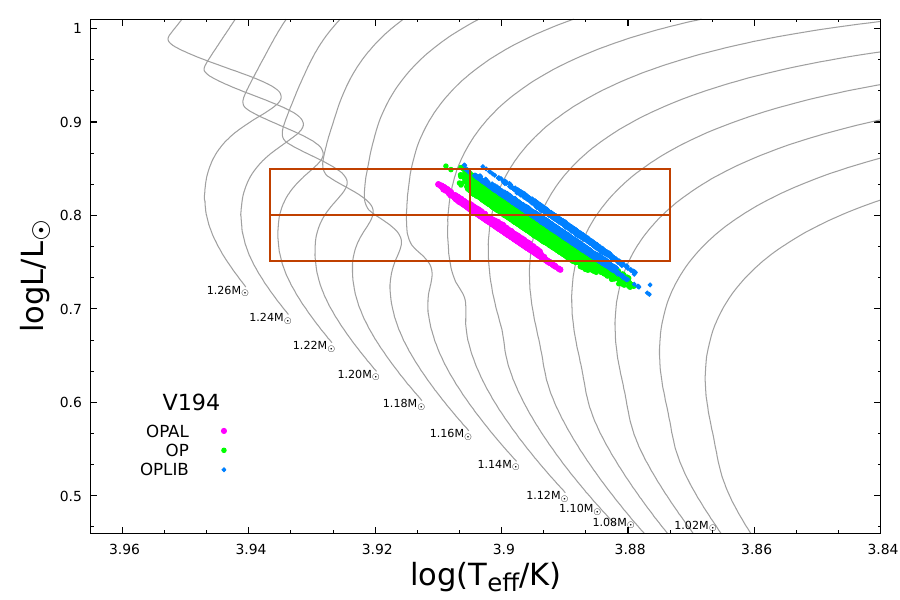}
	\includegraphics[width=88mm,height=6.7cm,clip]{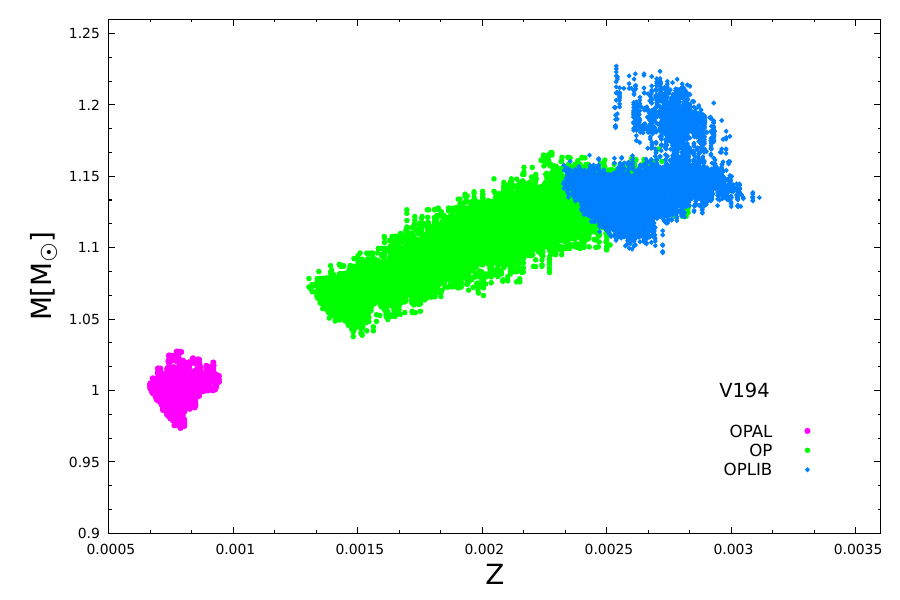}
	\caption{Left panel: the position of seismic models of V194 on the HR diagram computed with the three opacity tables: OPAL, OP and OPLIB and the helium abundance $Y=0.30$. All models reproduce the frequency of two radial modes: fundamental and first overtone. The evolutionary tracks are displayed for guidance only.
		Right panel: the position of these seismic models on the $M$ vs $Z$ plane. }
\label{_HR_5stars}
\end{figure*}
\begin{figure*}
	\includegraphics[width=88mm,height=6.7cm,clip]{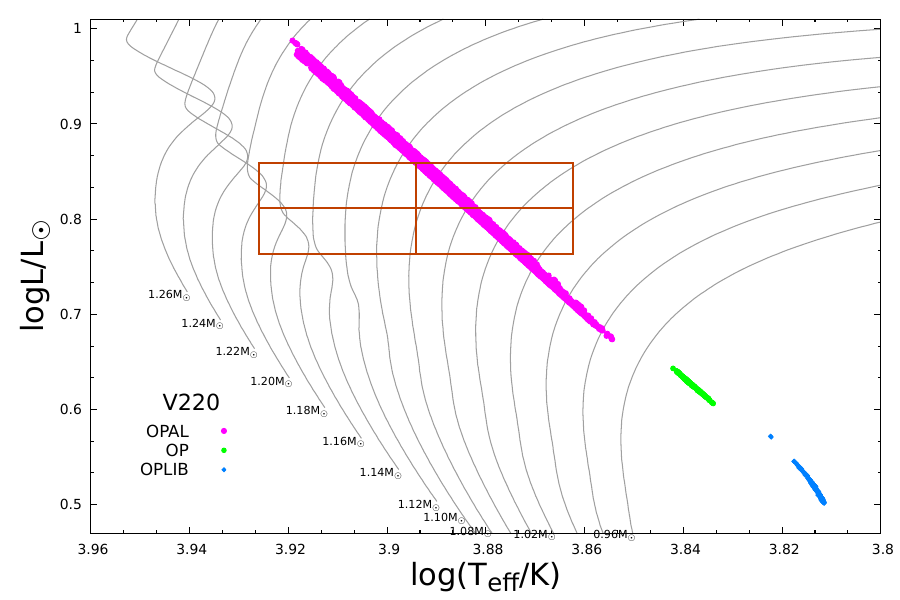}
	\includegraphics[width=88mm,height=6.7cm,clip]{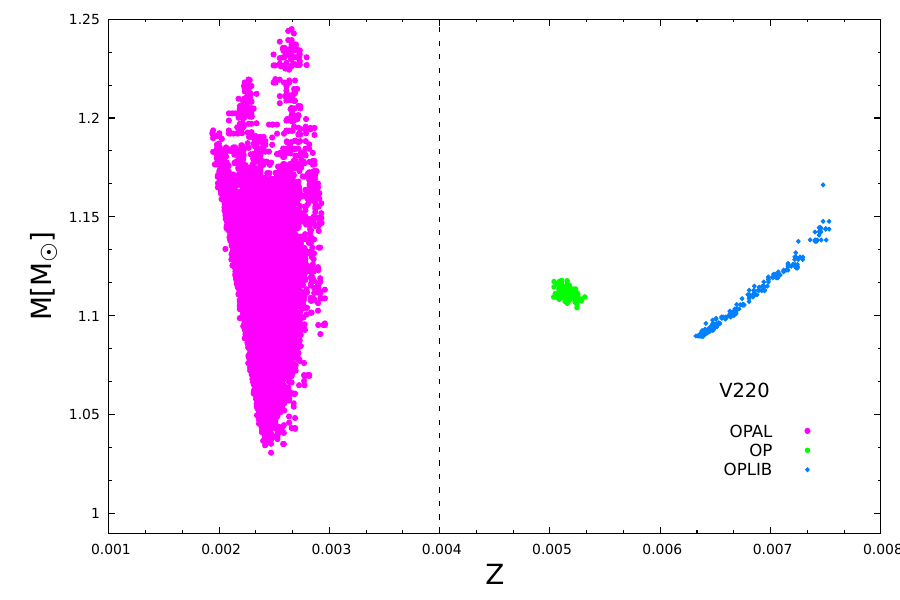}
	\caption{The same as in Fig.\,4 but for V220. The vertical line in the right panel marks the maximum metallicity derived for $\omega$ Centauri. }
	\label{_HR_5stars}
\end{figure*}
\begin{figure*}
	\includegraphics[width=88mm,height=6.7cm,clip]{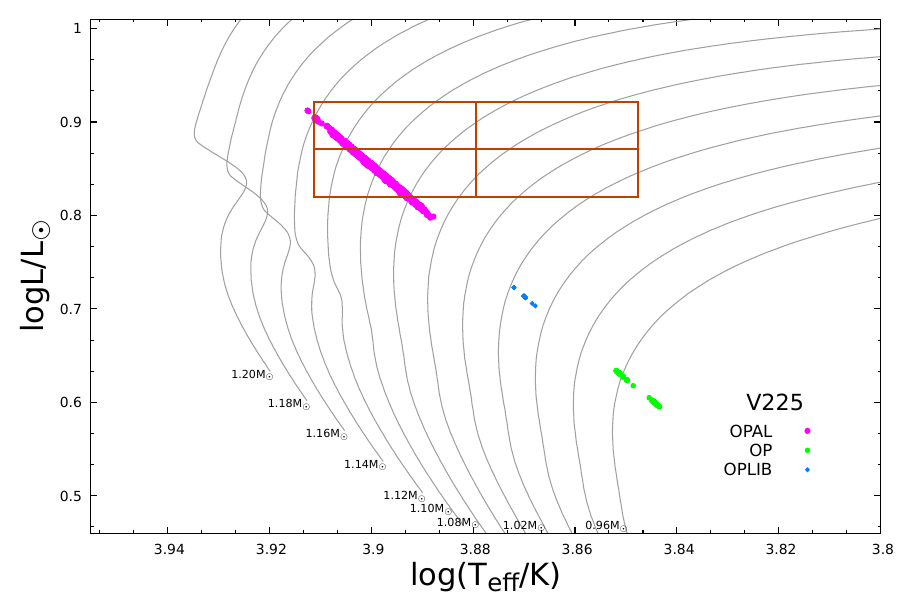}
	\includegraphics[width=88mm,height=6.7cm,clip]{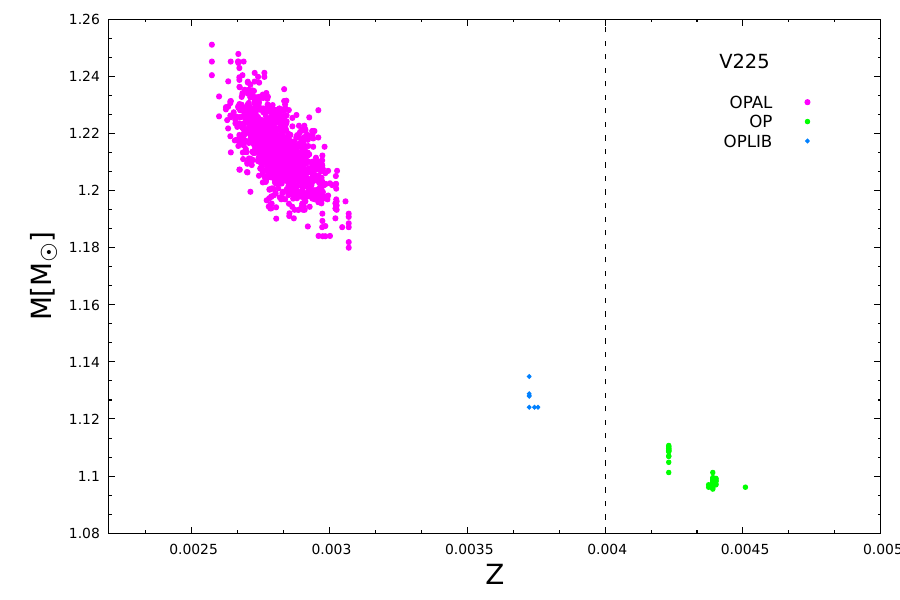}
	\caption{The same as in Fig.\,4 but for V225.}
	\label{_HR_5stars}
\end{figure*}
\begin{figure*}
	\includegraphics[width=88mm,height=6.7cm,clip]{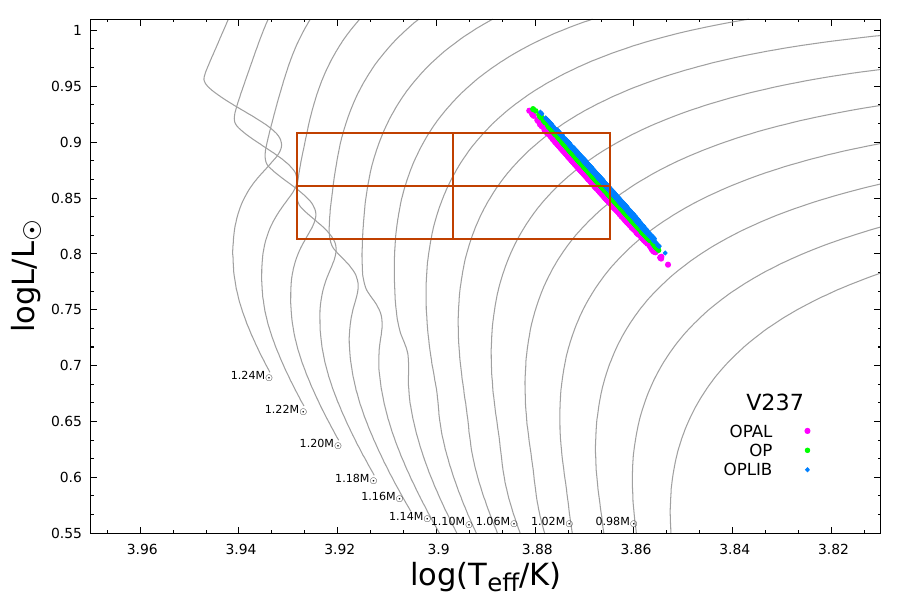}
	\includegraphics[width=88mm,height=6.7cm,clip]{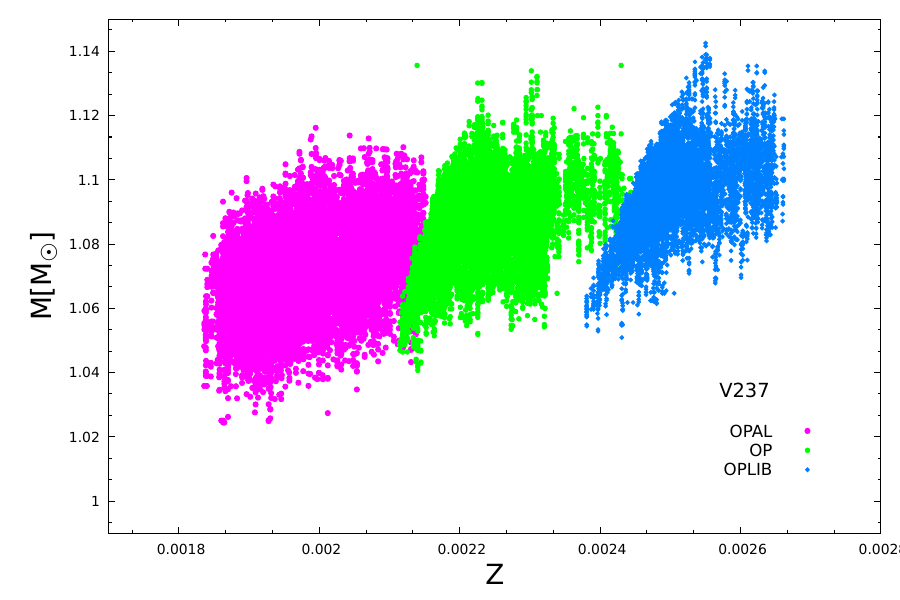}
	\caption{The same as in Fig.\,4 but for V237 and in this case the radial fundamental and second overtone were fitted.}
	\label{_HR_5stars}
\end{figure*}
\begin{figure*}
	\includegraphics[width=88mm,height=6.7cm,clip]{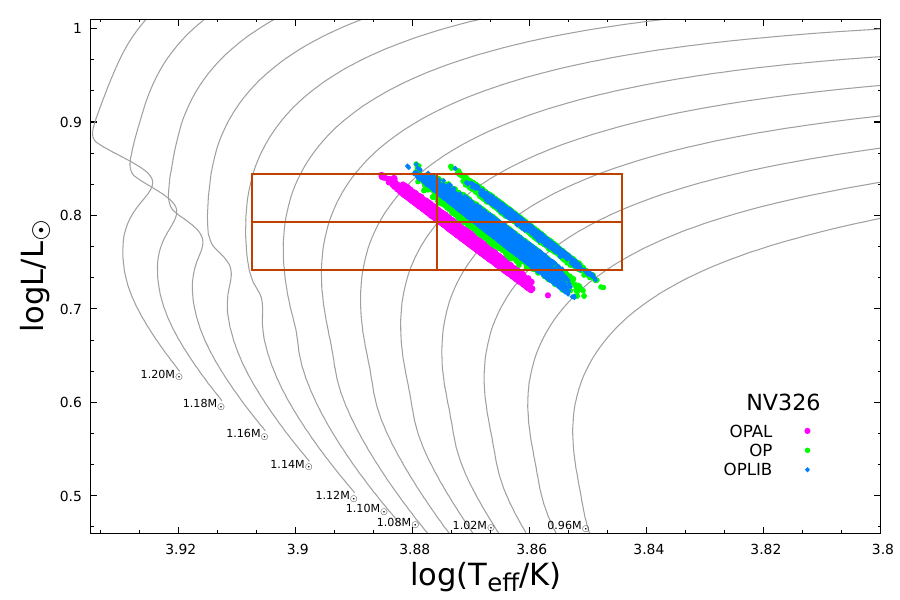}
	\includegraphics[width=88mm,height=6.7cm,clip]{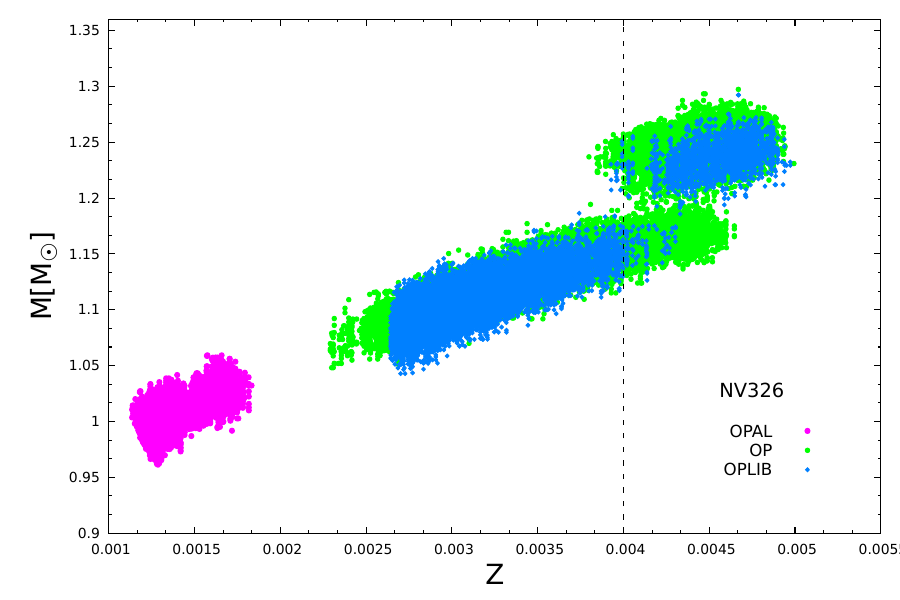}
	\caption{The same as in Fig.\,4 but for NV326.}
	\label{HR_5stars}
\end{figure*}

So, we were left with four parameters: mass $M$, helium abundance $Y$, metallicity $Z$ and initial rotational velocity $V_{\rm rot,0}$. 
We decided to fix the helium abundance and considered the two values $Y=0.30$
and $Y=0.33$. The higher helium abundance than the primordial seems plausible if the studied stars are merges.
To check the effect of lower value of $Y$, we also ran the simulations for $Y=0.27$ 
in the case of one star V194.
Then, for each randomly selected  set of parameters $(M,~Z,~V_{\rm rot,0})$, we calculated evolutionary and pulsational models.
Moreover, in the case of three stars, V220, V225 and NV326, for which the values of $V_{\rm rot}\sin i$ 
were derived from spectroscopic data, we could set the lower limit of the current rotational velocity, i.e., $V_{\rm rot}>22\,\kms$ for V220, $V_{\rm rot}>20\,\kms$ for V225
and $V_{\rm rot}>31\,\kms$ for NV326. One more observational constraint was the metallicity range derived for $\omega$\,Cen. We accepted seismic models
with the metallicity in the range $Z\in (0.0001, 0.004)$.

To study the effect of opacity data, we used the three table sets widely used in astrophysics: 
OPAL  \citep{Iglesias1996}, OP \citep{Seaton2005} and OPLIB \citep{Colgan2015,Colgan2016}.
Our previous studies of double-mode radial $\delta$ Scuti stars and the prototype SX Phe 
have showed a strong sensitivity of seismic models to the adopted opacity data, e.g., \citet{JDD2023}.
We got a perplexing result that only the OPAL seismic models had effective temperatures and luminosities consistent with the observational determinations 
for all studied double-radial pulsators. Therefore, we performed seismic modelling of the five SX Phe stars in $\omega$ Cen with all three opacity data sets.

We made a huge number of simulations, from about 150\,000 to 200\,000 depending on the star,
to maximize the likelihood function given in Eq.\,2 and to get stable solutions. Simulations were conducted for 
three opacity tables and two values of the helium abundance, $Y=0.30$ and $Y=0.33$. In the case of V194,
three values of $Y$ were considered, i.e., 0.30, 0.33 and 0.27.
The results for $Y=0.30$ are presented in Figs.\,4, 5, 6, 7 and 8. The left panels show the HR diagram with the position of seismic models which reproduce the two frequencies as radial modes;
the fundamental and first overtone (V194, V220, V225 and NMV326) or the fundamental and second overtone (V237). The right panels show the dependence between  mass and metallicity of  seismic models.  
The vertical line marks the maximum value of metallicity for $\omega$ Cen.

In the case of V194, we found seismic models within the error box with all three opacity tables (Fig.\,4).  All OPAL seismic models are in the hydrogen-shell burning (HSB) phase. With the OP and OPLIB data, we found  seismic models both in HSB and MS phase of evolution, however the content of hydrogen in the centre is already quite low, 
about 8\%.  The OP and OPLIB seismic models  are more massive and have higher metallicity. Comparing to 
the OPAL models, they have about 2 times (the OP models) or even 3 times (OPLIB models) higher $Z$ values.

For two stars, V220 and V225, we found seismic models within the error box only with the OPAL data (Figs.\,5 and 6). 
In the case of V220, all OP and OPLIB seismic models have much too low values of the effective temperature and luminosity,
whereas in the case of V225, they have much too low luminosities.
Besides in the case of V220, both, the OP and OPLIB models  have too high metallicities for objects belonging 
to the $\omega$ Cen cluster. In the case of V225,  the OP models have $Z>0.004$.
In addition, the OP and OPLIB models of both stars have very low rotation with the median values of 
about $1.6 - 3.6\,\kms$, which are in contradiction with the observed projected rotational velocities, i.e., 
$V_{\rm rot}\sin i=22$ and 20\,$\kms$ for V220 and V225, respectively. 
Therefore, the OP and OPLIB seismic models are rejected due to both the parameters $(T_{\rm eff},~L/{\rm L}_{\odot})$ 
and metallicity, and rotational velocity.
The OPAL seismic models of V220  are in the phase of hydrogen-shell burning whereas the OPAL seismic models
of V225 reach the end of main sequence evolution with a central hydrogen abundance of about 2\%.

V237 is the only SX Phe for which we reproduced the two frequencies as the radial fundamental and second overtone.
We found the seismic models of this star within the error box with each opacity data (Fig.\,7) and all these models burn hydrogen in the shell.   As for the previous stars, the OPAL seismic models have the lowest metallicity.

For the last  star, NV326, we also found seismic models within the error box with all three opacity data sets (Fig.\,8).
The OPAL models are in the HSB phase of evolution.  With the OP and OPLIB data two solutions were obtained, i.e., 
HSB phase with lower  $(M,~Z)$ and MS phase with higher $(M,~Z)$. However,  the MS seismic models 
have too high values of $Z$. 

To check the effect of helium abundance, we also performed simulations for $Y=0.33$ with the OPAL tables
and in the case of V194 for $Y=0.33$ and 0.27.
In Table\,3, we give the median values for the parameters of the OPAL seismic models 
{computed for the fixed value of $Y$} for all five studied stars. 
We chose median because this is more informative statistic for skewed distributions or distributions with outliers, which sometimes happens for some parameters. 
Medians of the following parameters are given:  mass $M$, metallicity $Z$, the central hydrogen content $X_c$, 
age, the current rotational velocity $V_{\rm rot}$, as well as logarithmic values of the effective temperature
$T_{\rm eff}$, luminosity $L$ and surface gravity $g$. 
The errors were calculated as the 0.84 quantile minus the median and the median minus the 0.16 quantile. These quantiles correspond to estimates of values separated by 1 standard deviation from the mean value in the case of a normal distribution.

The seismic models with higher helium abundance have rather similar metallicity. The exception is V220,
for which models with $Y=0.30$ are in the HSB phases with $Z\approx 0.00255$ whereas models 
with $Y=0.33$ are in the MS phase with metallicity more than two times higher, $Z\approx 0.005$,
and too high  for object belonging to $\omega$ Cen. 
In the case of V225, the seismic models with $Y=0.33$ are in the HSB phases whereas the seismic models with $Y=0.30$ are in the MS phase, however their metallicities are quite close.
The seismic models of V194 with $Y=0.27$  are more massive comparing to $Y=0.30$ and $0.33$
and have higher metallicity. Their median value of age reaches the lowest value.

The median values of the same parameters for the OP and OPLIB seismic models computed with $Y=0.30$ are give in Table\,4.  We did not list the parameters for V220 and V225 because their effective temperature and luminosity are far beyond the error boxes.
    \begin{table*}    	
	\begin{center}
		\caption{The median values of parameters of the OPAL seismic models of the five SX Phe stars from the Monte Carlo simulations. Results for the two values of helium abundance $Y$ are given. }
		\small
		\begin{tabular}{cccccccccc}
			\hline
			star & $Y$ & $M$   &  $Z$  &  $X_c$  &   age   &  $V_{\rm rot}$  &  $\log(T_{\rm eff}/{\rm K})$ & $\log L/{\rm L}_{\odot}$   &  $\log g$  \\
			&     & [M$_\odot$] &   &         &  [Gyr]  &     [$\kms$]    &           &          & $g$ [cm$\cdot$ s$^{-2}$]    \\
			\hline
			\hline
			&&&&&&&&&   \\
			{\bf V194} & 0.30 & $0.996^{+0.010}_{-0.008}$ &  $0.00075^{+0.00005}_{-0.00004}$ &  0   &  $3.79^{+0.11}_{-0.13}$ &    $14.5^{+13.1}_{-8.6}$ &  $3.8999^{+0.0042}_{-0.0033}$ &  $0.785^{+0.019}_{-0.015}$ &  $4.201^{+0.001}_{-0.001}$ \\
			&&&&&&&&&   \\
			\vspace{0.1cm}
			&  0.33 & $0.963^{+0.014}_{-0.010}$ &  $0.00085^{+0.00012}_{-0.00007}$ &  0  &  $3.55^{+0.14}_{-0.18}$ &  $23.2^{+16.1}_{-16.0}$ &  $3.9031^{+0.0034}_{-0.0038}$ &  $0.788^{+0.016}_{-0.017}$ &  $4.196^{+0.001}_{-0.001}$  \\
			&&&&&&&&&   \\
			\vspace{0.1cm}
			&  0.27 & $1.079^{+0.018}_{-0.012}$ &  $0.00102^{+0.00018}_{-0.00008}$ &  0  &  $3.40^{+0.14}_{-0.20}$ &  $56.4^{+9.4}_{-6.1}$ &  $3.8966^{+0.0034}_{-0.0044}$ &  $0.793^{+0.015}_{-0.020}$ &  $4.211{+0.002}_{-0.001}$  \\		
			\hline
			\hline
			&&&&&&&&&   \\
			\multicolumn{10}{l}{{\bf V220} -  models with $V_{\rm rot}>22\kms$}\\
			& 0.30 & $1.132^{+0.033}_{-0.028}$ &  $0.00255^{+0.00012}_{-0.00010}$ &  0 &  $2.61^{+0.25}_{-0.26}$ &    $29.9^{+6.5}_{-3.2}$ &  $3.8863^{+0.0099}_{-0.010}$ &  $0.827^{+0.048}_{-0.047}$ &  $4.159^{+0.004}_{-0.003}$\\
			&&&&&&&&&   \\
			\vspace{0.1cm}
			& 0.33 & $1.263^{+0.029}_{-0.021}$ &  $0.00558^{+0.00075}_{-0.00041}$ &  $0.225^{+0.041}_{-0.027}$ &  $1.38^{+0.09}_{-0.11}$ &  $37.2^{+16.0}_{-12.6}$ &  $3.8772^{+0.0042}_{-0.0036}$ &  $0.818^{+0.021}_{-0.018}$ &  $4.179^{+0.003}_{-0.002}$\\
			\hline
			\hline
			&&&&&&&&&   \\
			\multicolumn{10}{l}{{\bf V225} -  models with $V_{\rm rot}>20\kms$}\\
			& 0.30 & $1.214^{+0.010}_{-0.009}$ &  $0.00282^{+0.00005}_{-0.00007}$ &  $0.020^{+0.003}_{-0.002}$ &  $1.87^{+0.05}_{-0.05}$ &  $29.2^{+7.2}_{-5.7}$ &  $3.8983^{+0.0033}_{-0.0031}$ &  $0.846^{+0.015}_{-0.015}$ &  $4.219^{+0.002}_{-0.001}$\\
			&&&&&&&&&  \\
			\vspace{0.1cm}
			& 0.33 & $1.114^{+0.049}_{-0.034}$ &  $0.00245^{+0.00078}_{-0.00044}$ &  0 &  $2.16^{+0.27}_{-0.29}$ &  $43.5^{+23.7}_{-12.5}$ &  $3.9027^{+0.0039}_{-0.0046}$ &  $0.839^{+0.019}_{-0.017}$ &  $4.202^{+0.007}_{-0.004}$ \\
			\hline
			\hline
			&&&&&&&&&  \\
			{\bf V237} &  0.30 & $1.071^{+0.013}_{-0.012}$ &  $0.00197^{+0.00010}_{-0.00006}$ &  0  &  $3.37^{+0.13}_{-0.13}$ &   $40.5^{+8.7}_{-6.2}$ &  $3.8678^{+0.0035}_{-0.0036}$ &  $0.862^{+0.017}_{-0.017}$ &  $4.024^{+0.002}_{-0.002}$ \\
			&&&&&&&&&   \\
			\vspace{0.1cm}
			& 0.33 &  $1.009^{+0.012}_{-0.012}$ &  $0.00177^{+0.00006}_{-0.00003}$ &  $0$  &  $3.41^{+0.15}_{-0.13}$ &   $15.2^{+11.4}_{-11.2}$ &  $3.8708^{+0.0034}_{-0.0037}$ &  $0.858^{+0.017}_{-0.019}$ &  $4.017^{+0.002}_{-0.002}$\\
			\hline
			\hline
			&&&&&&&&&   \\
			\multicolumn{10}{l}{{\bf NV326} -  models with $V_{\rm rot}>31\kms$}\\
			& 0.30 & $1.018^{+0.014}_{-0.013}$ &  $0.00153^{+0.00013}_{-0.00016}$ &  $0$ &  $3.79^{+0.16}_{-0.16}$ &  $45.5^{+7.1}_{-11.4}$ &  $3.8718^{+0.0039}_{-0.0038}$ &  $0.783^{+0.019}_{-0.017}$ &  $4.100^{+0.001}_{-0.001}$\\
			&&&&&&&&&  \\
			\vspace{0.1cm}
			& 0.33 & $0.973^{+0.010}_{-0.009}$ &  $0.00151^{+0.00004}_{-0.00004}$ &  0 &  $3.69^{+0.13}_{-0.13}$ &  $34.6^{+3.2}_{-2.8}$ &  $3.8754^{+0.0035}_{-0.0034}$ &  $0.785^{+0.017}_{-0.016}$ &  $4.091^{+0.001}_{-0.001}$ \\
			\hline
			\hline
		\end{tabular}
	\end{center}
	\label{median_OPAL}
	%\footnotesize $^{*}$ - models with $V_{\rm rot}>22\kms$ for V220 and with $V_{\rm rot}>20\kms$ for V225.
\end{table*}	

    \begin{table*}
	\begin{center}
		\caption{The median values of parameters of the OP and OPLIB seismic models of the three SX Phe stars from the Monte Carlo simulations. The helium abundance was $Y=0.30$.}
		\small
		\begin{tabular}{cccccccccc}
			\hline
			star & models & $M$   &  $Z$  &  $X_c$  &   age   &  $V_{\rm rot}$  &  $\log(T_{\rm eff}/{\rm K})$ & $\log L/{\rm L}_{\odot}$   &  $\log g$  \\
			&     & [M$_\odot$] &   &         &  [Gyr]  &     [$\kms$]    &           &          & $g$ [cm$\cdot$ s$^{-2}$]    \\
			\hline
			\hline
			V194 &  OP--HSB  & $1.118^{+0.023}_{-0.024}$ &  $0.00215^{+0.00034}_{-0.00031}$ &           $0$             &  $2.53^{+0.21}_{-0.21}$ &  $45.1^{+8.3}_{-9.0}$  &  $3.8941^{+0.0037}_{-0.0040}$ &  $0.790^{+0.016}_{-0.018}$ &  $4.220^{+0.003}_{-0.003}$ \\
			\vspace{0.1cm}
			V194 &  OP--MS   & $1.196^{+0.008}_{-0.008}$ &  $0.00268^{+0.00010}_{-0.00011}$ & $0.083^{+0.024}_{-0.020}$ &  $1.85^{+0.05}_{-0.06}$ &  $61.8^{+2.0}_{-1.4}$  &  $3.8912^{+0.0025}_{-0.0035}$ &  $0.796^{+0.011}_{-0.016}$ &  $4.229^{+0.001}_{-0.001}$ \\
			
			V194 & OPLIB--HSB & $1.136^{+0.011}_{-0.011}$ &  $0.00257^{+0.00021}_{-0.00010}$ &           $0$             &  $2.29^{+0.08}_{-0.09}$ & $14.8^{+17.4}_{-10.3}$ &  $3.8926^{+0.0037}_{-0.0038}$ &  $0.790^{+0.017}_{-0.018}$ &  $4.224^{+0.001}_{-0.001}$ \\
			
			V194 & OPLIB--MS  & $1.189^{+0.013}_{-0.012}$ &  $0.00276^{+0.00009}_{-0.00013}$ & $0.089^{+0.019}_{-0.026}$ &  $1.82^{+0.07}_{-0.07}$ &  $39.5^{+1.9}_{-3.3}$  &  $3.8900^{+0.0035}_{-0.0035}$ &  $0.791^{+0.015}_{-0.016}$ &  $4.231^{+0.001}_{-0.001}$ \\
			
			\hline
			\hline
			\vspace{0.1cm}
			V237 &  OP   & $1.085^{+0.013}_{-0.012}$ &  $0.00222^{+0.00008}_{-0.00006}$ &  $0$  &  $3.29^{+0.13}_{-0.13}$ &  $19.7^{+12.1}_{-12.7}$ & $3.8669^{+0.0034}_{-0.0035}$ & $0.863^{+0.017}_{-0.017}$ & $4.028^{+0.002}_{-0.002}$ \\
			\vspace{0.1cm}
			V237 & OPLIB & $1.095^{+0.012}_{-0.011}$ &  $0.00249^{+0.00008}_{-0.00004}$ &  $0$  &  $3.10^{+0.11}_{-0.11}$ &  $15.0^{+11.9}_{-9.6}$  & $3.8659^{+0.0031}_{-0.0031}$ & $0.862^{+0.016}_{-0.015}$ & $4.030^{+0.002}_{-0.002}$ \\
			\hline
			\hline
			NV326 &    OP--HSB & $1.136^{+0.026}_{-0.028}$ &  $0.00352^{+0.00058}_{-0.00050}$ &         $0$                &  $2.73^{+0.24}_{-0.19}$ &  $61.7^{+9.4}_{-9.5}$ &  $3.8652^{+0.0038}_{-0.0038}$ &  $0.783^{+0.017}_{-0.017}$ &  $4.114^{+0.002}_{-0.003}$  \\
			
			\vspace{0.1cm}
			
			NV326 &    OP--MS  & $1.243^{+0.015}_{-0.015}$ &  $0.00447^{+0.00024}_{-0.00026}$ &  $0.111^{+0.021}_{-0.030}$ &  $1.88^{+0.09}_{-0.08}$ &  $67.9^{+2.4}_{-2.9}$ &  $3.8610^{+0.0037}_{-0.0038}$ &  $0.791^{+0.017}_{-0.019}$ &  $4.128^{+0.002}_{-0.002}$ \\

			NV326 & OPLIB--HSB & $1.118^{+0.019}_{-0.016}$ &  $0.00327^{+0.00034}_{-0.00015}$ &         $0$                &  $2.77^{+0.14}_{-0.16}$ &  $38.9^{+9.6}_{-5.5}$ &  $3.8656^{+0.0037}_{-0.0037}$ &  $0.784^{+0.018}_{-0.018}$ &  $4.114^{+0.002}_{-0.002}$ \\
			
			NV326 & OPLIB--MS  & $1.238^{+0.015}_{-0.016}$ &  $0.00462^{+0.00016}_{-0.00028}$ &  $0.119^{+0.013}_{-0.030}$ &  $1.83^{+0.09}_{-0.08}$ &  $61.7^{+2.3}_{-4.1}$ &  $3.8601^{+0.0034}_{-0.0033}$ &  $0.787^{+0.017}_{-0.016}$ &  $4.128^{+0.002}_{-0.002}$  \\
			\hline
			\hline
		\end{tabular}
	\end{center}
	\label{median_all_kappa}
	%\footnotesize $^{*}$ - models with $V_{\rm rot}>22\kms$ for V220 and with $V_{\rm rot}>20\kms$ for V225.
\end{table*}
Regardless of the adopted opacity tables, in all seismic models radial modes up to 3rd radial order were unstable (i.e., excited).
In the Appendix\,C, we show histograms for the following parameters of the OPAL seismic models: 
mass $M$,  metallicity $Z$, age, current rotation $V_{\rm rot}$, radius $R$ and the instability parameter $\eta$. 
The positive values of $\eta$ mean unstable, i.e., excited modes. This parameter will be discussed in more detail in Sect.\,3.3.
	
\subsection{Adding the non-radial mode for V194}

The second independent frequency which appeared in the $BV$ light curve of V194 
is $\nu_4=21.75057$\,d$^{-1}$.  Its amplitude in both passbands is larger than the amplitude 
of frequency 27.16331\,d$^{-1}$, considered as the radial first overtone.
After radial modes, dipole modes have the greatest visibility in photometry if 
intrinsic mode amplitudes are comparable.

Therefore,  in this subsection we considered $\nu_4$ as the $\ell=1$ mode and  checked
how adding a non-radial mode will narrow the range of parameters in our seismic modeling.
We considered all azimuthal orders $m=-1,~0,~+1$ and fitted the theoretical frequencies
resulting from the first-order splitting formulae:
$$\nu_{m}=\nu_0+m(1-C_{n\ell})\nu_{\rm rot}$$
where $\nu_{\rm rot}$ is the rotational frequency and $C_{n\ell}$ is the Ledoux constant, which depends on model parameters, the mode degree $\ell$ and radial order $n$. 

Having an additional observable, i.e., the frequency of a non-radial mode, we were able to determine one more parameter from seismic modelling. Now our adjustable parameters are: $M, ~Y, ~Z$ and  $V_{\rm rot}$, where the rotational velocity is related to the rotational  frequency, i.e.,  $\nu_{\rm rot}=V_{\rm rot}/(2\pi R)$.

We performed about 100\,000 MC simulations for each azimuthal order, i.e., $m=-1,~  0,~+1$, assuming the OPAL opacities. We found  seismic models 
that have effective temperatures and luminosities consistent with the observed 
error box for all three azimuthal orders $m$. All seismic models are undergoing hydrogen shell burning.

The median values of the seismic models fitting the frequencies of three modes, i.e.,
two radial modes and one dipole mode are  given in Table\,5.
\begin{table*}    	
	\begin{center}
		\caption{Median values of the parameters of the OPAL seismic models of V194 which
			reproduce the three frequencies: of two radial modes and of one dipole mode with different azimuthal order $m$ indicated in the first column.}
		\small
		\begin{tabular}{cccccccccc}
			\hline
 $m$ &  $M$   &$Y$ &  $Z$  &  $X_c$  &   age   &  $V_{\rm rot}$  &  $\log(T_{\rm eff}/{\rm K})$ & $\log L/{\rm L}_{\odot}$   &  $\log g$  \\
			 &  [M$_\odot$] &  & &         &  [Gyr]  &     [$\kms$]    &           &          & $g$ [cm$\cdot$ s$^{-2}$]    \\
			\hline
			\hline
%			&&&&&&&&   \\			
			\vspace{0.1cm}
			%      \hline
         	$-1$ & $0.978^{+0.002}_{-0.003}$ & $0.3009^{+0.0013}_{-0.0008}$ &    $0.000795(3)$ &  0 &  $3.993^{+0.014}_{-0.015}$ &
			$13.3^{+0.7}_{-1.1}$ & $3.8931(3)$ &  $0.752^{+0.002}_{-0.001}$ & $4.1991(4)$  \\
			\vspace{0.1cm}
			$~~~0$ & $0.983^{+0.002}_{-0.004}$ & $0.2988^{+0.0003}_{-0.0004}$ &    $0.000786(1)$ &  0 &  $3.974^{+0.047}_{-0.037}$ &
			$13.4^{+0.6}_{-0.4}$ & $3.8940(2)$ &  $0.757^{+0.003}_{-0.008}$ & $4.1998(5)$  \\
			\vspace{0.1cm}
			$+1$ & $1.010^{+0.004}_{-0.001}$ & $0.3018^{+0.0012}_{-0.0023}$ &    $0.000695(1)$ &  0 &  $3.533^{+0.034}_{-0.012}$ &
			$20.8^{+0.9}_{-1.9}$ & $3.9094(10)$ &  $0.827^{+0.006}_{-0.003}$ & $4.2031(6)$  \\			
			\hline
			\hline
		\end{tabular}
\end{center}
\label{median_OPAL_l1}
\end{table*}	

As can be seen, adding a dipole mode to the seismic modelling allowed us to constrain the parameters much more tightly. The uncertainties of determined parameters decreased by one order of magnitude. 
For example, with the $\ell=1$ mode added the mass is determined to the third decimal place, the metallicity 
to the sixth decimal place and the age (expressed in Gyr) to the second decimal place.
However, we cannot make a difference between different azimuthal order $m$ 
of the dipole mode as the seismic models have acceptable parameters in each case.
Nevertheless,  independently of $m$ the abundance of helium amounts to about $Y\approx 0.30$, i.e., 
it is  higher that primordial.
Of course, treating $\nu_4=21.75057$\,d$^{-1}$  as a dipole mode is only a hypothesis, although 
it is quite plausible. However, our goal was to demonstrate the enormous potential of non-radial modes in seismic modelling.

The HR diagram with the position of seismic models is given in the Appendix\,C. As an example, we also showed
histograms for the parameters $M,~Z, ~Y$ and age for the seismic models fitting the two radial modes and dipole retrograde mode ($\ell=1,~m=-1$).

\subsection{Properties of pulsational modes in SX Phe star models}

As mentioned in Sect. 3.1, the radial modes considered for the five SX Phe stars are globally excited
in the seismic models computed with each opacity data. All OPAL models burn hydrogen in the shell. The frequency range of unstable modes for 
a representative seismic model of V194 is presented in the top panel of Fig.\,9, where the instability 
parameter $\eta$ is plotted as the frequency of modes with the harmonic degree $\ell=0,~1,~2$. 
The V194 seismic model reproduces the two radial modes, fundamental and first overtone, and has the following parameters: 
$Y=0.30$, $Z=0.0008$, $M=0.978$ M$_{\odot}$, $\log (T_{\rm eff}/K)=3.8913$, $\log L/{\rm L}_{\odot}=0.747$ and $V_{\rm rot}=18.3~\kms$.  
The parameter $\eta$ is a normalized work integral computed over the whole stellar interior. For globally excited modes the values of $\eta$ are greater than zero.

As one can see, the pulsational instability in the considered model occurs in the range (18,~53)\,d$^{-1}$, for all mode degrees.
The median values of $\eta$ for fundamental, first and second overtones are in the range of about (0.05,~0.10)
for all seismic models of the analysed SX Phe stars. In Appendix C, we have included the $\eta$ histograms for radial fundamental modes in Figs\,C1-C5. 
Histograms for the first overtone (V194, V220, V225, NV326) and the second overtone (V237) look similar.
Additionally, in Figs.\,C6 and C7 we show how the parameter $\eta$ depends on mass and age for seismic models of two example stars, V220 and V237.
The left panels are for the radial fundamental modes whereas the right panels for the first (V220) or second overtone modes (V237). The metallicity is coded with colours.
As one can see, the instability of radial fundamental and first overtone decreases with the mass and increases with the age of seismic models.
On the other hand, the second overtone instability has an inverse dependence, i.e. it increases with mass and decreases with age.
%Seismic models of V220 are younger, hotter and more massive.
%
 \begin{figure}
	\includegraphics[width=88mm,height=11.5cm,clip]{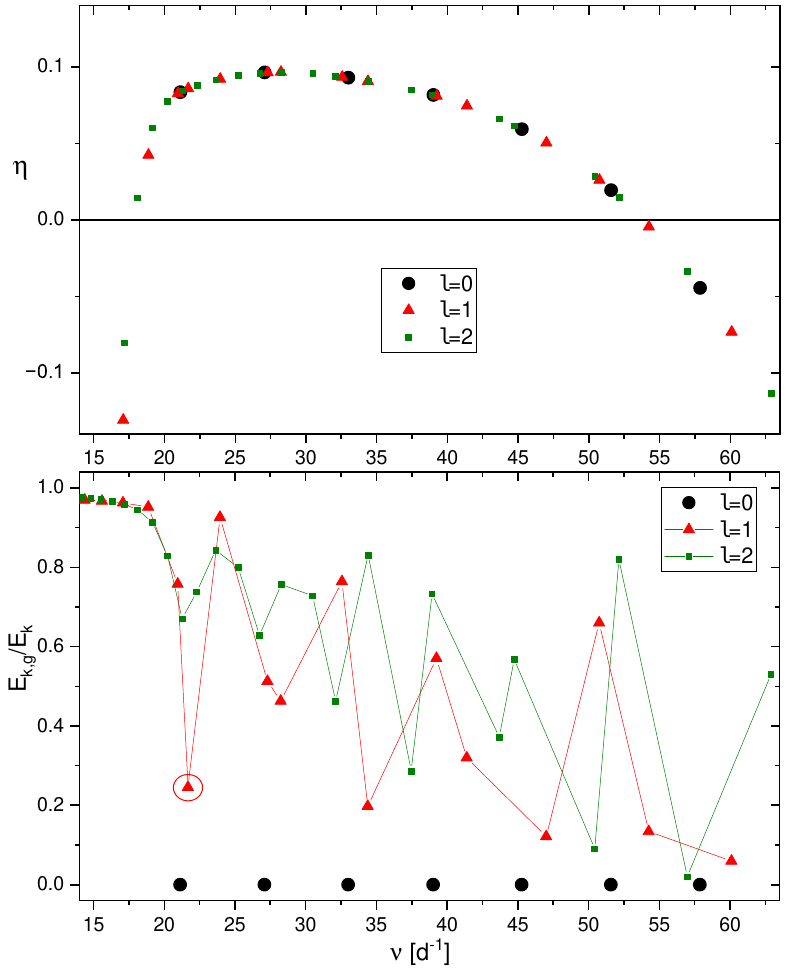}
	\caption{Top panel: the normalized instability parameter $\eta$ as a function of the frequency for modes with $\ell=0,1,2$, for the OPAL seismic model of V194.
	The model parameters are: $Y=0.30$, $Z=0.0008$, $M=0.978$ M$_{\odot}$, $\log (T_{\rm eff}/K)=3.8913$, $\log L/{\rm L}_{\odot}=0.747$ and $V_{\rm rot}=18.3~\kms$.
	 Bottom panel: the corresponding values of the ratio $E_{\rm k,g}/E_{\rm k}$.}
	\label{fig:eta_Ekg}
\end{figure}

In the bottom panel of Fig.\,9, we provided the corresponding values of the ratio of kinetic energy in gravity propagation zone to the total kinetic energy 
of the mode, $E_{\rm k,g}/E_{\rm k}$. Modes with $E_{\rm k,g}/E_{\rm k}=0.0$ are pure pressure and those with $E_{\rm k,g}/E_{\rm k}=1.0$ are pure gravity. 
As one can see, all nonradial modes with $\nu>20$\,d$^{-1}$ have a mixed character and those with $\nu<20$\,d$^{-1}$ are pure gravity. This is because the model 
is in the hydrogen-shell burning phase.
The dipole mode corresponding to the observed frequency $\nu_4=21.75057$\,d$^{-1}$ of V194, used in the seismic modelling in the previous subsection,
is marked by circle in the bottom panel of Fig.\,9. This is the $g_6$ mode which is globally unstable $\eta=0.086$ and has $E_{\rm k,g}/E_{\rm k}=0.25$. 
Its energetic properties are depicted in Fig.\,10. The left Y axis contains the values of pressure eigenfunction, $\delta P/P$, which should be large 
and slowly varying in the driving zone for a given mode to be unstable. The right Y axis shows the differential work integral, $d\log W/d\log T$.
The main driving of the $\ell=1,~g_6$ mode occurs in the HeII partial ionization zone with a small contribution from the H ionization zone and the Z-bump. 
However, below the Z-bump a strong damping takes place. The inset in Fig.\,10 shows a distribution of the kinetic energy of the mode, $d\log E_{\rm k}/d\log T$. 
A lot of kinetic energy accumulates in the central parts, where a steep gradient of chemical composition has developed and
the Brunt-V\"ais\"al\"a frequency reaches very high values.
 \begin{figure}
	\includegraphics[width=88mm,height=7cm,clip]{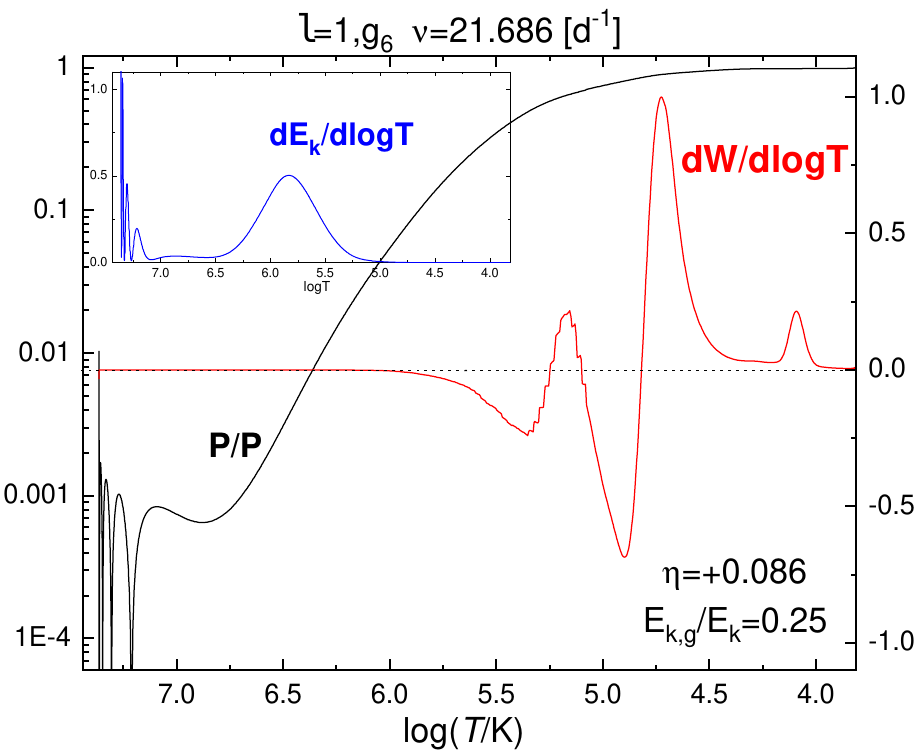}
	\caption{Properties of the $\ell=1,~g_6$ mode in the same OPAL seismic model of V194 as in Fig.\,9. The left Y axis shows the values of pressure eigenfunction 
		and the right Y axis - the values of differential work integral. The inset shows differential kinetic energy.}
	\label{fig:l1g6}
\end{figure}
\section{Discussion}
\begin{figure}
	\includegraphics[width=88mm,height=10.7cm,clip]{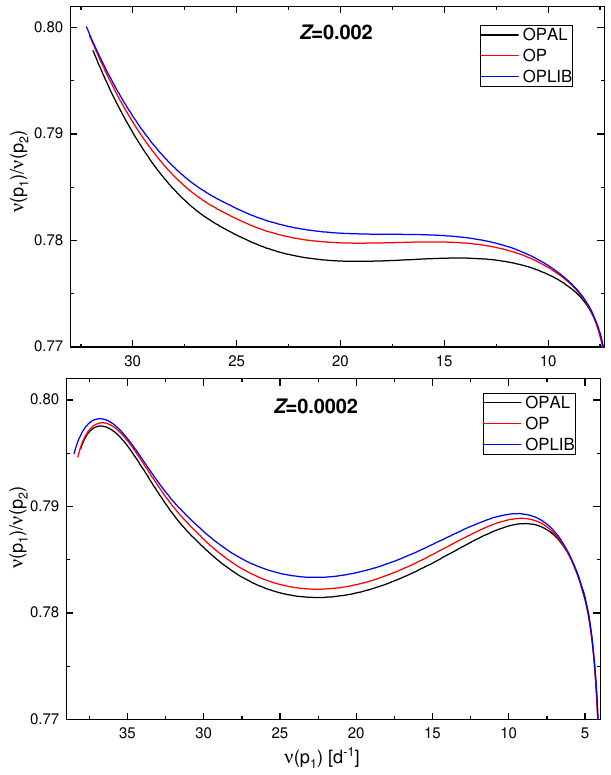}
	\caption{The effect of the adopted opacity tables on the Petersen diagram showing the frequency ratio of radial fundamental to first overtone as a function of the fundamental mode frequency. The models were computed for a mass $M=1.05$\,M$_{\odot}$, mixing length parameter $\alpha_{\rm MLT}=1.0$ and AGSS09 mixture assuming the zero-rotation. 
		Two values of metallicity were considered: 	$Z=0.002$ (top panel) and $Z=0.0002$ (bottom panel). The scale on OY axes is the same on both plots whereas the scale on OX axes is adapted to the frequency range.}
	\label{fig:Petersen_kappaZ}
\end{figure}
\begin{figure}
	\includegraphics[width=88mm,height=7cm,clip]{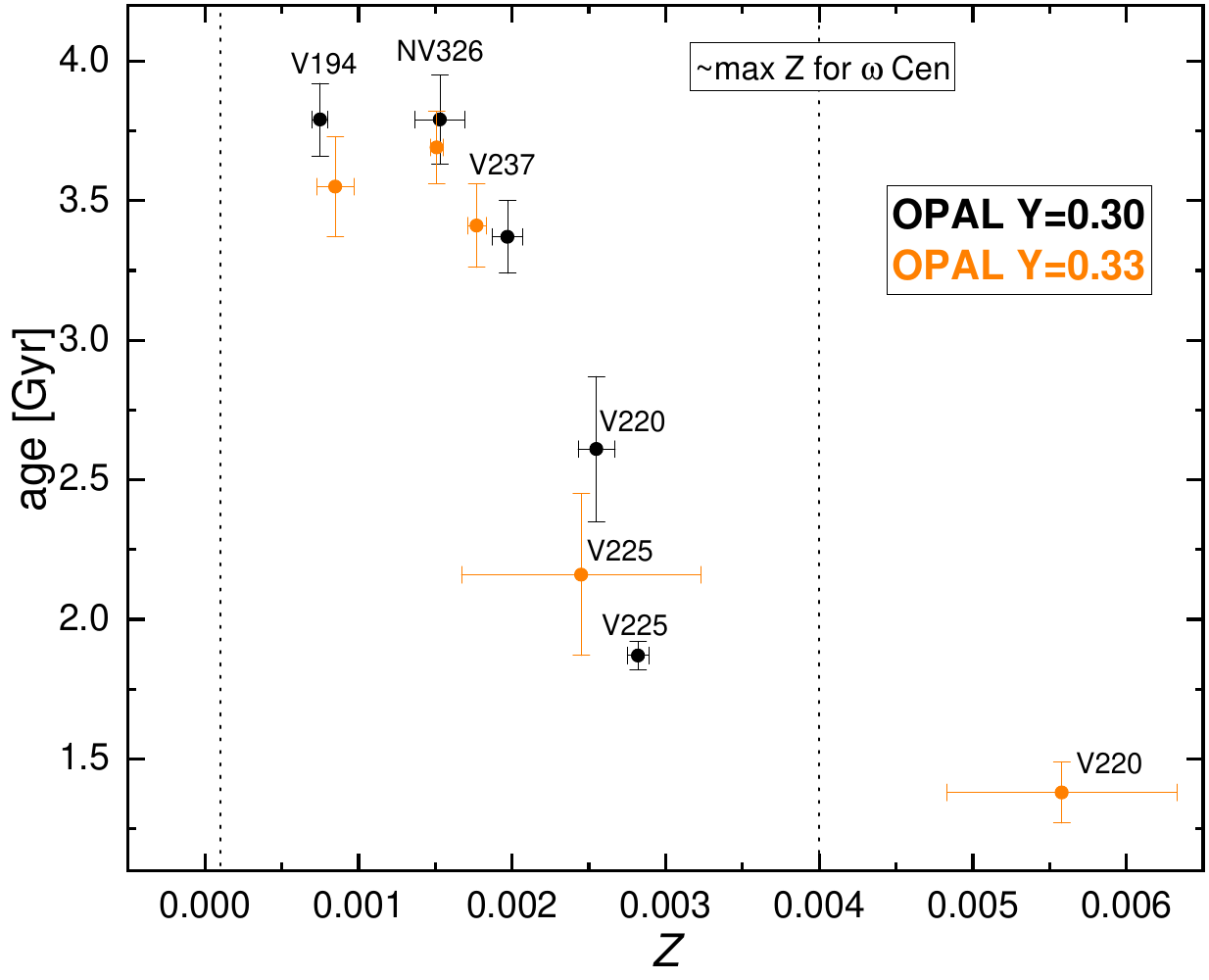}
	\caption{The relation between median values of the age and metallicity derived from the OPAL seismic models of the five SX Phe stars. The vertical dotted lines show the range of metallicity observed in $\omega$ Cen. Results for two values of the helium abundance are shown $Y=0.30$ and 0.33.}
	\label{age_Z_OPAL}
\end{figure}

\begin{figure*}
	\includegraphics[width=88mm,height=7cm,clip]{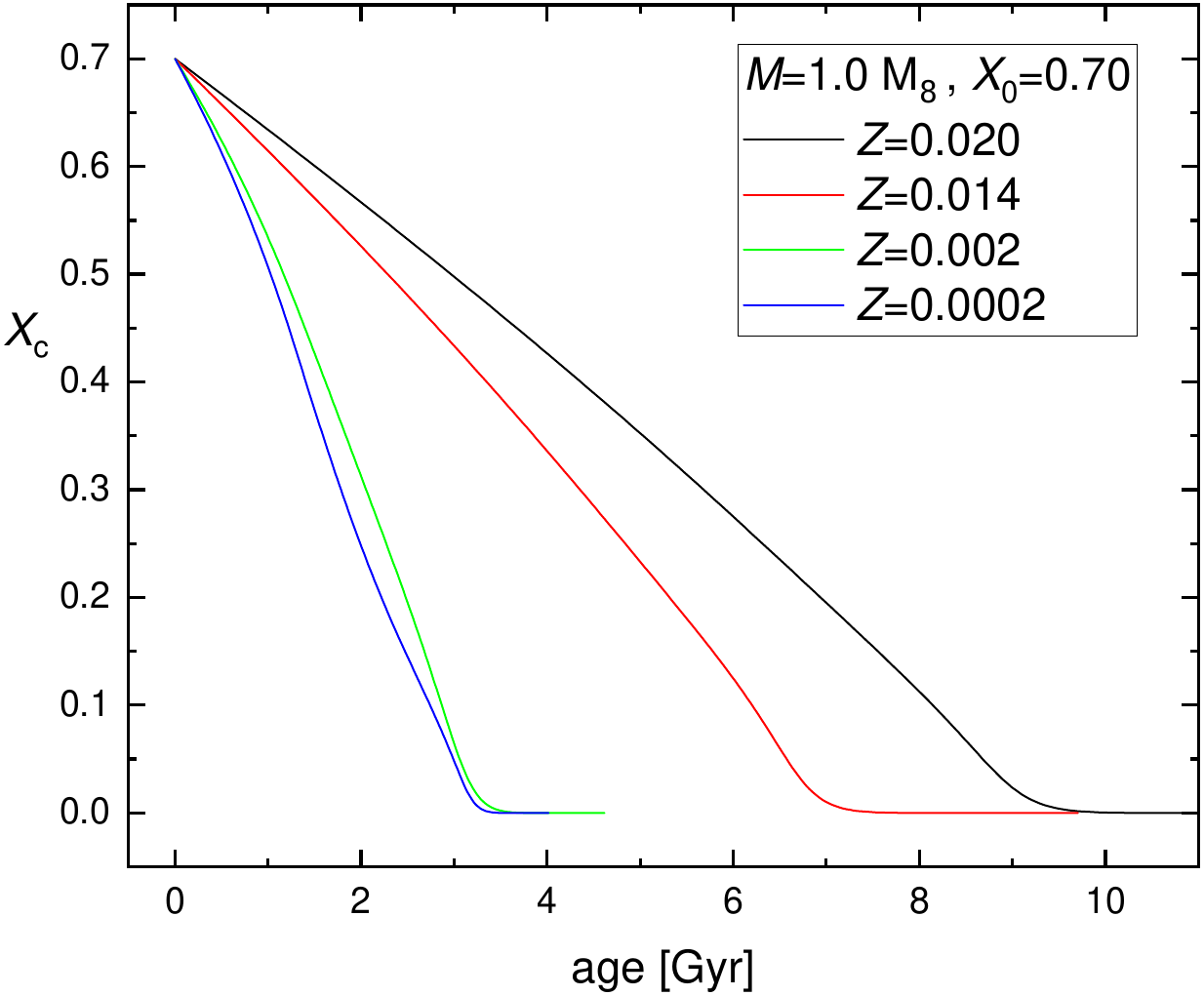}
	\includegraphics[width=88mm,height=7cm,clip]{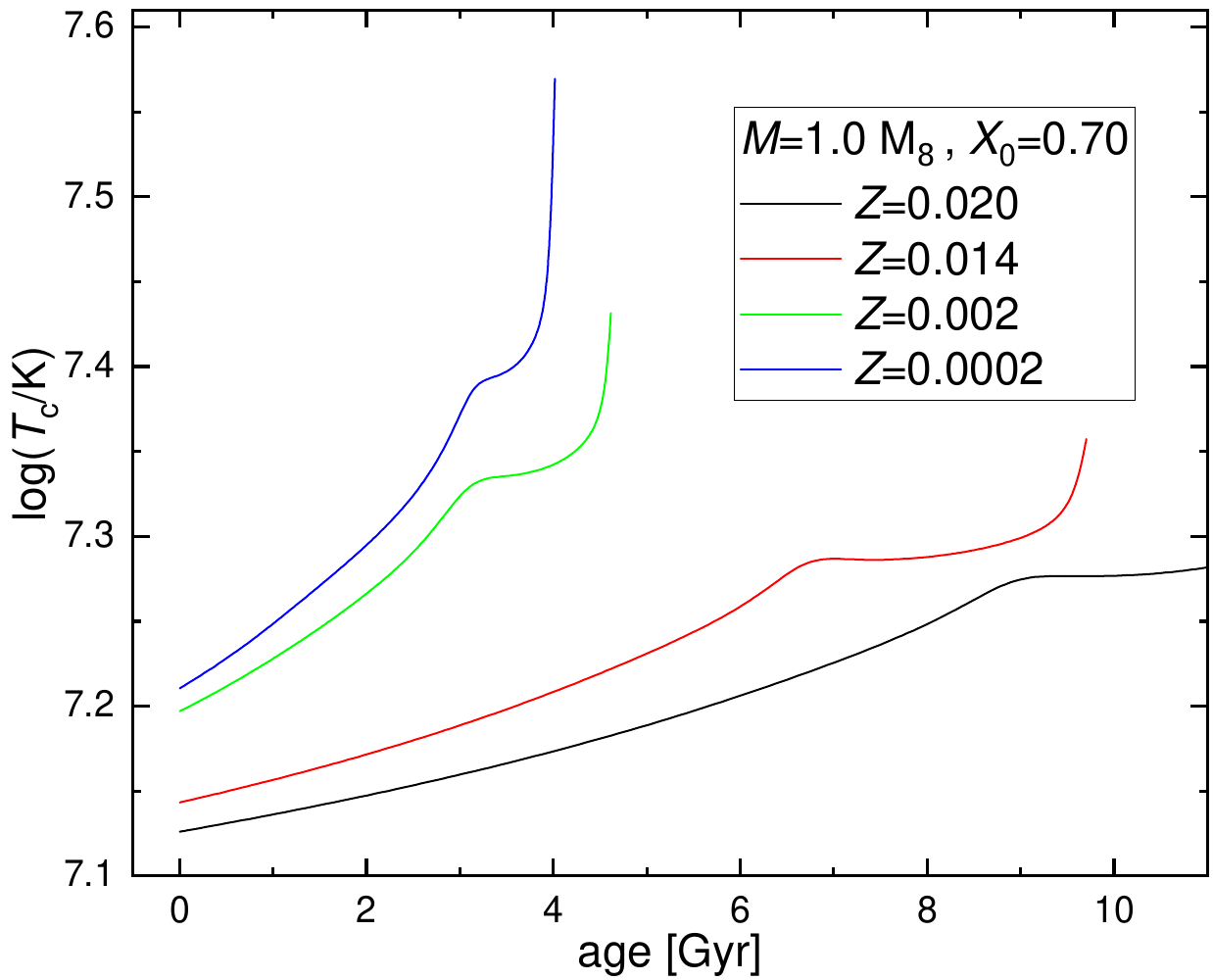}
	\caption{Left panel: the change of the central hydrogen abundance with age for the four models with $M=1.0$\,M$_{\odot}$ but different metallicity. The initial hydrogen abundance $X_0=0.70$, OPAL opacities and AGSS09 mixture were adopted. Right panel. the corresponding changes of the central temperature with age.}
	\label{Xc_Tc_OPAL}
\end{figure*}

For all five SX Phe stars, we managed to find seismic models within the observed ranges of effective temperature 
and luminosity. However, for each of these stars, the seismic models  are strongly dependent on the adopted 
opacity tables.
In two cases, V220 and V225, only the OPAL seismic models have the parameters $(T_{\rm eff},~L)$ consistent with observational determinations. 
Results for these two stars are the same as in the case of studied by us high-amplitude $\delta$ Sct stars and the prototype SX Phe \citep{JDD2023}.
For the other three stars, V194, V237 and NV326, there are seismic models within the error box computed with 
each opacity table.
Perhaps, this opacity effect is related to the metal abundance because for V220 and V225 we obtained the higher content of $Z$ than for 
the other three. Indeed, the effect of $\kappa$ on the Petersen diagrams decreases with decreasing $Z$, as shown in Fig.\,11.
To delve into this issue, thorough insight and testing of opacity data are needed, which is beyond the scope of this paper.

As mentioned in the Introduction, in the case of $\omega$ Cen, the age-metallicity relation is well known in the literature.
Because with the OPAL data we got seismic models within the error boxes for all stars,
we plotted their median values of $({\rm age}, Z)$ in Fig.\,12. We included the values for $Y=0.30$ and 0.33.
The vertical dotted lines mark the range of metallicity determined for $\omega$ Cen \citep{Nitschai2023}. 
As can be seen, in general, for smaller metallicity the stars are older.
The youngest seismic models are the models of V220 with $Y=0.33$ which are in the MS phase. However, they have too high metallicity for $\omega$ Cen.
Thus, within the allowed range of $Z$, the five SX Phe stars have the metallicity in the range of about $Z\in(0.0007,~0.0028)$ and ages in the range of about (1.87,\,3.79)\,Gyr. The oldest stars are those that have the lowest metallicity, i.e., V194 and NV326.
The youngest star V225 has the highest metallicity and it still contains a small amount of hydrogen in its core,
approximately 2\%,  

Another, quite surprising result, is the evolutionary stage of seismic models, i.e., most of them are in the phase
of hydrogen shell burning.
This may seem at odds with their age, since our HSB models with masses around 1\,$M_{\odot}$ 
are less than 4 Gyr old. However, the rate of hydrogen depletion strongly depends on the metallicity, i.e.,
the lower the metallicity, the faster the hydrogen depletion in the centre. This is because  
lower metallicity results in lower opacity and higher temperature which in turn increases the energy generation rate.
 %$(\epsilon \propto T^n)$.  
The effect of metallicity on the relation between central hydrogen abundance
$X_c$ and age is shown in the left panel of  Fig.\,13. The models with $M=1.0$\,$M_{\odot}$ and initial hydrogen abundance $X_0=0.70$ were considered. 
For a  1\,$M_{\odot}$ star, the burning time of hydrogen in the centre at metallicity $Z=0.002$ is more than twice shorter than at metallicity $Z=0.020$.
The corresponding changes of the central values of temperature are depicted in the right panel of Fig.\,13.  Already at ZAMS, models with $Z\leq 0.002$
have the central temperature exceeding about 15.6\,million\,K and at the age of about 2\,Gyr it reaches 18\,million\,K. However, the CNO cycle is not activated yet
because the transition between pp chain and CNO cycle strongly depends on the CNO abundance.
%Fig - the run of $\delta p /p$ for the first overtone for the model with OPAL, OP and OPLIB,
%the shift of the node in the case of OPAL model. The node is located about 16\,000\,K deeper
%comparing to the OP and OPLIB modes. 

%Fig - differences in the Rosseland mean  opacity in selected ranges of $\log T$

\section{Summary}

We presented extensive seismic modelling of the five SX Phe stars belonging to the globular cluster $\omega$ Cen.
To this end, we used the Bayesian analysis based on Monte Carlo simulations. 
We studied the effect of helium abundance and opacity tables.
%OPAL OP, OPLIB  and two values of helium abundance $Y=0.30$ and ).33.
In our computations we relied on 1D single-star evolution. As mentioned in the Introduction, if these pulsating blue stragglers are mergers, i.e., they were
formed via collision or coalescence in binaries, their evolution after reaching thermal equilibrium should rather resemble a single star evolution.
Stellar pulsations were computed using the non-adiabatic linear code of \citet{Dziembowski1977}.

A huge number of simulations were performed for each star (about 150\,000), adopting two values of helium abundance $Y$ (three in the case of V194), three opacity data and with several starting values of searched parameters. This amounted to over 5.5 million simulations in total and took about 9 months of calculations.

We derived reasonable range of metallicity and age for the five pulsating blue stragglers in $\omega$ Cen. Despite the very small sample, 
we obtained indications that the age-metallicity relation may be fulfilled by the studied BSSs.
Seismic masses of all stars are significantly higher than the turn-off mass (0.8\,M$_{\odot}$) and their rotational velocities range 
from a dozen to several dozen of $\kms$. The values of surface gravity agree with observational determinations for SX Phe stars.
One of the most important results is the finding that seismic models are very sensitive to the adopted opacity tables.
In the case of three SX Phe stars, V194, V237, NV326, we found seismic models within the observed $(T_{\rm eff},~L)$ range with all three sets of opacity data.
However, the OP and OPLIB seismic models have significantly higher  metallicity and higher masses.
In the case of V220 and V225, which also have the highest metallicity, we managed to find the seismic models within the error box only with the OPAL tables.

We are aware that non-standard formation channels of BSSs require more sophisticated modelling. 
However, considering the significant uncertainties in modelling stellar collisions, mass transfer, coalescence in binary systems, or triple system evolution, our seismic modelling based on 1D single-star evolution provides 
a reasonable approach.
In addition, extensive seismic modelling based on MC simulations would not be feasible for time-consuming modelling of 3D processes.
We therefore believe that our seismic modelling of blue straggler pulsators in $\omega$ Cen 
is a useful approximation and can be considered as the first step toward further, more advanced studies of these intriguing objects. This is our task that we plan to undertake in the near future.

%This first attempt of seismic modelling of pulsating blue straggles is a good starting point for more advanced studies also considering formation 
%by collisions or in binary systems. This is our task for the near future.  

\section*{Acknowledgements}
We would like to thank the anonymous referee for valuable and constructive comments.
The work was financially supported by the Polish National Science Centre grant 2023/50/A/ST9/00144.
Calculations have been partly carried out using resources provided by Wroclaw Centre for Networking and Supercomputing (http://www.wcss.pl), grant No. 265.

\section*{Data Availability}
The CASE data are available in the paper of \citet{Kaluzny2004}.
Theoretical computations of stellar evolution and pulsations will be shared 
on reasonable request to the corresponding author.

\bibliographystyle{mnras}
\bibliography{JDD_biblio3} % if your bibtex file is called example.bib

\appendix

\section{Effective temperatures and luminosities of SX Phe stars in $\omega$ Cen}

The parameters of 58 SX Phe stars in $\omega$ Centauri.
The third and fourth columns of Table\,A1 contain the new values of dereddened colour $(B-V)_0$ and absolute visual magnitude $M_V$ resulting from the assumed reddening $E(B-V)=0.185\pm 0.040$ for the central region of $\omega$ Cen and the true distance modulus $(V-M_V)_0=13.672\pm 0.019$\,mag, respectively \citep{Clontz2024}.  In the fifth and sixth columns, we give the values of effective temperature and luminosities derived using Kurucz model atmospheres.
The uncertainties in $(T_{\rm eff},~ L)$ are due to errors in $E(B-V)$ and  $(V-M_V)_0$ as well as due to the unknown values of metallicity and $\log g$. 
Since $\omega$ Cen is chemically inhomogeneous, we had to consider the four values of [m/H]$=-2, -1.5, -1.0, -0.5$
and three values of $\log g=3.5, 4.0, 4.5$, for which the Kurucz atmosphere models are tabulated.

\begin{table*}
%	\small
	\centering
		\caption{The new values of $(B-V)_0$ and $M_V$, as well as derived logarithmic values of $(T_{\rm eff},L/ \rm L_{\odot})$ for 58 SX Phe stars in $\omega$ Cen.}
	\label{tab:TL_all_stars}
	\begin{tabular}{|c|c|c|c|c|c|}
		\hline
		& Star & $(B-V)_0$ [mag] & $M_V$ [mag]& $\log(T_{\rm eff}/\rm K)$ & $\log(L/ \rm L_{\odot})$ \\
		& &  $ \pm 0.04$ &   $ \pm 0.059$ &    & \\
		\hline
		1 & V194 & 0.145 & 2.771 & 3.905(32) & 0.801(49) \\		
		2 & V195 & 0.186 & 2.535 & 3.889(32) & 0.895(49) \\		
		3 & V197 & 0.361 & 2.605 & 3.816(28) & 0.887(55) \\		
		4 & V198 & 0.138 & 3.288 & 3.908(32) & 0.595(51) \\		
		5 & V199 & 0.148 & 2.444 & 3.904(32) & 0.931(49) \\		
		6 &  V200 & 0.356 & 2.323 & 3.818(28) & 0.999(55) \\		
		7 &  V204 & 0.178 & 2.636 & 3.892(32) & 0.854(49) \\		
		8 &  V217 & 0.244 & 2.793 & 3.864(31) & 0.797(53) \\		
		9 &  V218 & 0.145 & 2.850 & 3.905(32) & 0.769(49) \\
		10 &  V219 & 0.152 & 3.058 & 3.902(32) & 0.685(48) \\
		11 &  V220 & 0.172 & 2.741 & 3.894(32) & 0.811(48) \\
		12 &  V221 & 0.266 & 2.435 & 3.855(30) & 0.944(54) \\
		13 &  V225 & 0.208 & 2.600   & 3.879(32) & 0.871(51) \\
		14 &  V226 & 0.228 & 3.053 & 3.871(31) & 0.692(52) \\
		15 &  V227 & 0.212 & 3.027 & 3.878(32) & 0.701(51) \\
		16 &  V228 & 0.218 & 2.954 & 3.875(32) & 0.730(52) \\
		17 &  V229 & 0.109 & 3.162 & 3.920(33) & 0.653(57) \\
		18 & V231 & 0.156 & 3.174 & 3.901(32) & 0.638(48) \\
		19 & V233 & 0.168 & 2.965 & 3.896(32) & 0.721(48) \\
		20 &  V237 & 0.166 & 2.616 & 3.897(32) & 0.861(48) \\
		21 &  V238 & 0.164 & 3.110  & 3.897(32) & 0.663(47) \\
		22 &  V249 & 0.176 & 3.190 & 3.893(32) & 0.632(48) \\
		23 &  V250 & 0.195 & 3.188 & 3.885(32) & 0.634(50) \\
		24 &  V252 & 0.220 & 3.200 & 3.874(32) & 0.632(52) \\
		25 &  V253 & 0.123 & 2.987 & 3.914(32) & 0.719(54) \\
		26 &  NV294 & 0.143 & 3.047 & 3.906(32) & 0.691(50) \\
		27 &  NV295 & 0.172 & 3.038 & 3.894(32) & 0.692(48) \\
		28 &  NV296 & 0.180  & 2.701 & 3.891(32) & 0.828(49) \\
		29 &  NV297 & 0.116 & 2.383 & 3.917(32) & 0.962(56) \\
		30 &  NV298 & 0.169 & 3.165 & 3.895(32) & 0.641(48) \\
		31 &  NV299 & 0.247 & 3.080 & 3.863(31) & 0.683(53) \\
		32 &  NV300 & 0.178 & 3.239 & 3.892(32) & 0.613(49) \\
		33 &  NV301 & 0.254 & 2.729 & 3.860(31) & 0.824(54) \\
		34 &  NV302 & 0.177 & 2.835 & 3.892(32) & 0.774(48) \\
		35 &  NV303 & 0.137 & 2.686 & 3.908(32) & 0.836(51) \\
		36 &  NV304 & 0.236 & 2.991 & 3.867(31) & 0.717(53) \\
		37 &  NV305 & 0.206 & 3.139 & 3.880(32) & 0.655(51) \\
		38 &  NV306 & 0.185 & 3.282 & 3.889(32) & 0.596(49) \\
		39 &  NV307 & 0.334 & 2.823 & 3.827(28) & 0.796(55) \\
		40 &  NV308 & 0.197 & 3.033 & 3.884(32) & 0.697(50) \\
		41 &  NV309 & 0.177 & 2.346 & 3.892(32) & 0.970(48) \\
		42 &  NV310 & 0.230  & 2.546 & 3.870(31) & 0.895(52) \\
		43 &  NV312 & 0.205 & 2.149 & 3.881(32) & 1.051(51) \\
		44 &  NV313 & 0.163 & 3.433 & 3.898(32) & 0.534(47) \\
		45 &  NV314 & 0.201 & 2.841  & 3.882(32) & 0.774(50) \\
		46 &  NV315 & 0.333 & 2.147 & 3.827(28) & 1.067(55) \\
		47 &  NV316 & 0.141 & 3.081 & 3.907(32) & 0.677(50) \\
		48 &  NV317 & 0.268 & 2.723 & 3.854(30) & 0.829(54) \\
		49 &  NV318 & 0.287 & 2.554 & 3.846(30) & 0.899(55) \\
		50 & NV319 & 0.280  & 2.994 & 3.849(30) & 0.721(55) \\
         	51 &  NV320 & 0.255 & 3.049 & 3.859(31) & 0.696(54) \\
        	52 & NV321 & -0.049 & 2.164 & 4.024(66) & 1.209(163) \\
		53 &  NV323 & 0.251 & 2.393 & 3.861(31) & 0.958(53) \\
		54 &  NV324 & 0.112 & 2.157 & 3.919(33) & 1.054(57) \\
		55 &  NV325 & 0.316 & 2.165 & 3.834(29) & 1.057(55) \\
		56 &  NV326 & 0.216 & 2.796 & 3.876(32) & 0.793(51) \\
		57 &  NV327 & 0.089 & 2.397 & 3.929(35) & 0.966(63) \\
		58 &  NV328 & 0.413 & 2.809 & 3.798(27) & 0.812(53) \\
		\hline
	\end{tabular}
\end{table*}

\section{Frequencies from the CASE light curves of the five SX Phe stars}

In Table\,B1, we give the values of all frequencies of the five SX Phe stars extracted from the combined $BV$ CASE light curves at a $S/N>4$ level criterion.
There are given also the values of amplitudes and phases in the $BV$ filters.
The penultimate column gives information whether the significant frequency is independent or it is a harmonic/combination.
In the last column we compare our determinations with those in \citet{Olech2005}.
Fig.\,B1 shows an example of the spectral window and periodograms for the star NV326.
\begin{table*}
	\centering
	\caption{Significant frequencies of the five SX Phe stars extracted from the combined $BV$ data.}
	\subtable[\textbf{V194}]{
		\begin{tabular}{ccccccccc}
			\hline
			ID & frequency  & $A_B$  & $A_V$  & $\phi_B$ & $\phi_V$ & $S/N$ & Comb. & Olech2005 \\
			& [d$^{-1}$] &  [mag] &  [mag] &   [rad]  &  [rad]   &       &       &           \\
			\hline
			$\nu_{1}$ & 21.1964338(47) & 0.3268(15) & 0.24180(88) & 3.6412(52) & 3.6363(53) & 13.67 & indep. & $f_{0}$  \\
			%		&  &  &  &  &  & & & \\
			$\nu_{2}$ & 42.392871(15) & 0.0974(17) & 0.07307(86) & 3.013(16) & 3.038(18) & 15.22    & $2\nu_1$    & \\		
			$\nu_{3}$ & 63.589322(47) & 0.0273(15) & 0.02326(84) & 3.156(58) & 3.054(57) & 12.63    & $3\nu_1$    & \\
			$\nu_{4}$ & 21.75057(15) & 0.0139(15) & 0.00706(84) & 0.94(12) & 0.76(18) & 6.26        & indep. & $f_{3} + \frac{1}{\mathrm d} +\frac{1}{\mathrm{yr}}$\\
			$\nu_{5}$ & 27.16331(14) & 0.0100(16) & 0.00774(83) & 4.16(15) & 4.29(17) & 6.42        & indep. & $f_{1}$\\
			$\nu_{6}$ & 84.78566(17) & 0.0098(16) & 0.00594(83) & 3.05(16) & 3.20(21) & 6.90        & $4\nu_1$    & \\
			$\nu_{7}$ & 42.94683(27) & 0.0075(15) & 0.00397(85) & 0.33(23) & 0.12(31) & 4.39        &$\nu_1+\nu_4$& \\
			$\nu_{8}$ & 26.50446(31) & 0.0075(14) & 0.00297(85) & 6.03(24) & 5.65(40) & 4.45        & indep. & \\
			\hline
		\end{tabular}
	}
	\subtable[\bf{V220}]{
		\begin{tabular}{ccccccccc}
			\hline
			ID & frequency  & $A_B$  & $A_V$  & $\phi_B$ & $\phi_V$ & $S/N$ & Comb. & Olech2005 \\
			& [d$^{-1}$] &  [mag] &  [mag] &   [rad]  &  [rad]   &       &       &           \\
			\hline
			$\nu_{1}$ & 18.908260(15) & 0.0827(12) & 0.05590(62) & 5.740(15) & 5.750(18 )& 13.22 & indep. & $f_{0}$ \\
			$\nu_{2}$ & 24.316097(72) & 0.0173(11) & 0.01115(63) & 1.956(72) & 1.842(86) & 9.94  & indep. & $f_{1}$ \\
			$\nu_{3}$ & 37.81652(10) & 0.0092(11) & 0.00840(63) & 0.71(13) & 0.92(12) & 9.30     & $2\nu_1$    & \\
			$\nu_{4}$ & 23.84469(19) & 0.0069(11) & 0.00406(63) & 5.41(18) & 5.94(23) & 4.49     & indep. & $f_{2}$ \\
			$\nu_{5}$ & 42.22166(29) & 0.0060(11) & 0.00237(63) & 2.79(22) & 2.32(37) & 4.28     & $\nu_1+\nu_2 - \frac{1}{\mathrm d} - \frac{1}{\mathrm{yr}}$ & \\
			\hline				
		\end{tabular}
	}	
	
	\subtable[\textbf{V225}]{
		\begin{tabular}{ccccccccc}
			\hline
			ID & frequency  & $A_B$  & $A_V$  & $\phi_B$ & $\phi_V$ & $S/N$ & Comb. & Olech2005 \\
			& [d$^{-1}$] &  [mag] &  [mag] &   [rad]  &  [rad]   &       &       &           \\
			\hline
			$\nu_{1}$ & 20.560027(14) & 0.1483(16) & 0.09785(93) & 1.158(13) & 1.136(15) & 13.66 & indep. & $f_{0}$ \\   	
			$\nu_{2}$ & 41.120032(65) & 0.0320(17) & 0.01984(96) & 4.463(59) & 4.373(73) & 9.84  & $2\nu_1$    & \\
			$\nu_{3}$ & 26.414478(86) & 0.0240(17) & 0.01495(97) & 5.114(81) & 5.238(97) & 9.21  & indep. & $f_{1}$ \\
			$\nu_{4}$ & 26.82131(11) & 0.0162(18) & 0.01138(97) & 2.08(11) & 2.01(13) & 7.77     & indep. & $f_{2}$ \\
			$\nu_{5}$ & 5.85442(16) & 0.0127(17) & 0.00819(96) & 2.05(15) & 2.18(18) & 5.81      & $-\nu_1+\nu_3$ & \\
			$\nu_{6}$ & 45.97174(19) & 0.0105(17) & 0.00634(96) & 1.60(17) & 1.39(22) & 5.47     & $\nu_1+\nu_3- \frac{1}{\mathrm d} - \frac{1}{\mathrm{yr}}$ & \\
			$\nu_{7}$ & 47.38193(32) & 0.0063(17) & 0.00408(94) & 4.87(29) & 4.74(36) & 4.16     & $\nu_1+\nu_4$ & \\
			$\nu_{8}$ & 34.41053(32) & 0.0090(17) & 0.00372(93) & 5.09(22) & 4.61(38) & 4.06     & indep. & \\
			$\nu_{9}$ & 6.26138(23) & 0.0108(17) & 0.00547(97) & 5.28(19) & 5.15(26) & 4.05      & $-\nu_1+\nu_4$ & \\
			\hline
		\end{tabular}
	}   
	
	\subtable[\textbf{V237}]{
		\begin{tabular}{ccccccccc}
			\hline
			ID & frequency  & $A_B$  & $A_V$  & $\phi_B$ & $\phi_V$ & $S/N$ & Comb. & Olech2005 \\
			& [d$^{-1}$] &  [mag] &  [mag] &   [rad]  &  [rad]   &       &       &           \\
			\hline
			$\nu_{1}$ & 15.243343(14) & 0.1699(12) & 0.11586(68) & 2.1632(89) & 2.1512(92) & 13.04 & indep. & $f_0$ \\
			$\nu_{2}$ & 30.486668(41) & 0.0565(10) & 0.03985(64) & 0.116(25) & 0.101(26) & 14.11   & $2\nu_1$ & \\
			$\nu_{3}$ & 45.729947(99) & 0.0207(10) & 0.01515(63) & 4.476(66) & 4.460(65) & 12.12   & $3\nu_1$ & \\
			$\nu_{4}$ & 60.97344(28) & 0.0069(11) & 0.00563(64) & 2.06(18) & 2.42(18) & 6.75       & $4\nu_1$ & \\
			$\nu_{5}$ & 24.11603(35) & 0.0056(11) & 0.00510(65) & 6.12(22) & 6.17(22) & 5.40       & indep. & $f_1$ \\
			$\nu_{6}$ & 15.20130(36) & 0.0056(12) & 0.00516(68) & 1.39(25) & 1.42(22) & 5.23       & indep. & $f_2$ \\
			\hline
		\end{tabular} 
	}
	\subtable[\textbf{NV326}]{
		
		\begin{tabular}{ccccccccc}
			\hline
			ID & frequency  & $A_B$  & $A_V$  & $\phi_B$ & $\phi_V$ & $S/N$ & Comb. & Olech2005 \\
			& [d$^{-1}$] &  [mag] &  [mag] &   [rad]  &  [rad]   &       &       &           \\
			\hline
			$\nu_{1}$ & 17.572793(47) & 0.1098(24) & 0.0803(16) & 5.192(27) & 5.136(29) & 10.51 & indep.   & $f_0$\\
			$\nu_{2}$ & 22.537475(65) & 0.0790(23) & 0.0530(16) & 3.039(38) & 2.969(42) & 12.03 & indep.   & $f_1$\\
			$\nu_{3}$ & 40.11000(19) & 0.0359(24) & 0.0171(15) & 3.826(91) & 4.05(13) & 7.53    & $\nu_1+\nu_2$ & \\
			$\nu_{4}$ & 4.96418(20) & 0.0340(25) & 0.0168(15) & 1.831(91) & 2.12(13) & 7.53     & $-\nu_1+\nu_2$& \\
			$\nu_{5}$ & 35.14589(31) & 0.0155(24) & 0.0132(15) & 5.92(18) & 5.77(19) & 6.67     & $2\nu_1$      & \\
			$\nu_{6}$ & 20.37930(64) & 0.0130(23) & 0.0059(14) & 5.64(27) & 5.91(41) & 4.31     & indep.   & \\
			\hline
		\end{tabular}
	}
	\label{tab:frequencies}
\end{table*}

\begin{figure*}
	\centering
	\includegraphics[clip,width=12.5cm,height=22.3cm]{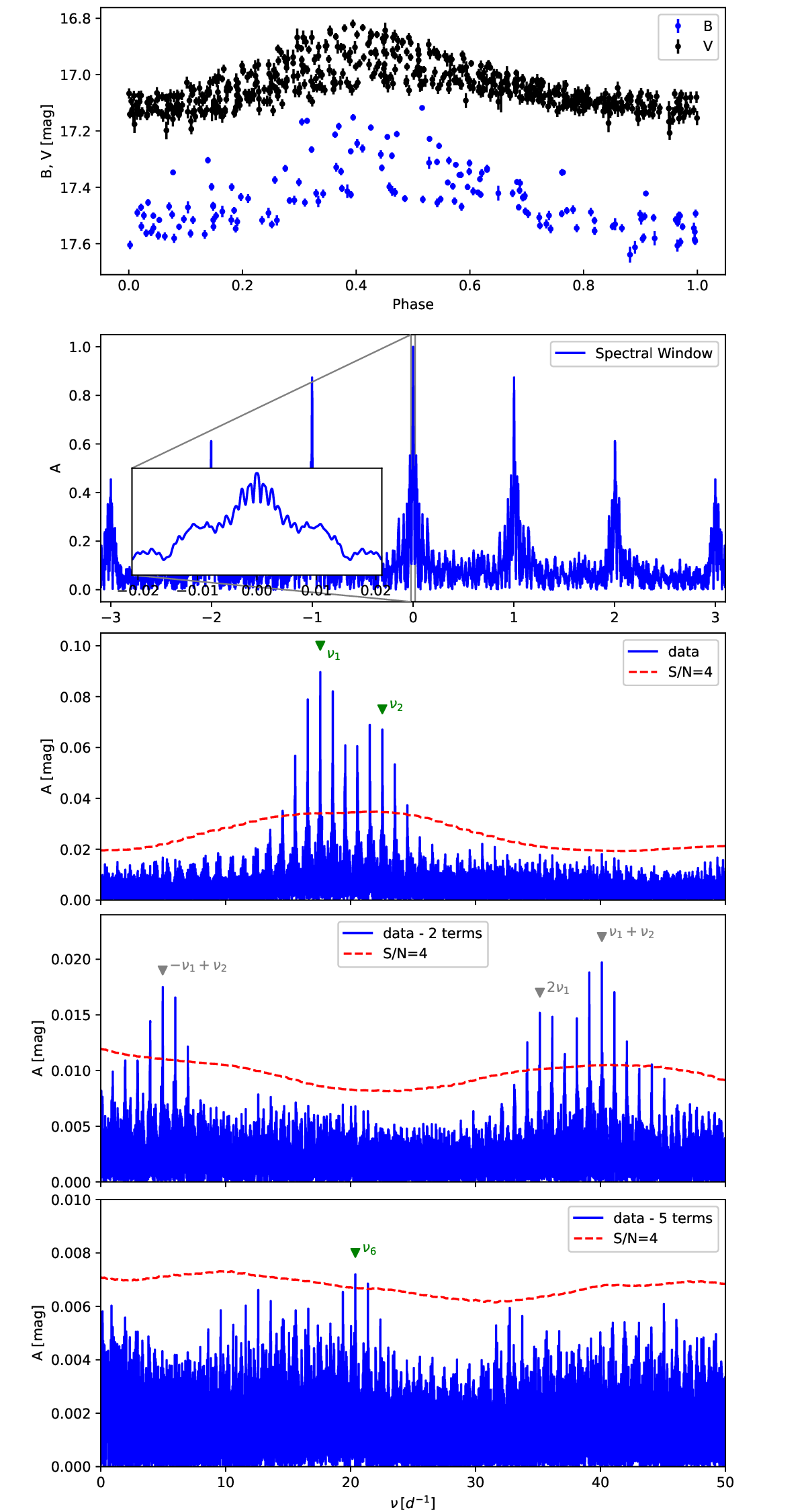}
	\caption{The top panel shows the observations of NV326 in the $B$ and $V$ filters phased 
	with the dominant frequency, $\nu_1 = 17.572793$\,$d^{-1}$.
	The second panel below presents the spectral window with a zoom-in on the central peak.
	The next panels below display the periodogram for the original data, the periodogram after subtracting 
	two frequencies, and the periodogram after subtracting five frequencies. The red line marks our detection 
	threshold at $S/N = 4$. Green and gray triangles indicate independent frequencies and 
	combination/harmonic frequencies, respectively.}
	\label{Periodograms_NV326}
\end{figure*}

\section{Histograms for various parameters from seismic modelling}

Figs.\, C1, C2, C2, C4 and C5 show histograms for the seismic values of mass $M$ in M$_{\odot}$, metallicity $Z$, age in Gyr, {\bf current} rotational velocity $V_{\rm rot}$ 
in $\kms$ and  radius in $R$ in R$_{\odot}$. The parameters were derived from fitting the two radial modes as well as the effective  temperature and luminosity 
$(T_{\rm eff},~L)$, using OPAL tables, AGSS09 chemical mixture and helium abundance $Y=0.30$. All models are in the HSB phase of evolution.
The right-bottom histogram in each figure contains the values of the instability parameter $\eta$ for the radial fundamental mode. The dependence of $\eta$
on mass and age for seismic models of the stars V220 and V237, as an example, are shown in Figs.\,C6 and C7. The left panels are for the radial fundamental modes
whereas the right panels for the first (V220) or second overtone modes (V237). The metallicity is coded with colours.

Figs.\,C8 and C9 present results from seismic modelling of V194 consisting of fitting the frequencies of two radial modes and, additionally, the dipole mode.
The OPAL tables and AGSS09 mixture were adopted.
Seismic models with $(T_{\rm eff},~L)$ consistent with observational determinations were found for all azimuthal orders of the dipole mode, i.e., $m=-1, 0, +1$.
Fig,\,C8 shows their positions on the HR diagrams. In Fig.\,C9, we present, as an example, histograms for the mass $M$ in M$_{\odot}$, metallicity $Z$,  helium abundance and age in Gyr for the seismic models reproducing two radial mode and one retrograde dipole mode.

\begin{figure*}
	\centering
	\includegraphics[clip,width=0.495\linewidth,height=65mm]{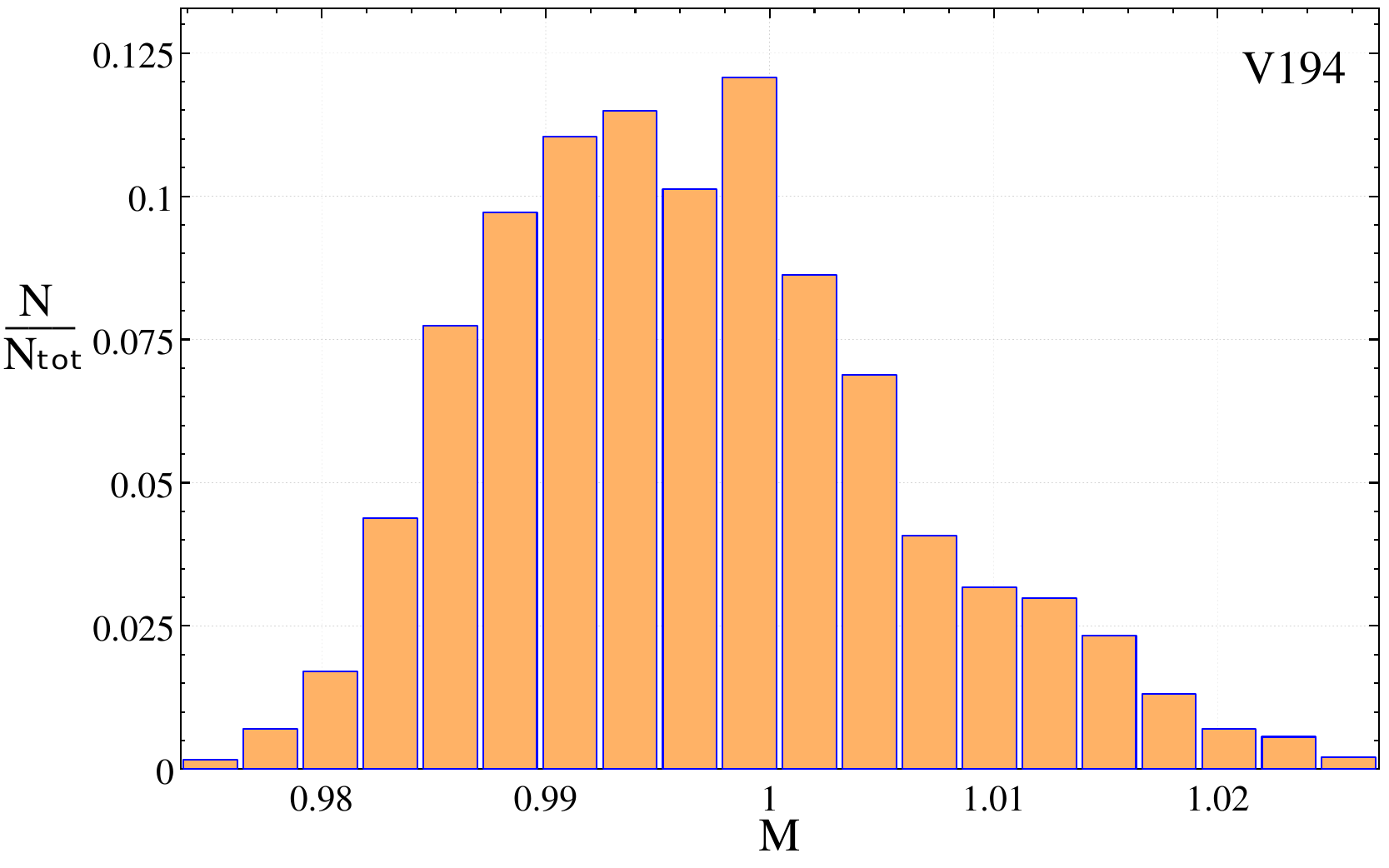}
	\includegraphics[clip,width=0.495\linewidth,height=65mm]{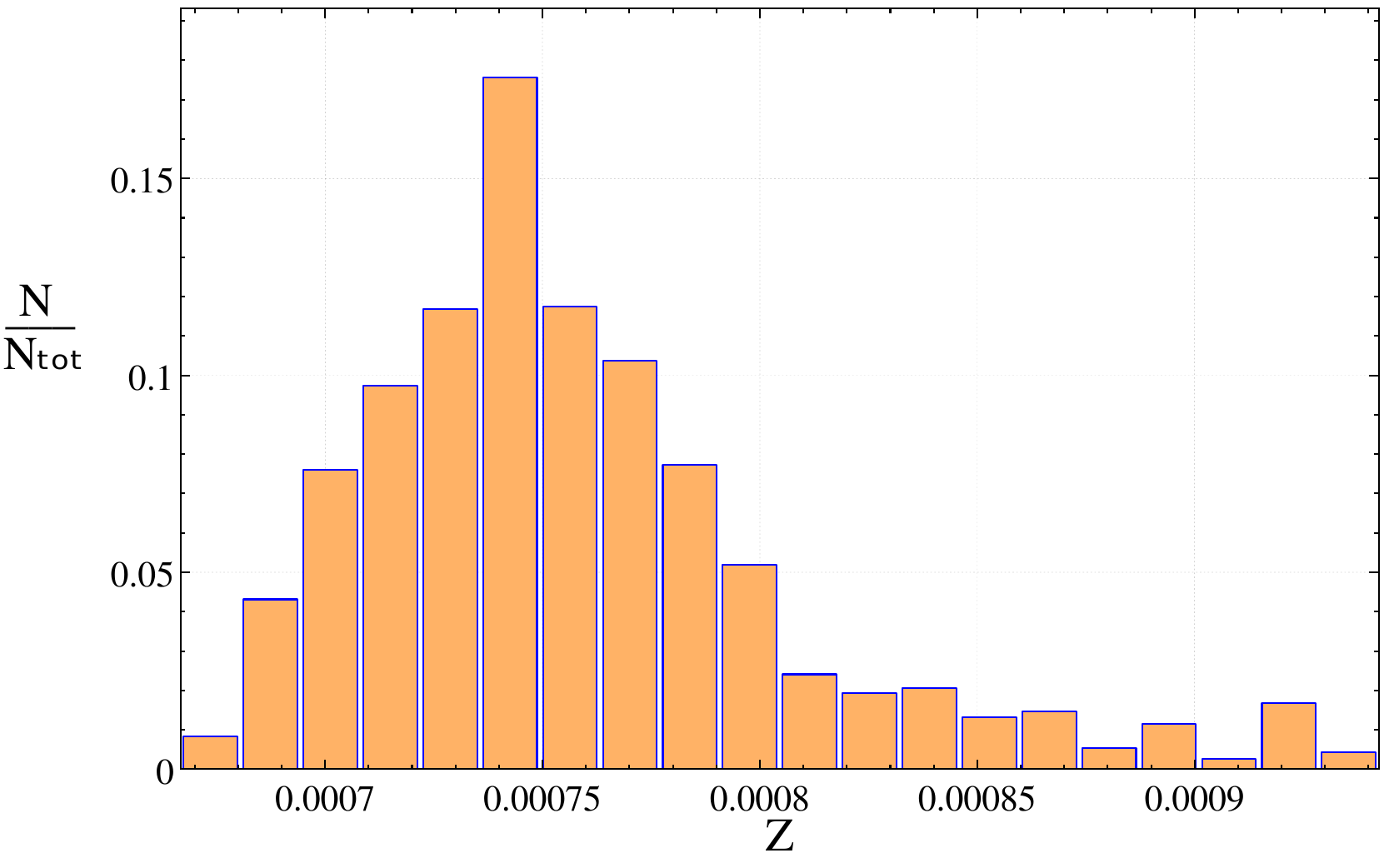}
	\includegraphics[clip,width=0.495\linewidth,height=65mm]{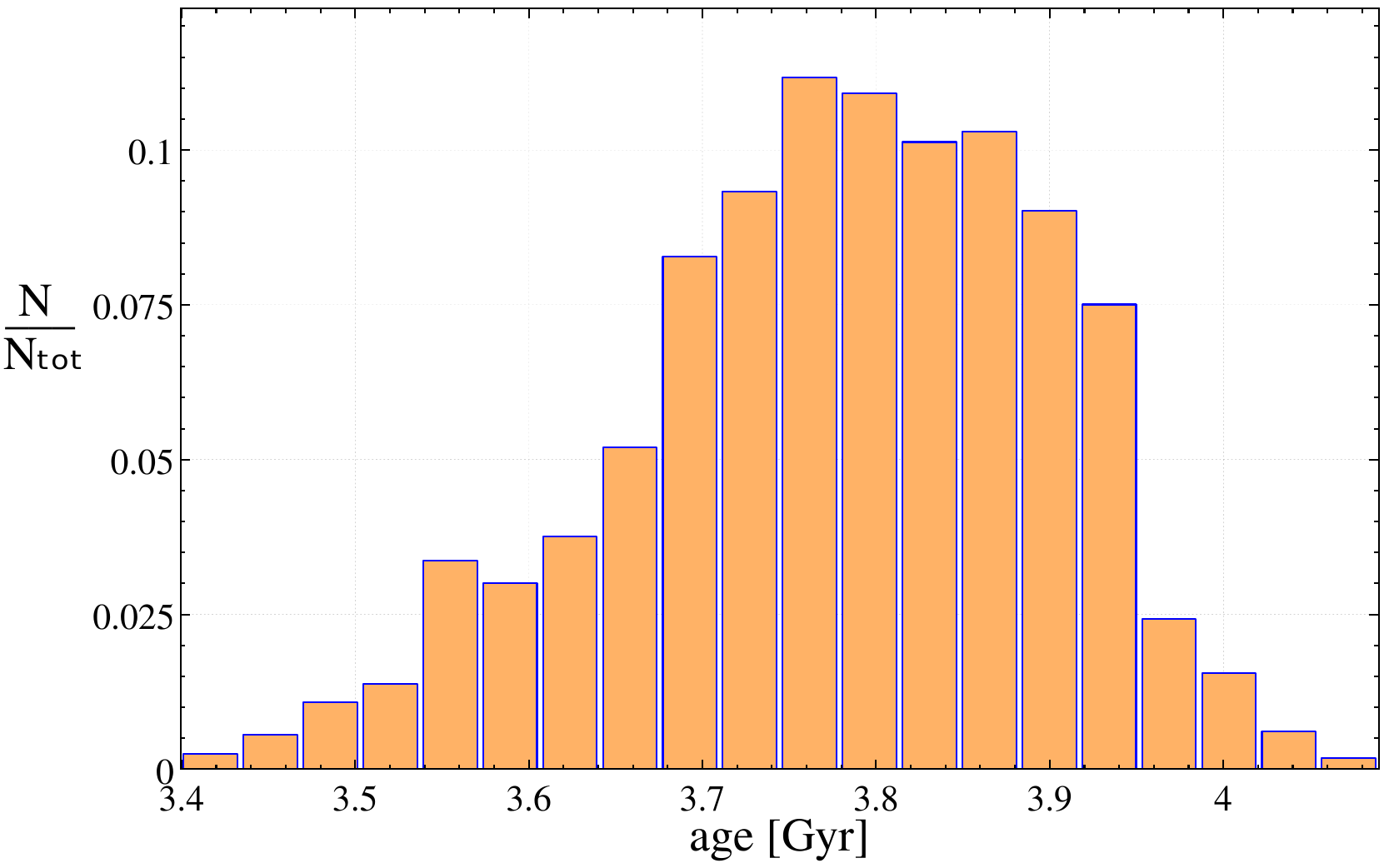}
	\includegraphics[clip,width=0.495\linewidth,height=65mm]{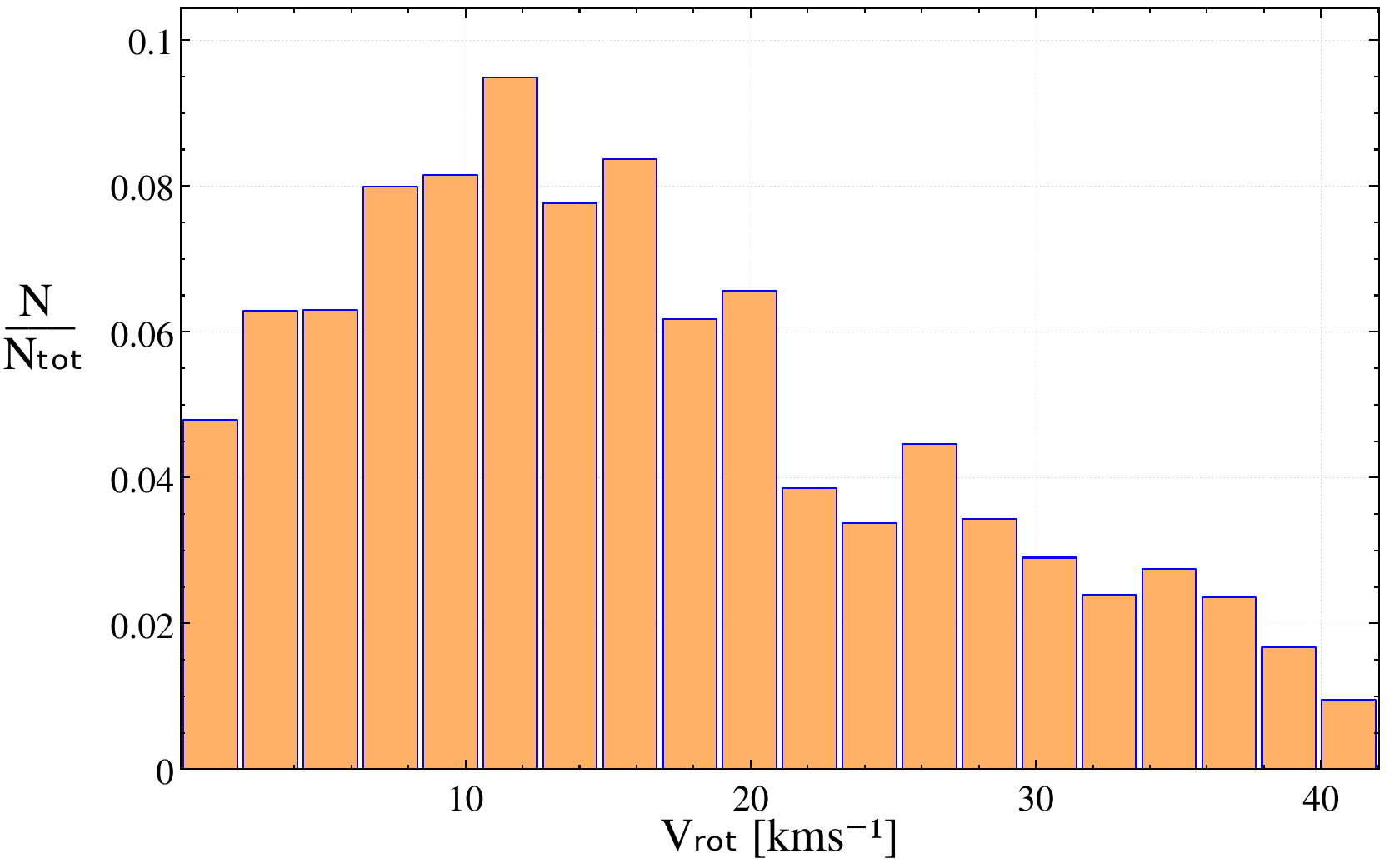}
	\includegraphics[clip,width=0.495\linewidth,height=65mm]{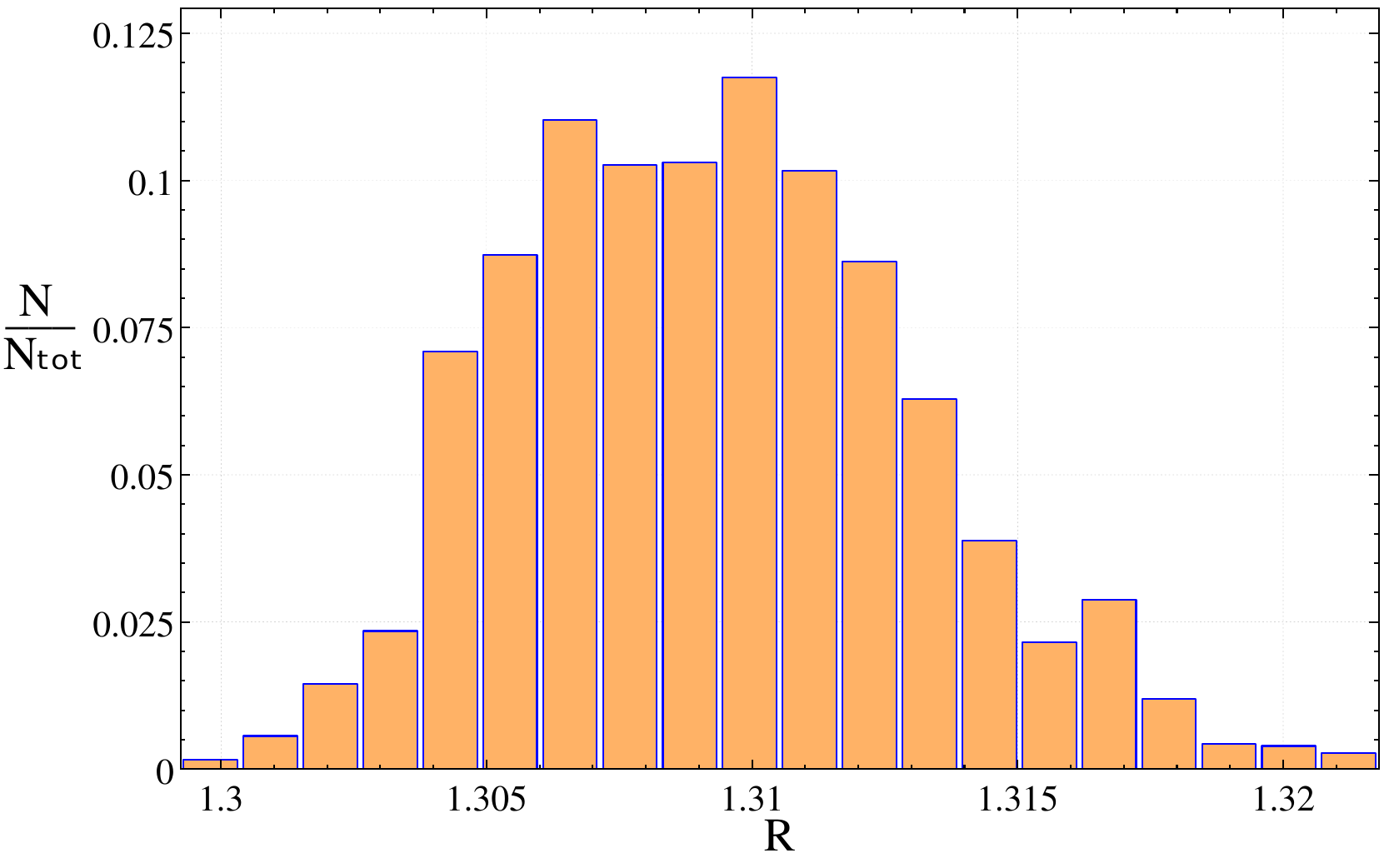}
    \includegraphics[clip,width=0.495\linewidth,height=65mm]{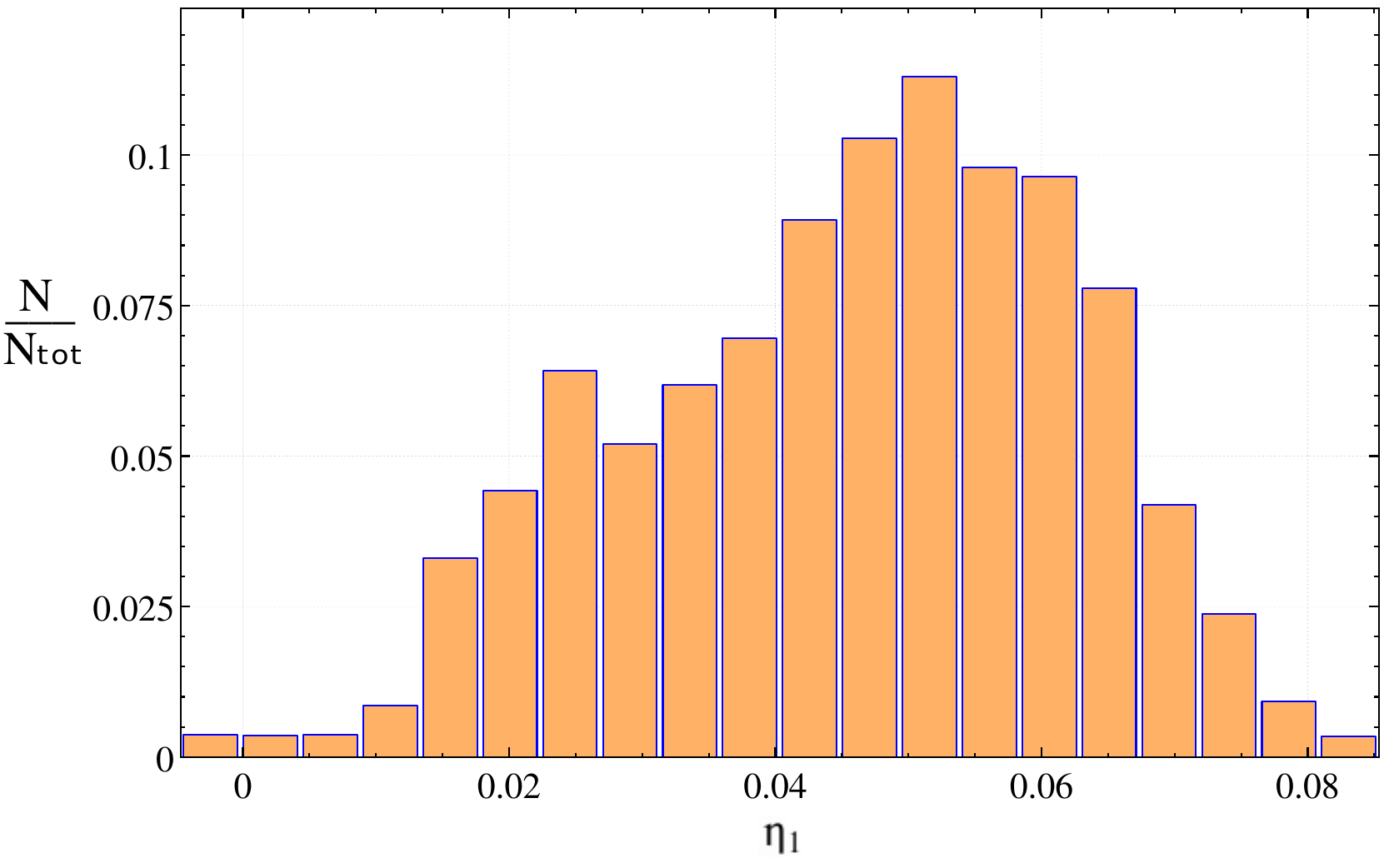}
	\caption{The normalized histograms for the parameters of OPAL seismic models of V194 which reproduce the two radial modes as fundamental and first overtone.
		 The AGSS09 chemical mixture and  $Y=0.30$ were assumed. The right-bottom panel contains the instability parameter for the radial fundamental mode $\eta_1$.}
	\label{histograms_V194}
\end{figure*}

\begin{figure*}
	\centering
	\includegraphics[clip,width=0.495\linewidth,height=65mm]{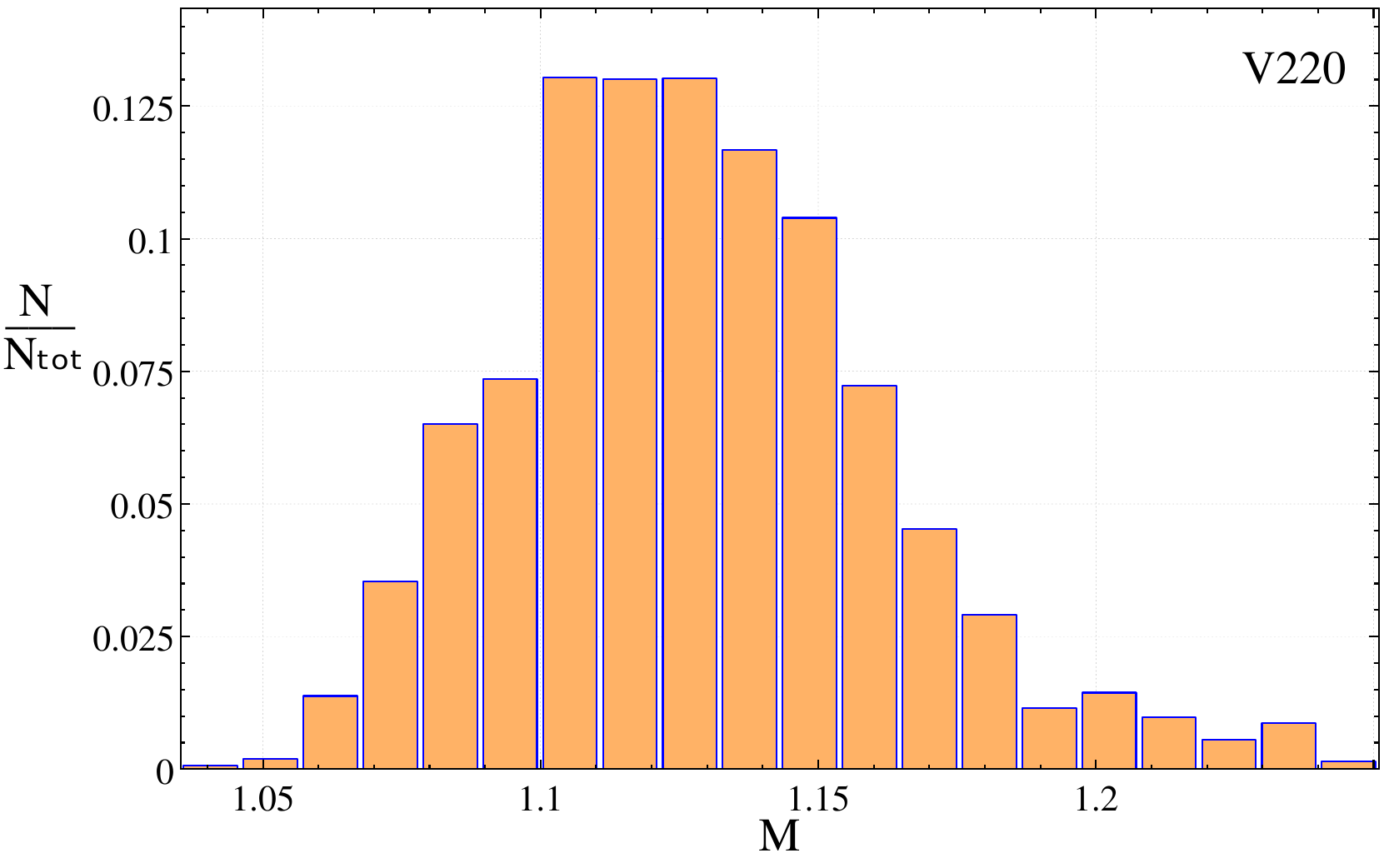}
	\includegraphics[clip,width=0.495\linewidth,height=65mm]{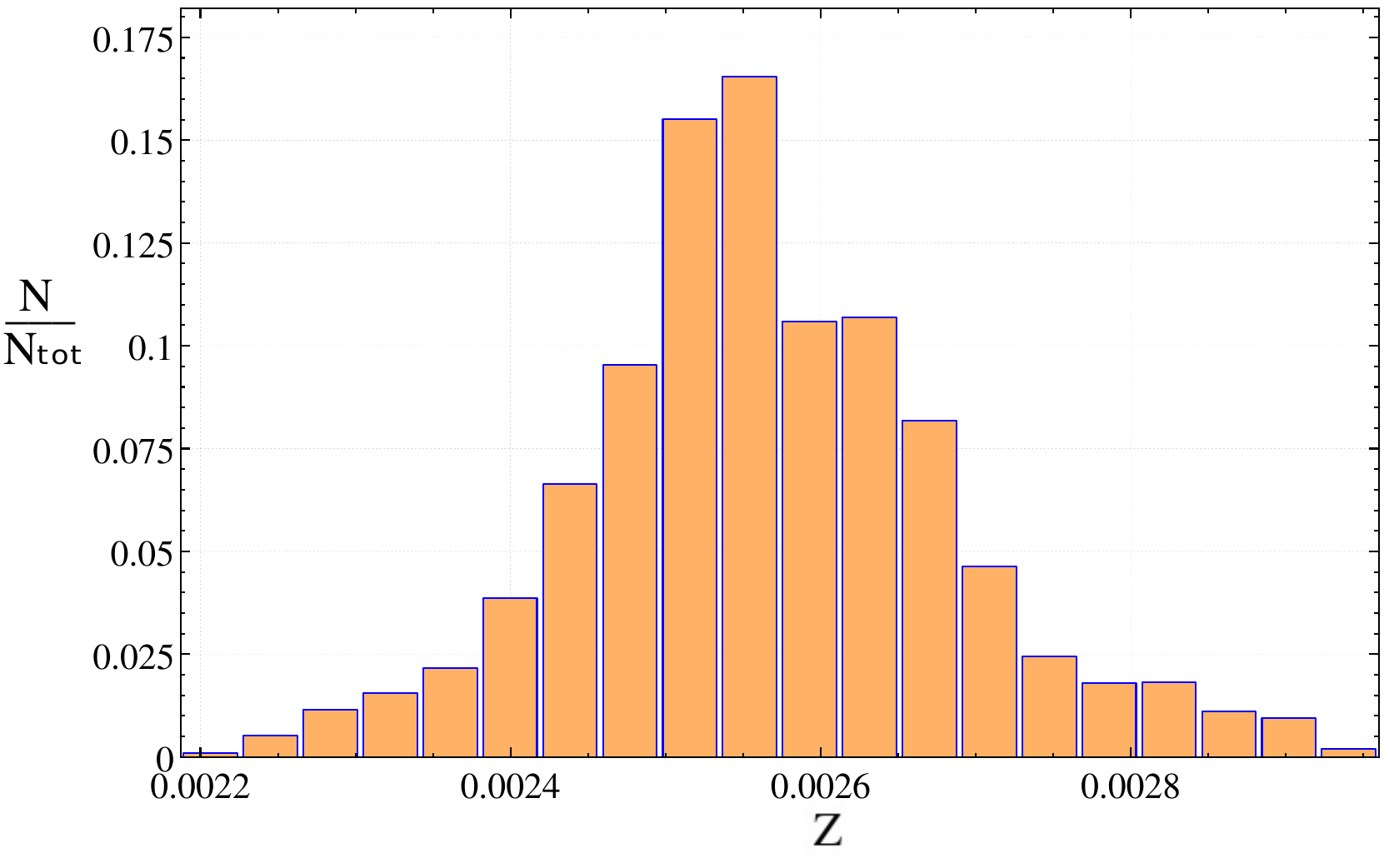}
	\includegraphics[clip,width=0.495\linewidth,height=65mm]{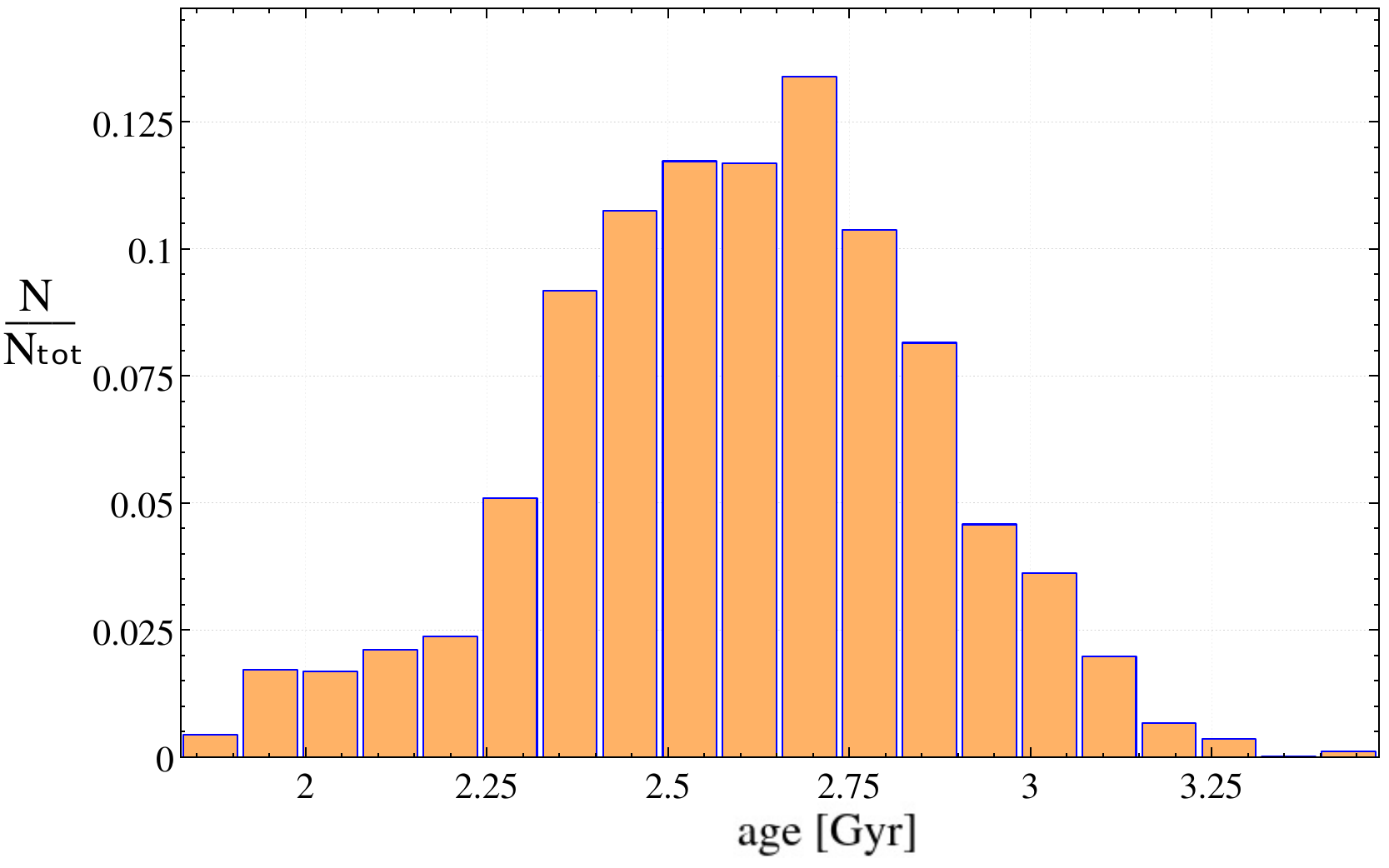}
	\includegraphics[clip,width=0.495\linewidth,height=65mm]{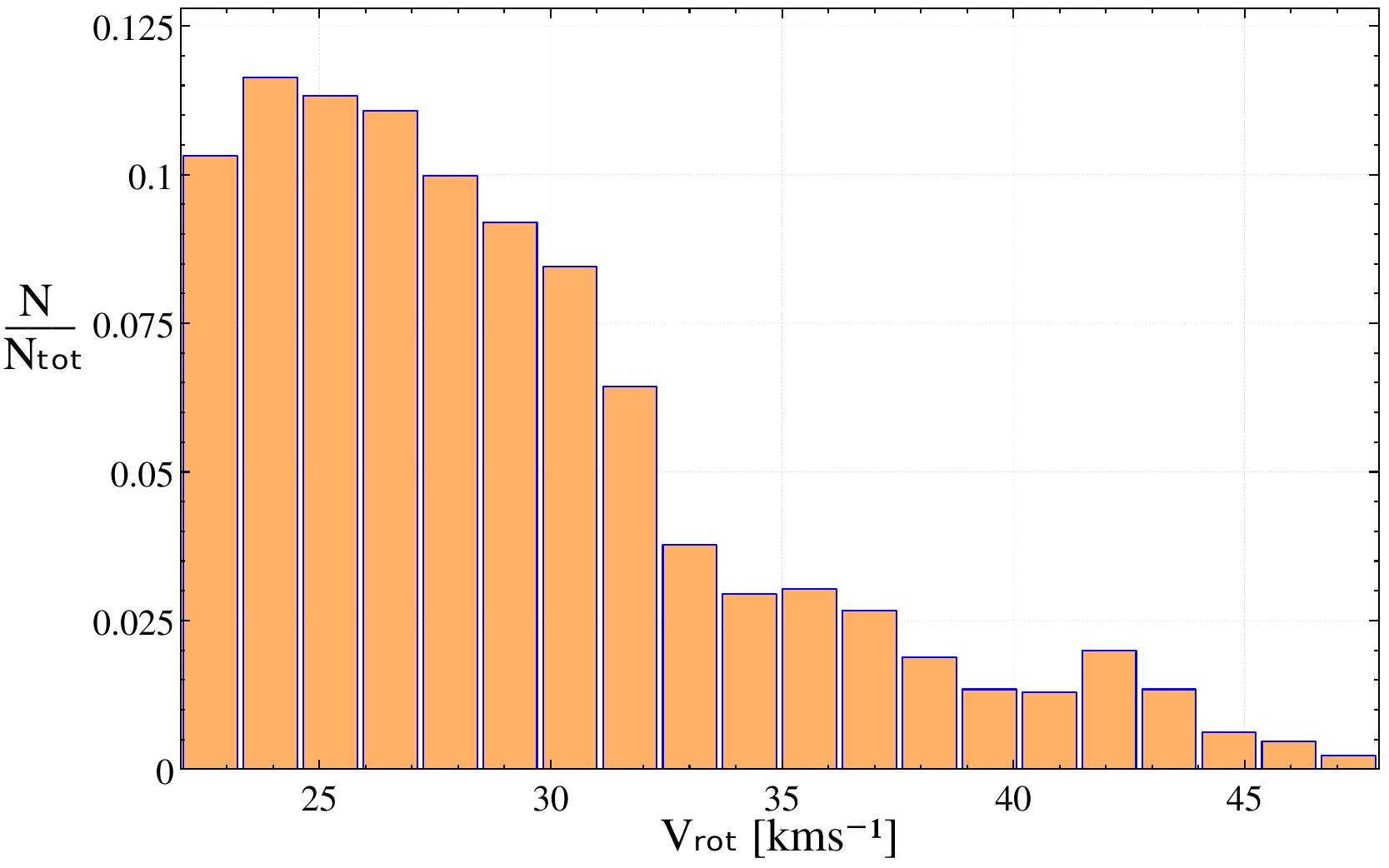}
	\includegraphics[clip,width=0.495\linewidth,height=65mm]{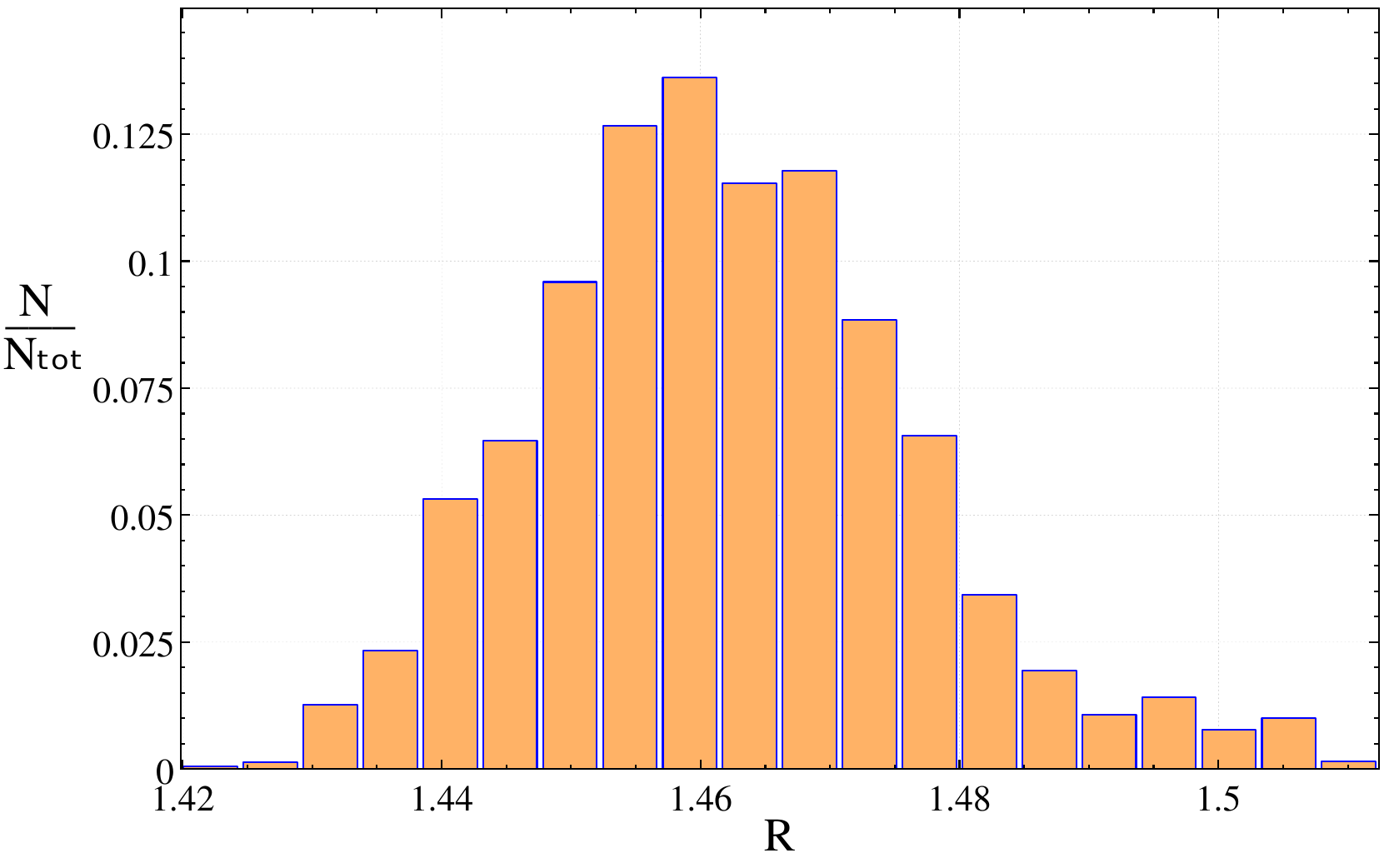}
    \includegraphics[clip,width=0.495\linewidth,height=65mm]{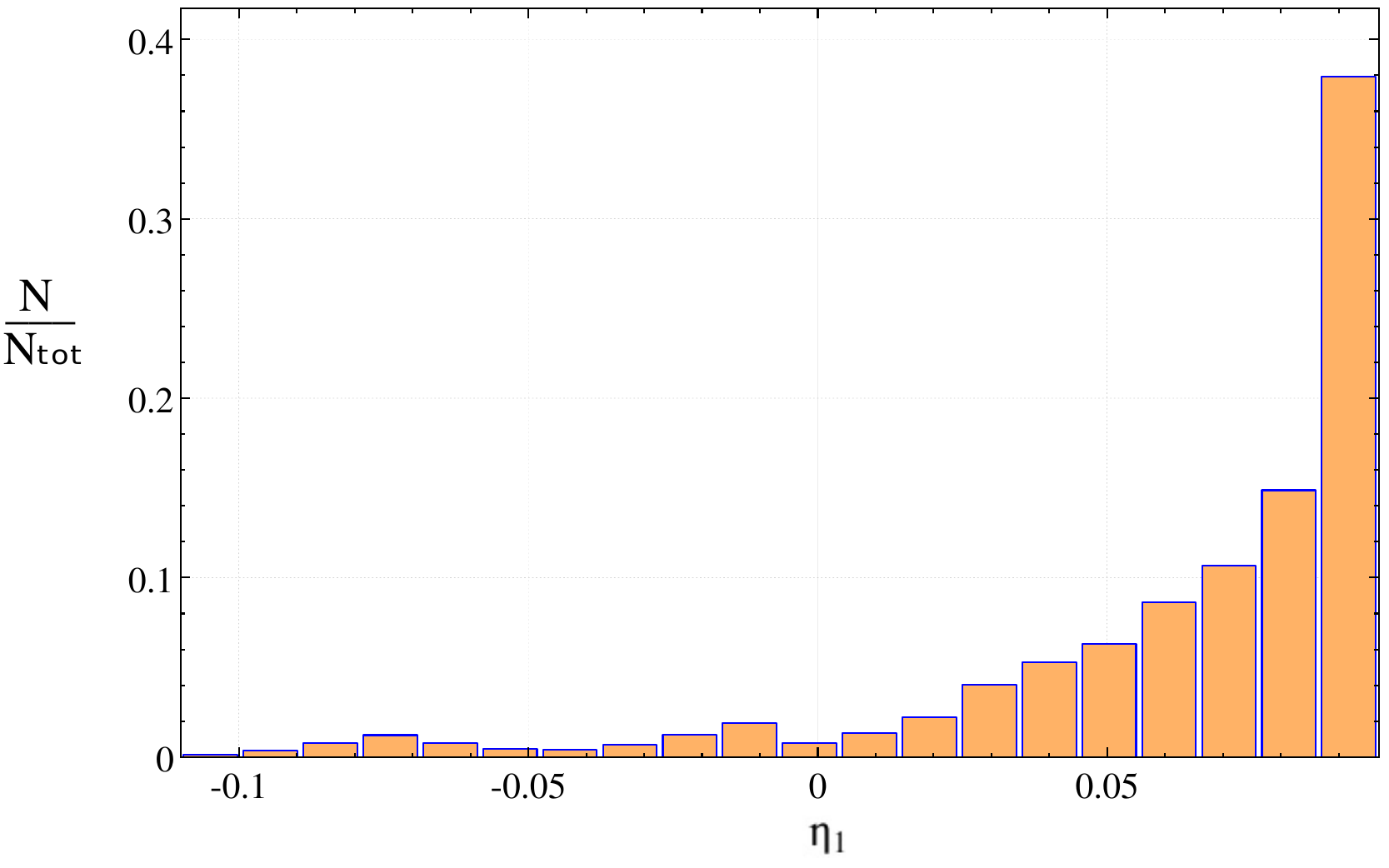}
	\caption{The same as in Fig.\,C1 but for V220. Only seismic models with the rotational velocities $V_{\rm rot}>22\,\kms$ were included}
	\label{histograms_V194}
\end{figure*}

\begin{figure*}
	\centering
	\includegraphics[clip,width=0.495\linewidth,height=65mm]{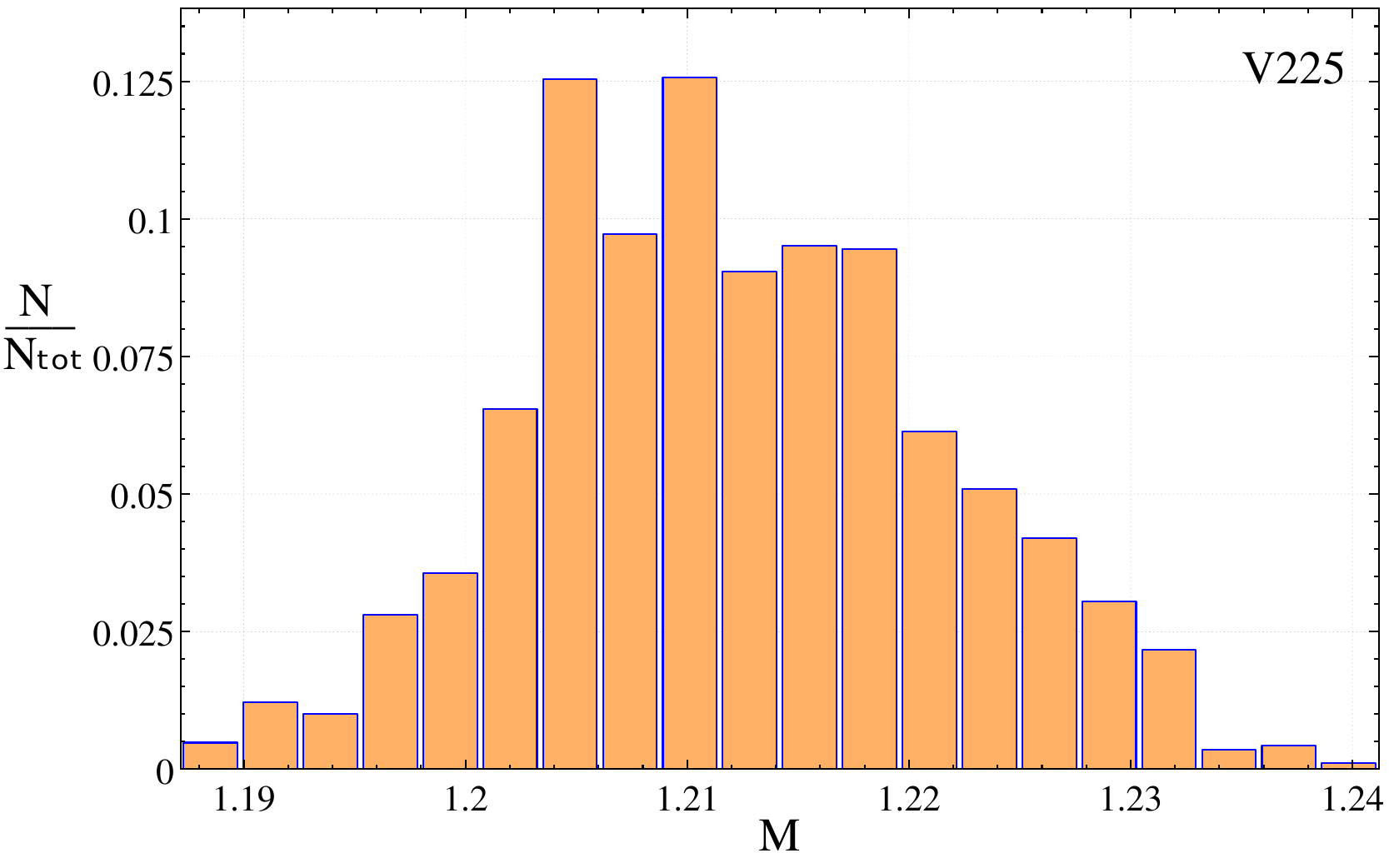}
	\includegraphics[clip,width=0.495\linewidth,height=65mm]{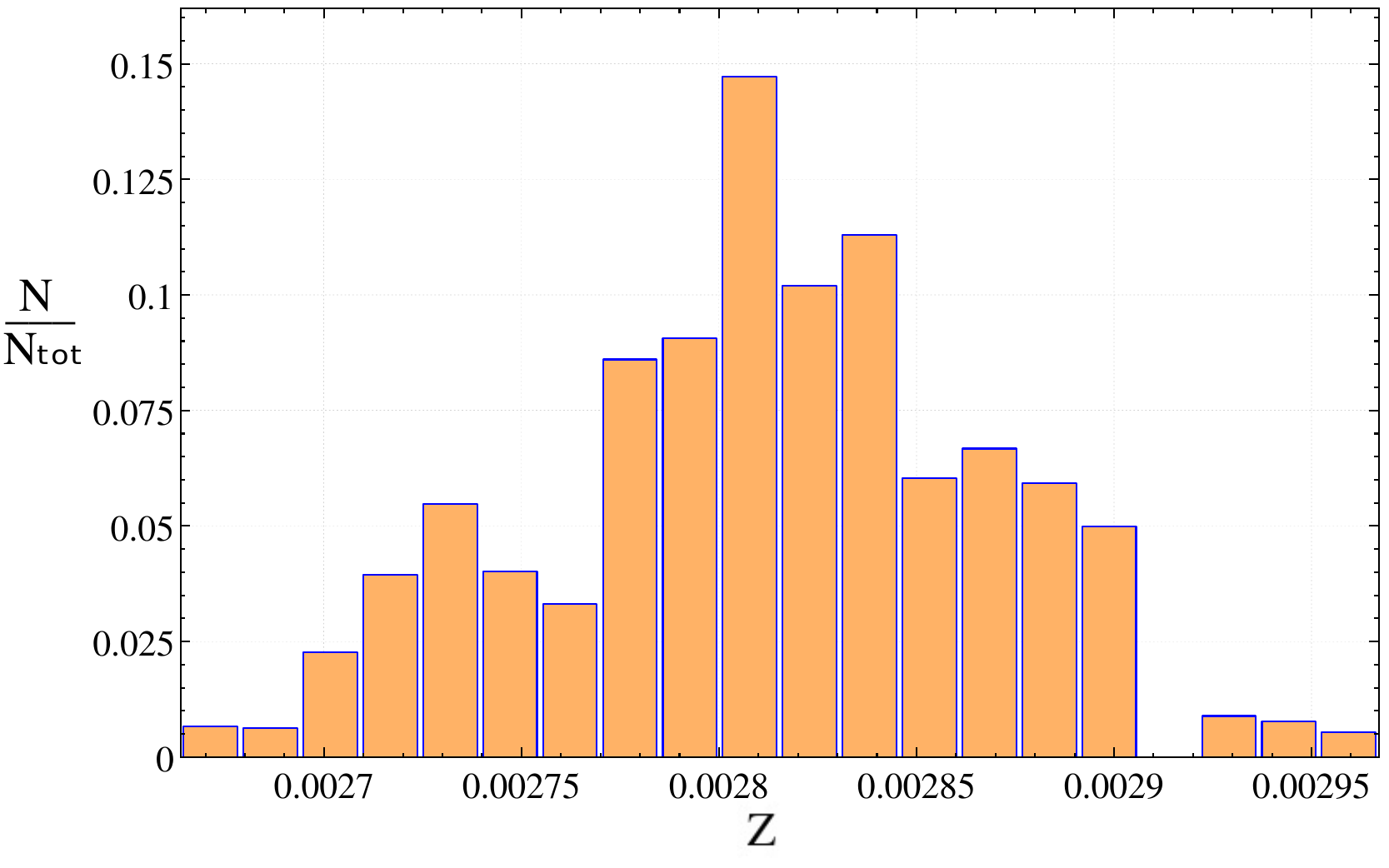}
	\includegraphics[clip,width=0.495\linewidth,height=65mm]{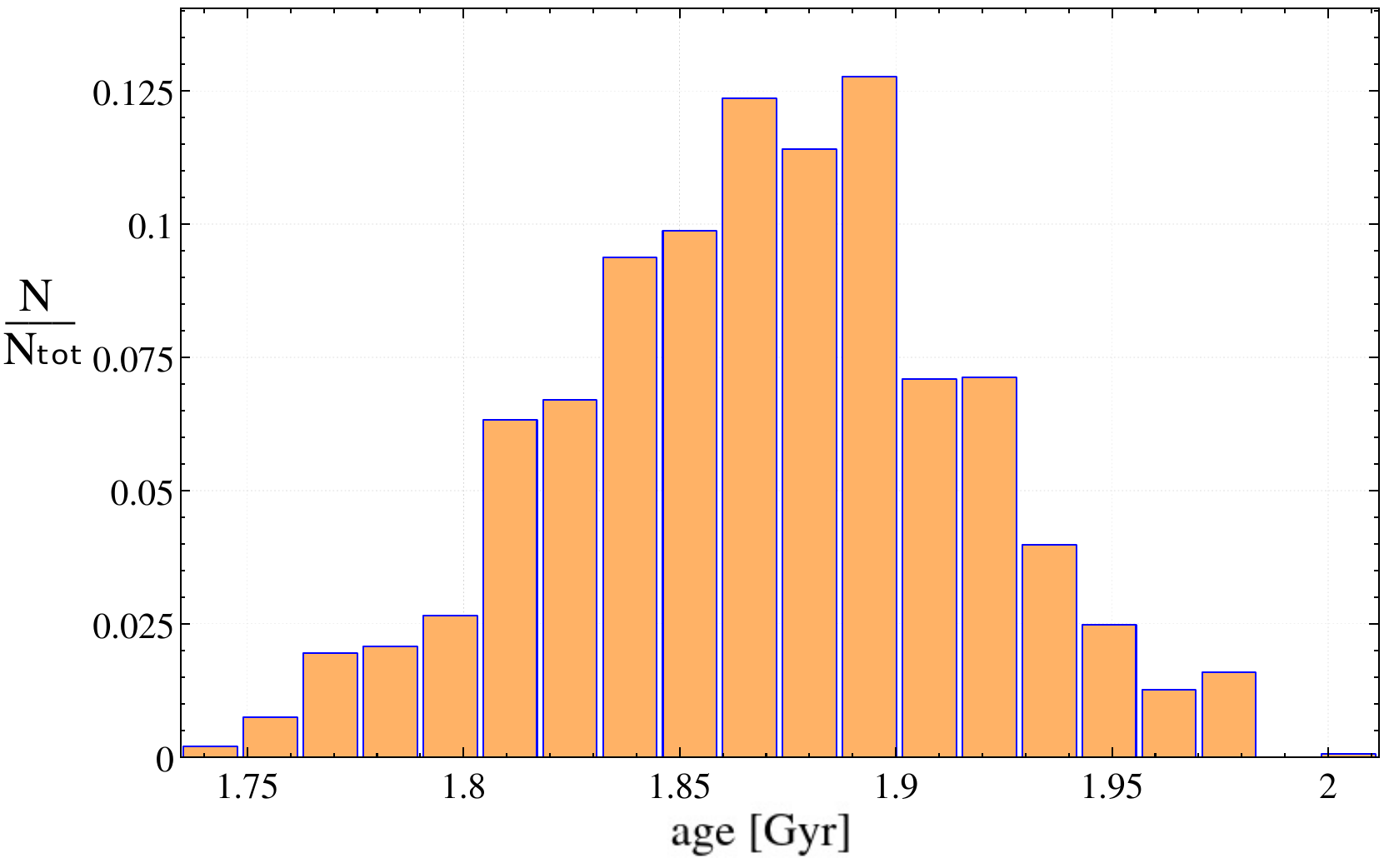}
	\includegraphics[clip,width=0.495\linewidth,height=65mm]{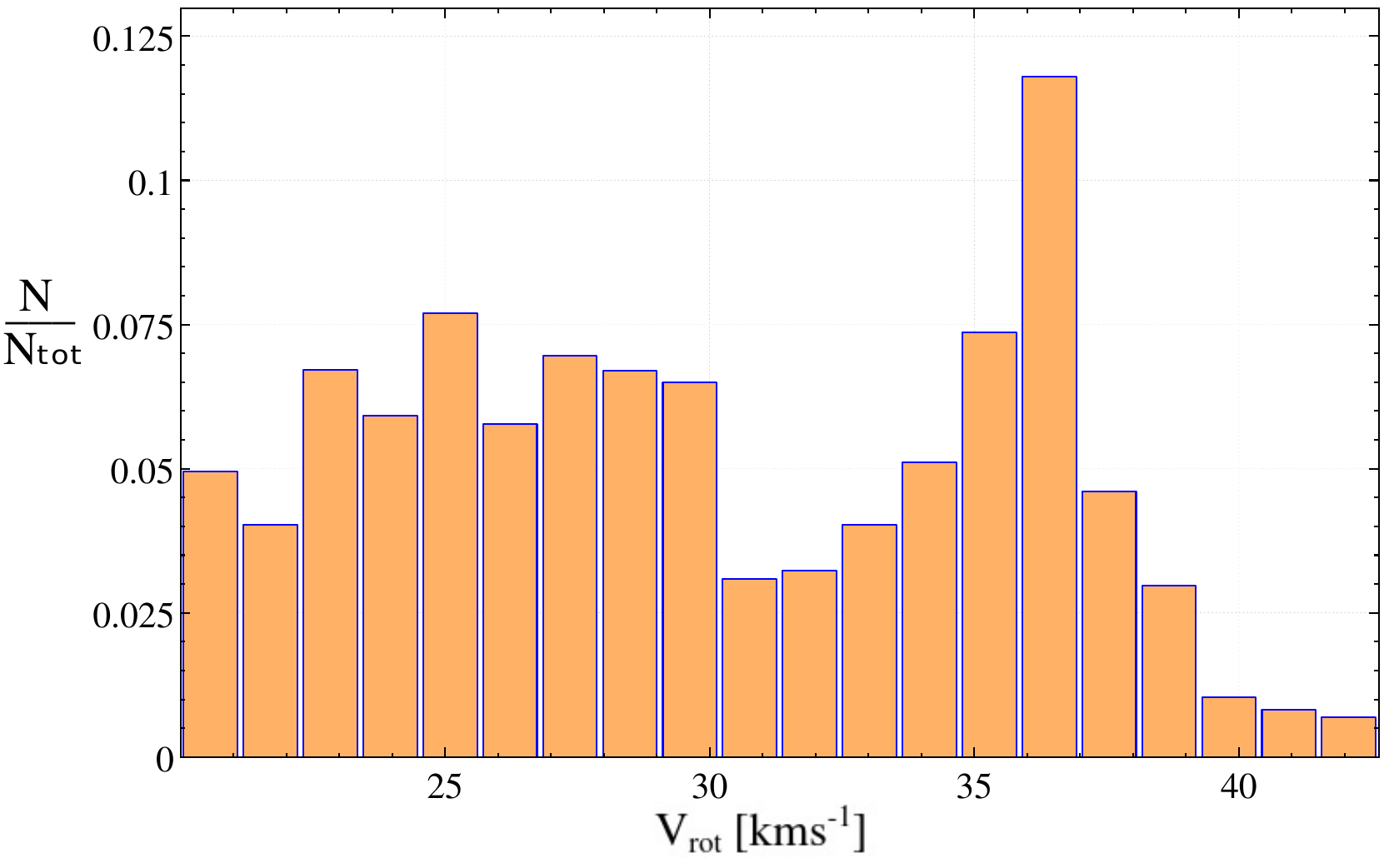}
	\includegraphics[clip,width=0.495\linewidth,height=65mm]{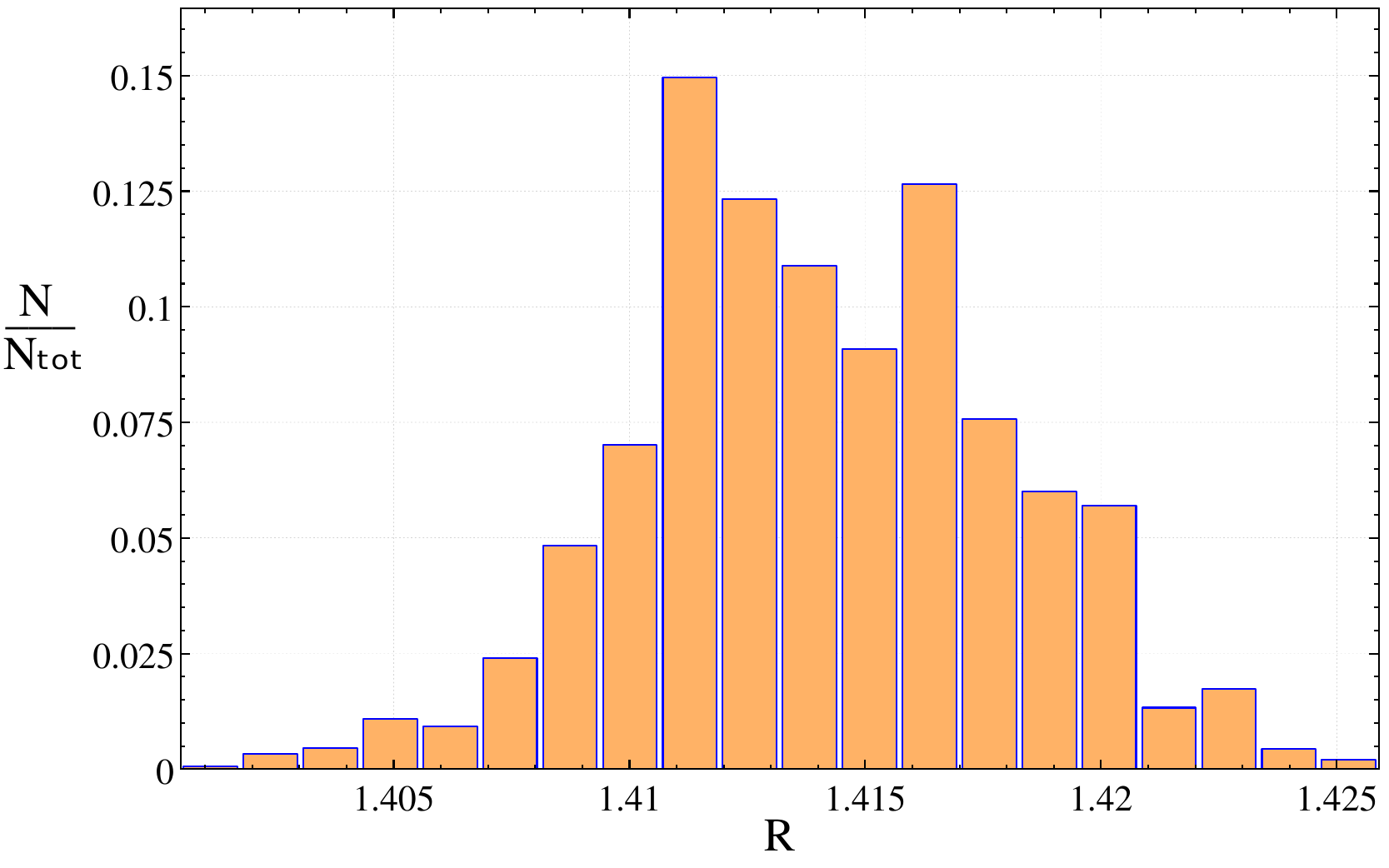}
    \includegraphics[clip,width=0.495\linewidth,height=65mm]{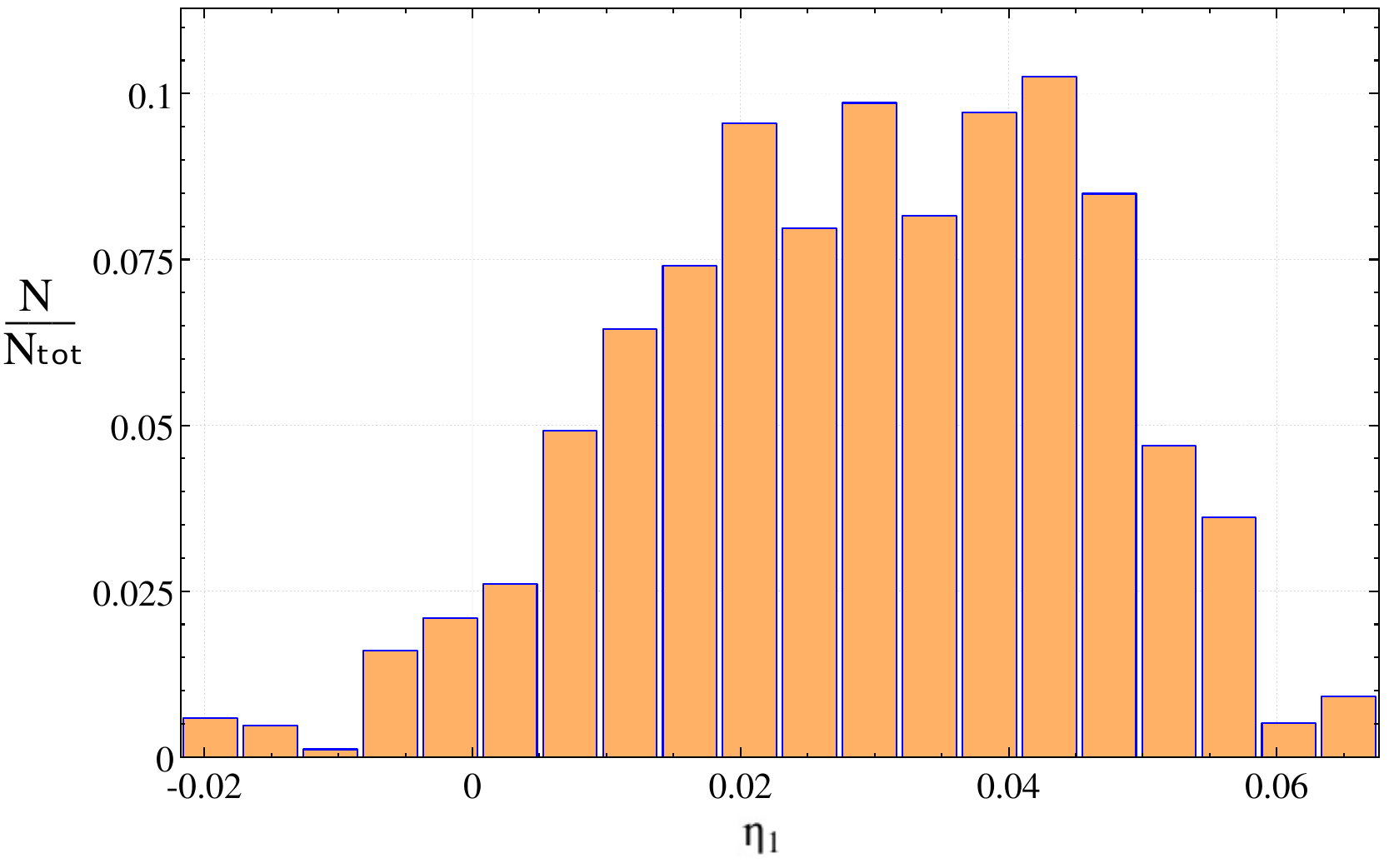}
	\caption{The same as in Fig.\,C1 but for V225. Only seismic models with the rotational velocities $V_{\rm rot}>20\,\kms$ were included}
	\label{histograms_V194}
\end{figure*}

\begin{figure*}
	\centering
	\includegraphics[clip,width=0.495\linewidth,height=65mm]{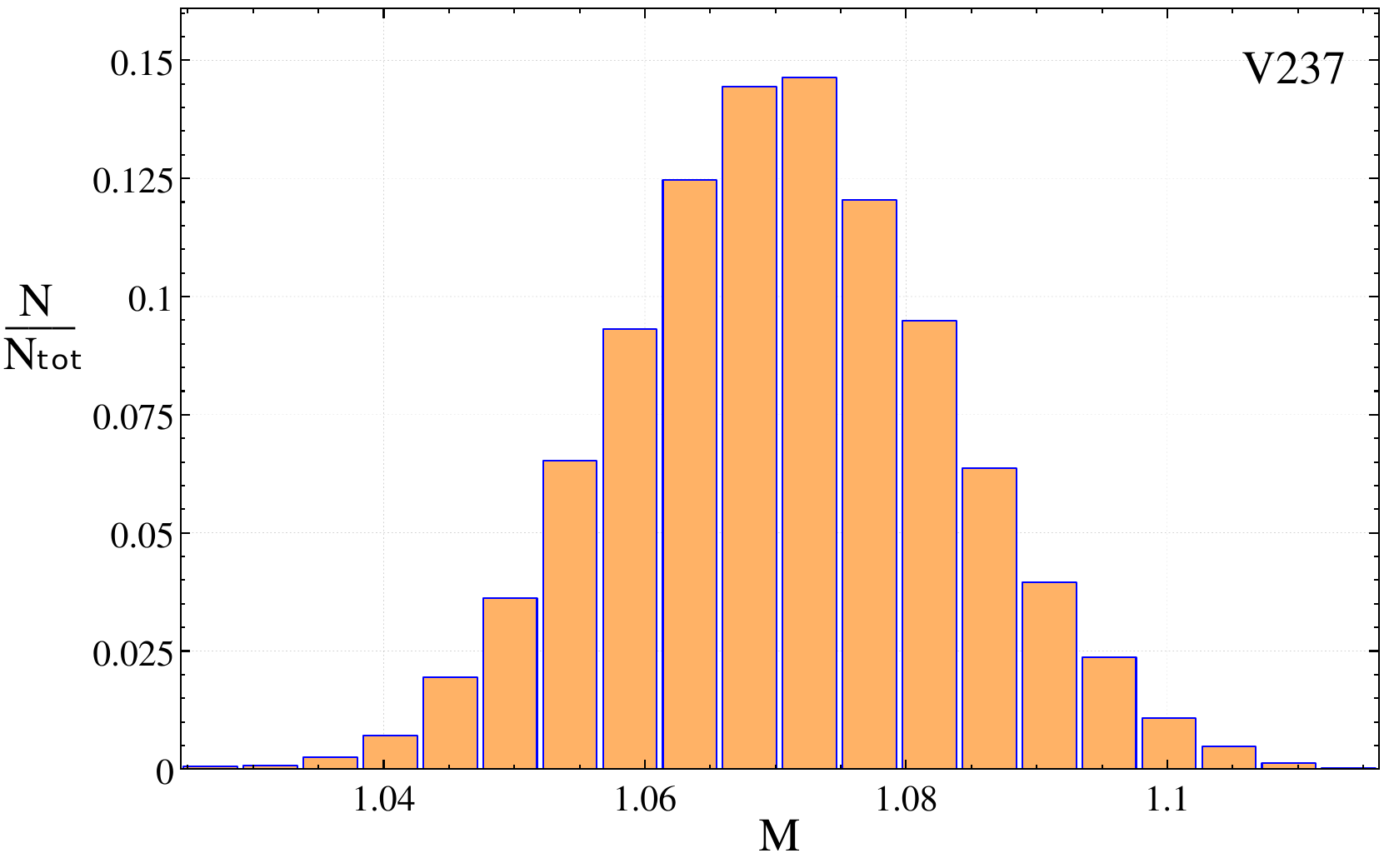}
	\includegraphics[clip,width=0.495\linewidth,height=65mm]{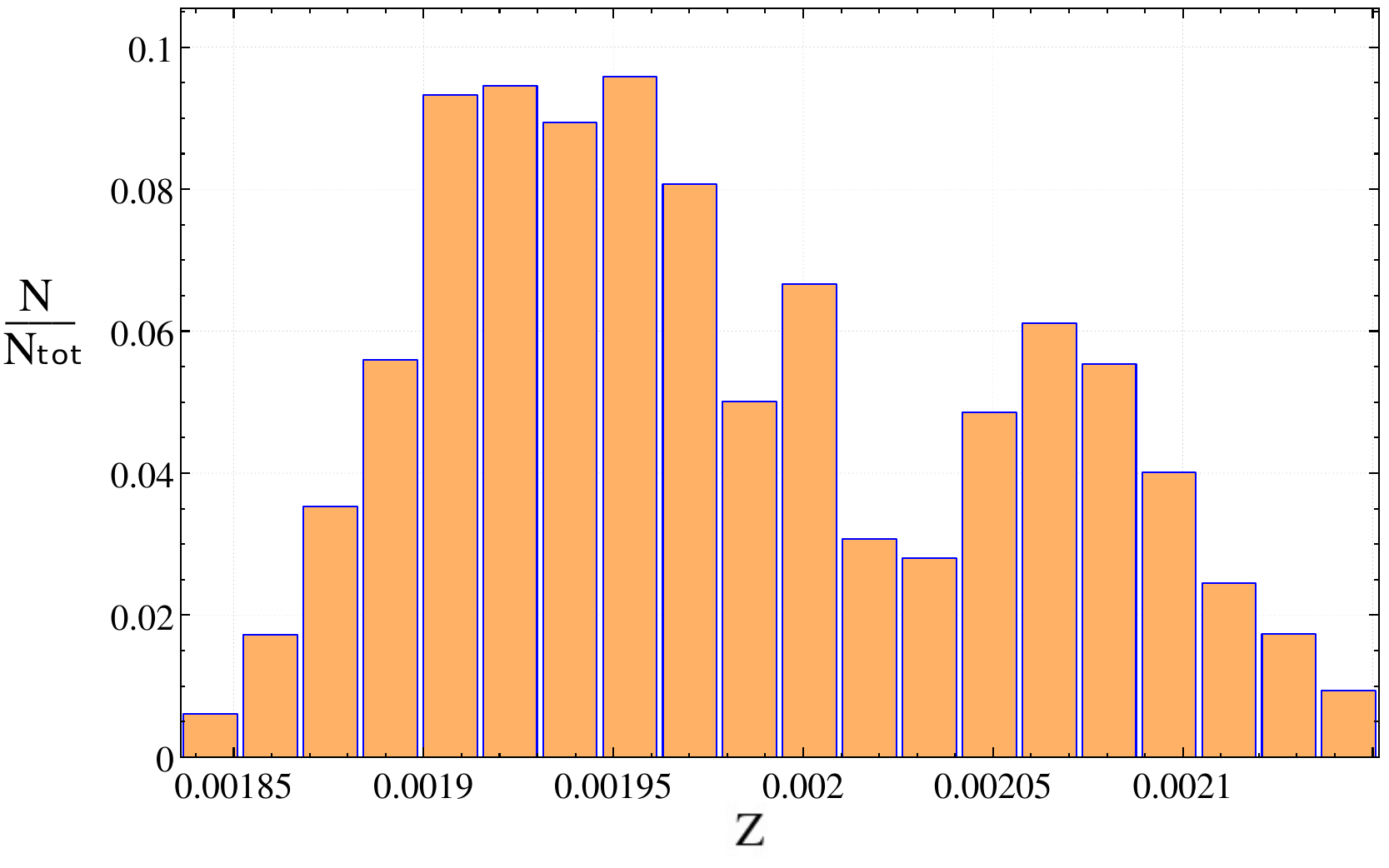}
	\includegraphics[clip,width=0.495\linewidth,height=65mm]{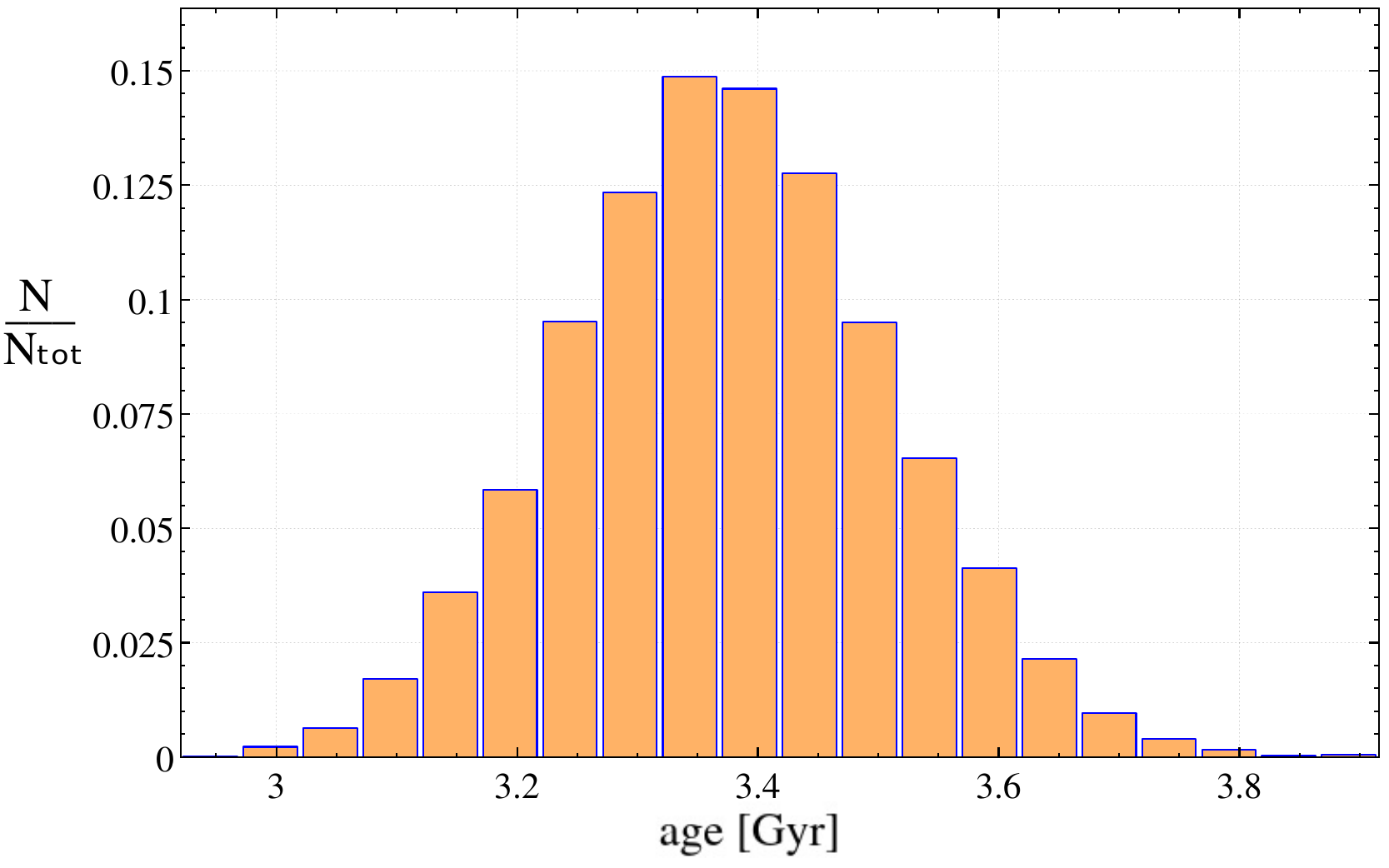}
	\includegraphics[clip,width=0.495\linewidth,height=65mm]{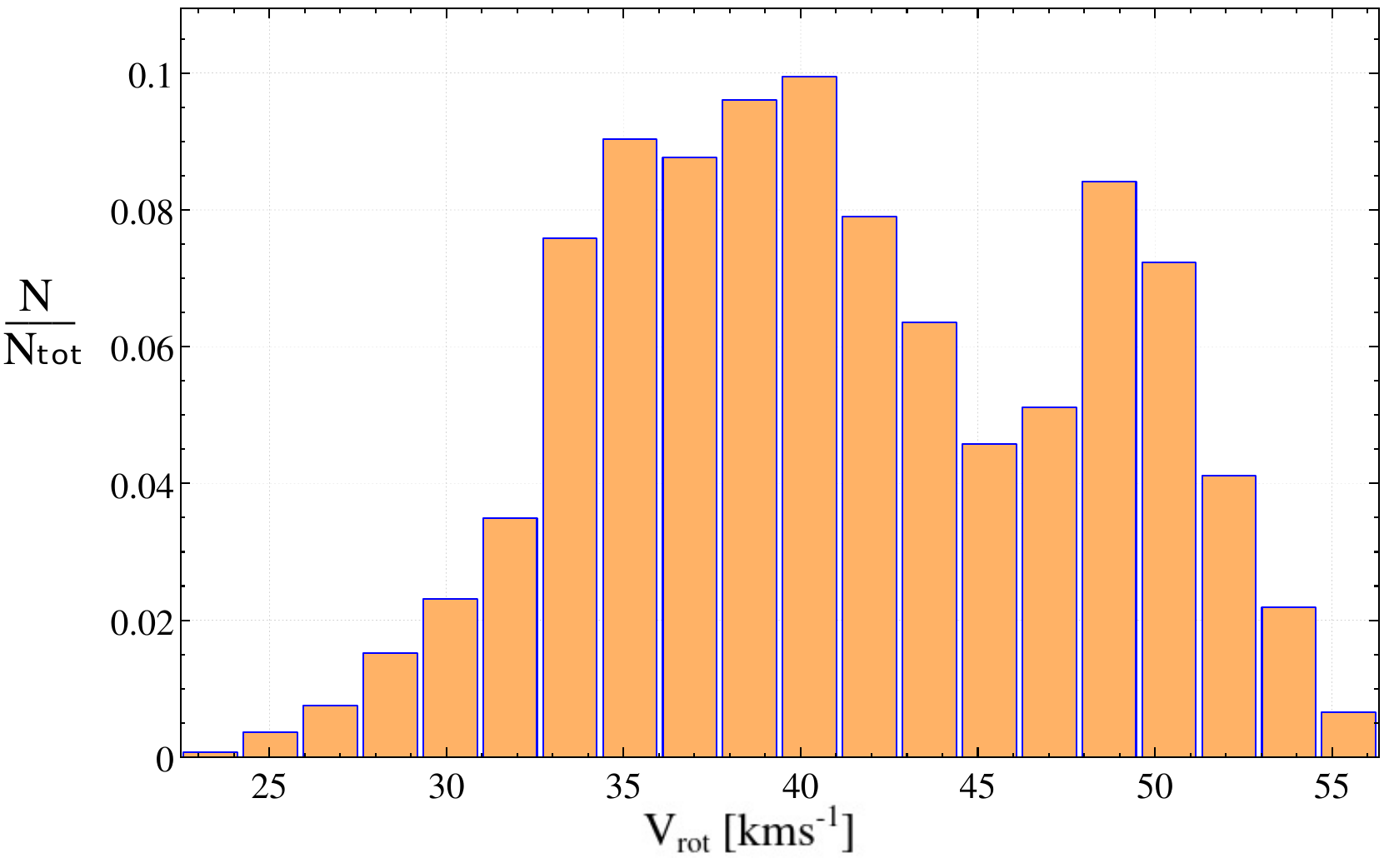}
	\includegraphics[clip,width=0.495\linewidth,height=65mm]{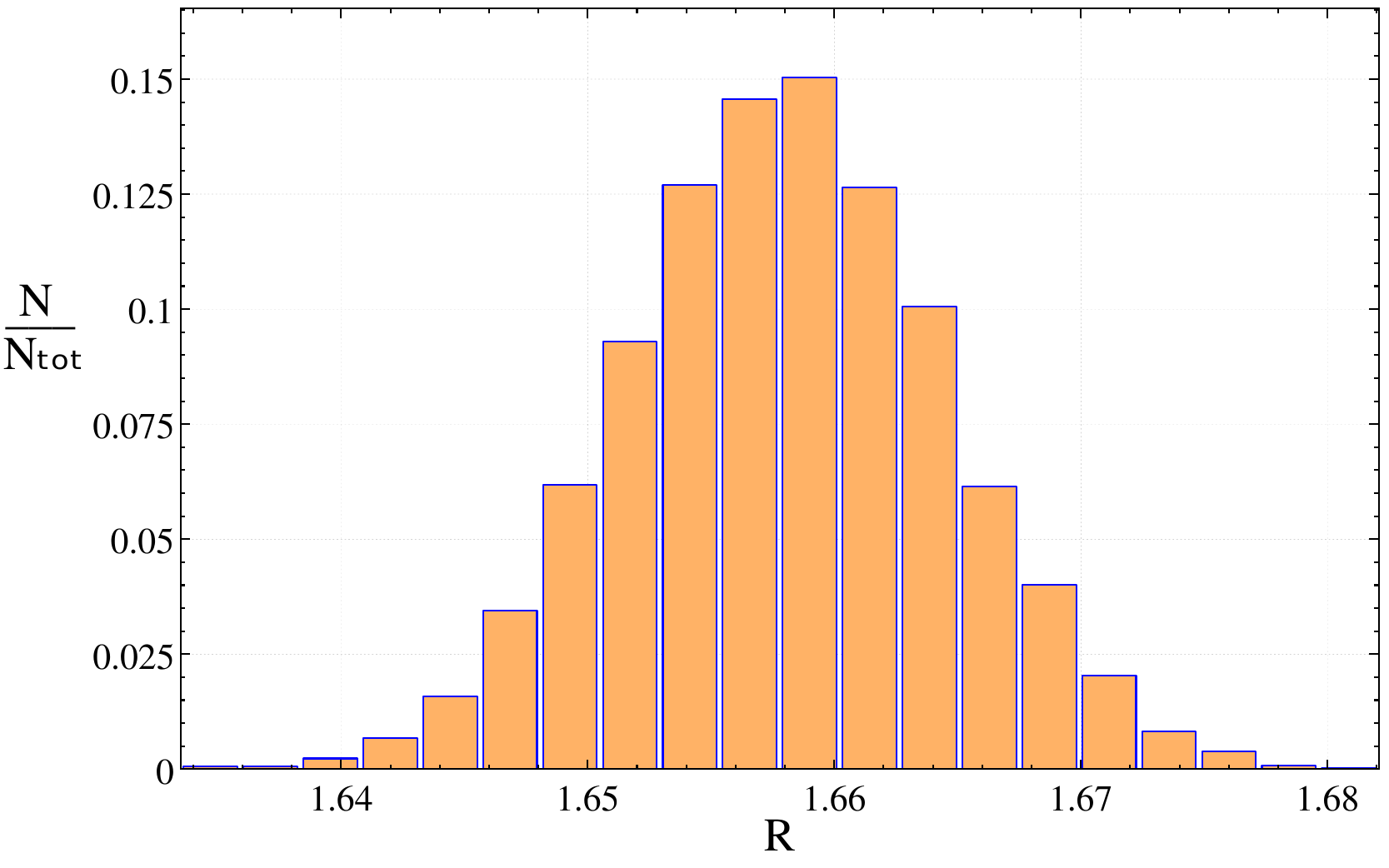}
	\includegraphics[clip,width=0.495\linewidth,height=65mm]{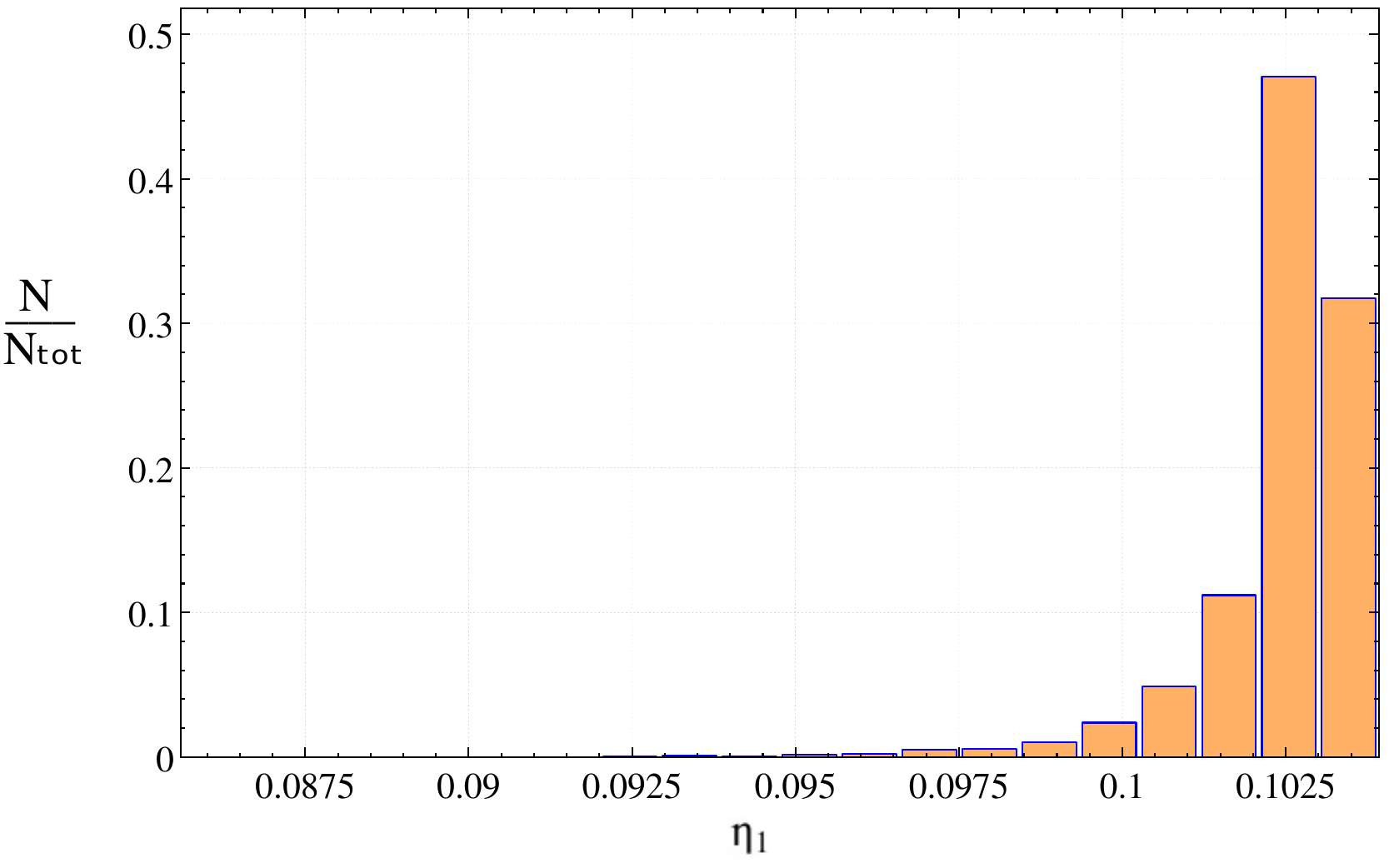}
	\caption{The same as in Fig.\,C1 but for V237 and the seismic models reproduce the fundamental and second overtone radial modes.}
	\label{histograms_V194}
\end{figure*}

\begin{figure*}
	\centering
	\includegraphics[clip,width=0.495\linewidth,height=65mm]{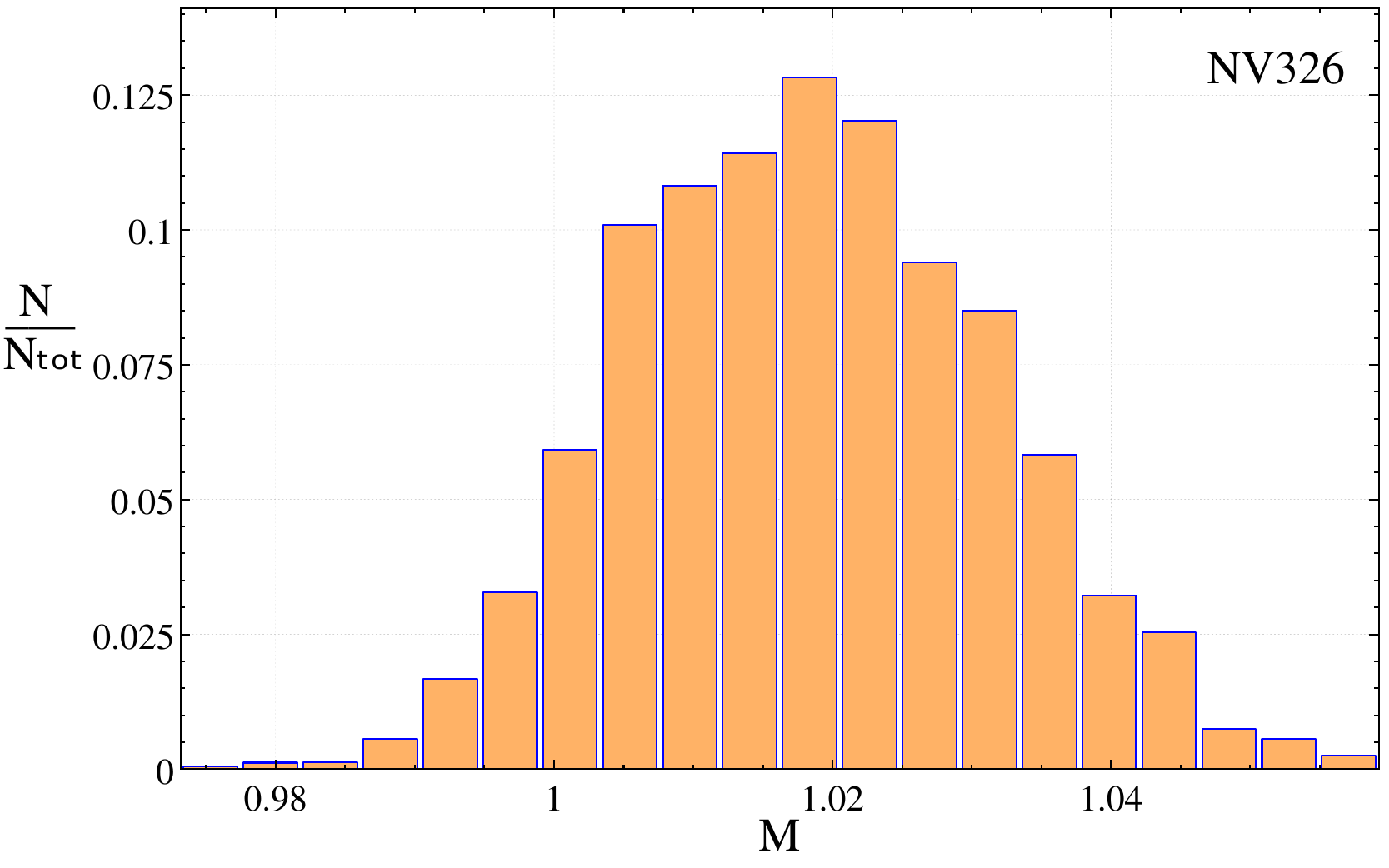}
	\includegraphics[clip,width=0.495\linewidth,height=65mm]{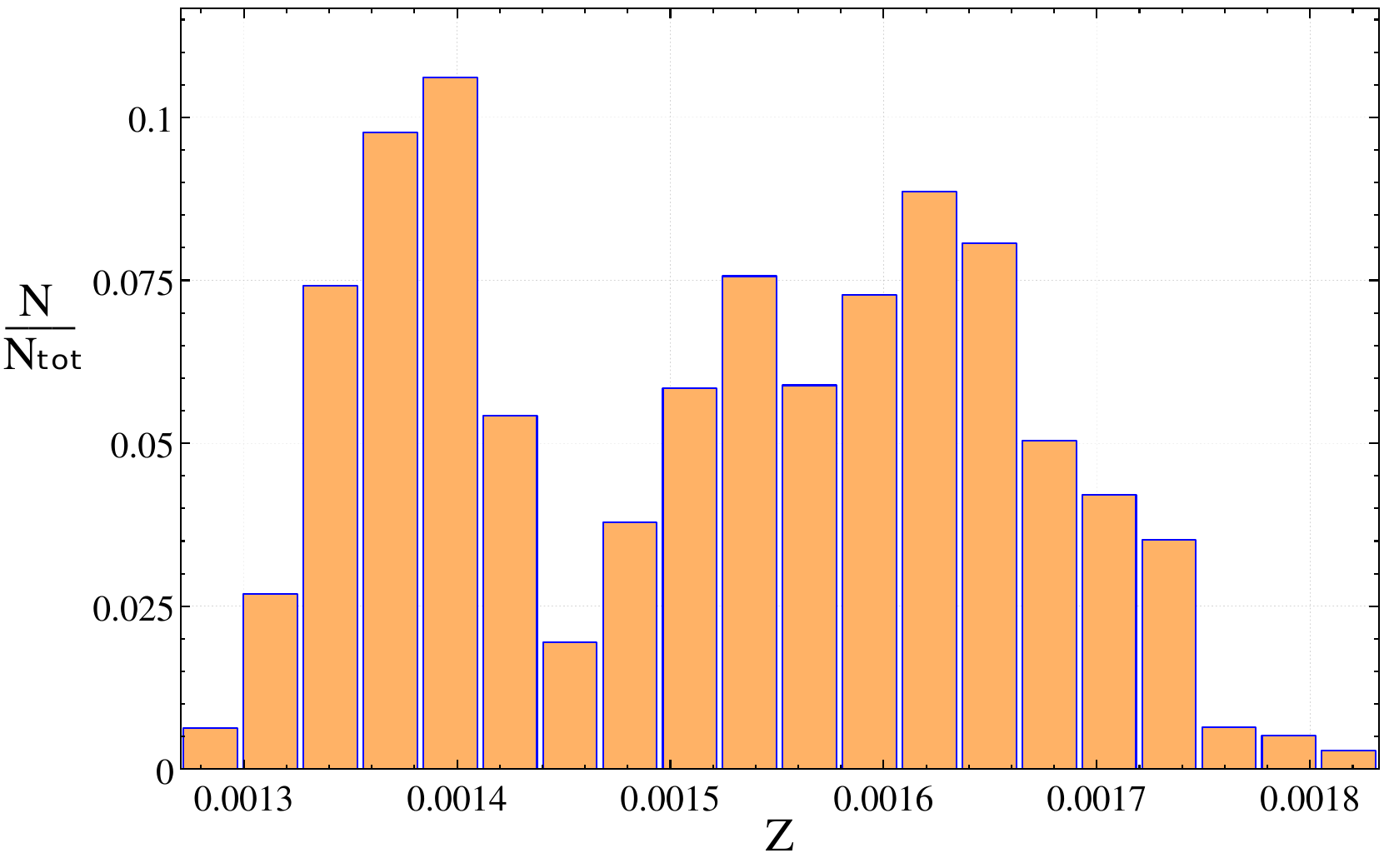}
	\includegraphics[clip,width=0.495\linewidth,height=65mm]{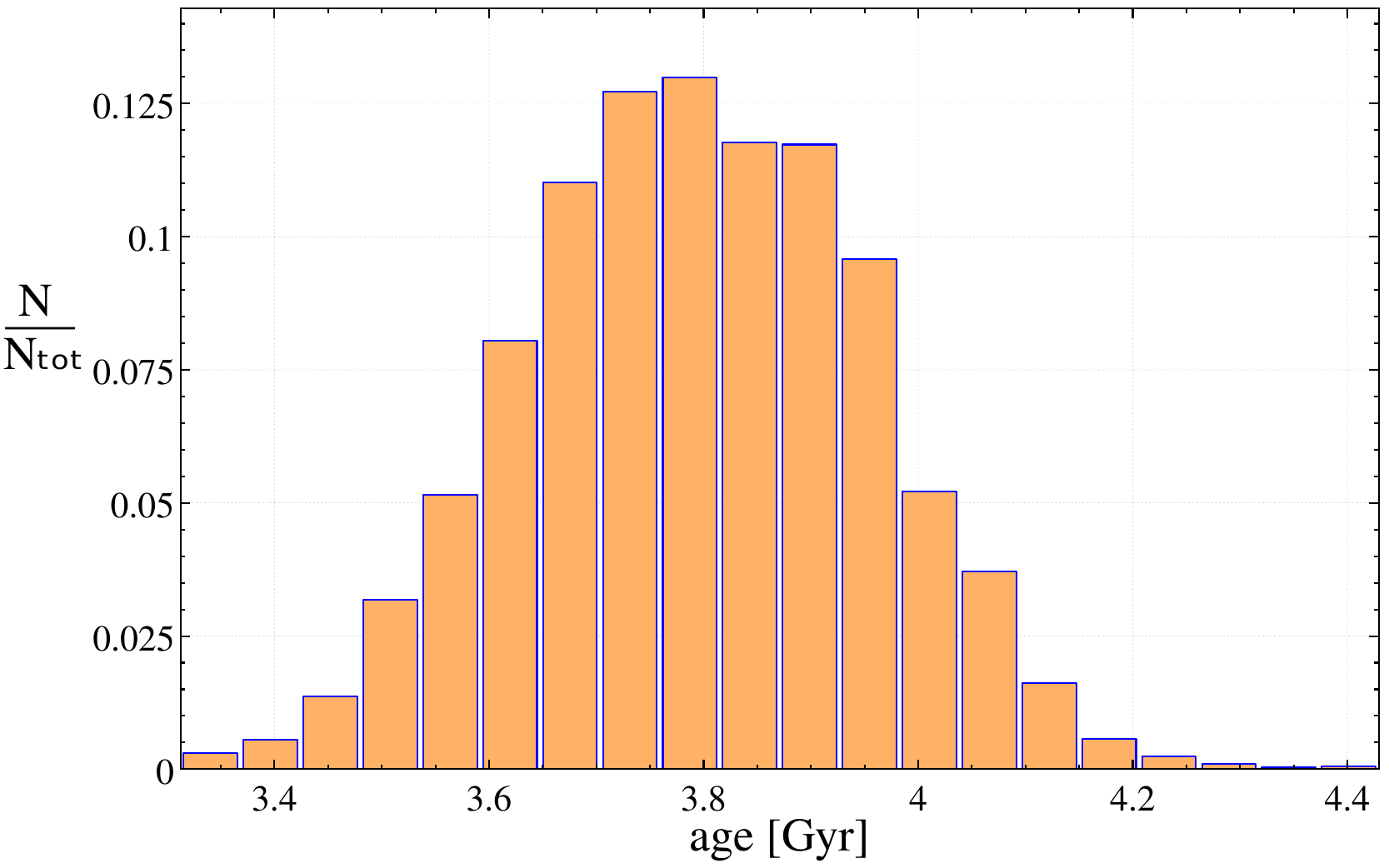}
	\includegraphics[clip,width=0.495\linewidth,height=65mm]{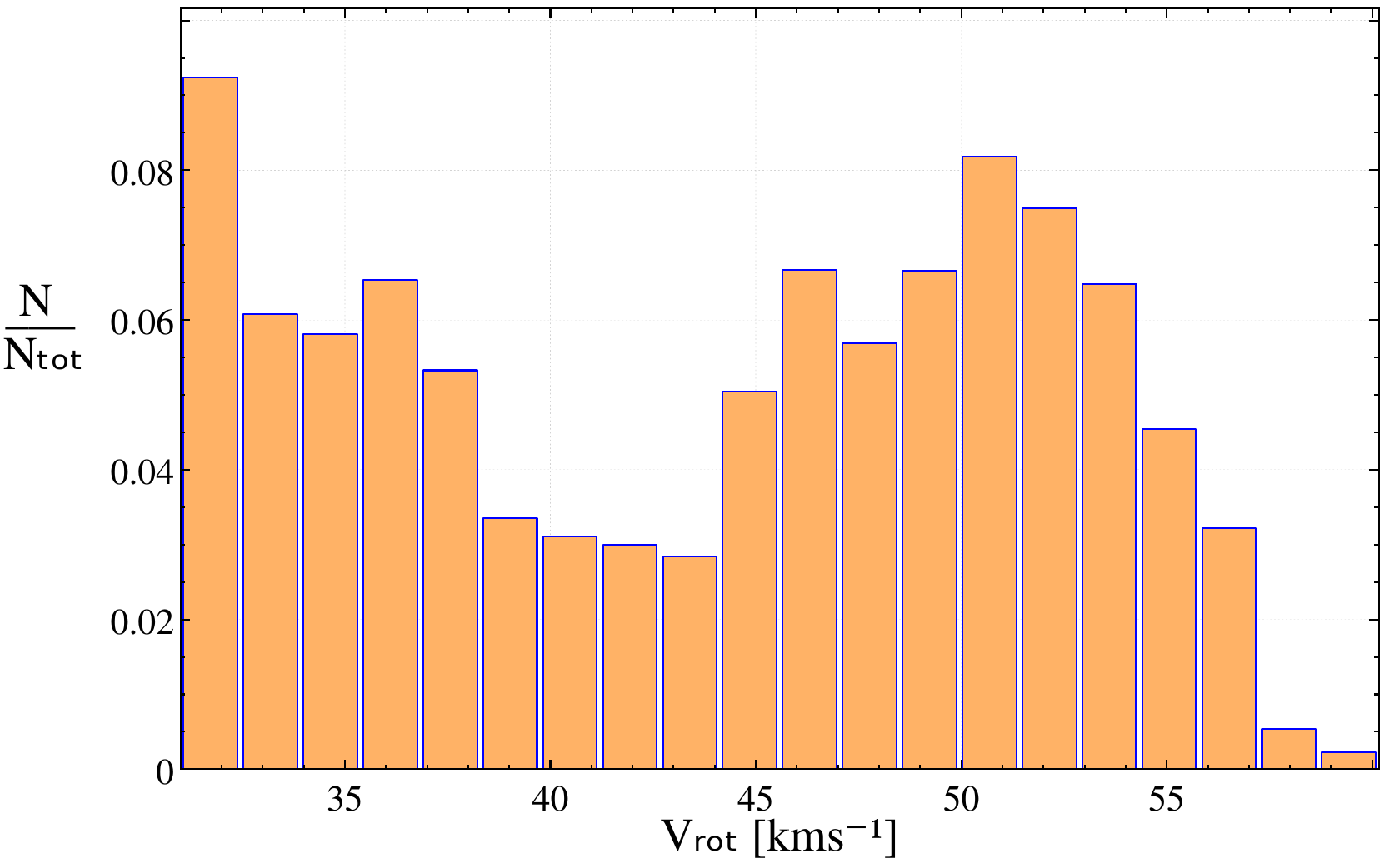}
	\includegraphics[clip,width=0.495\linewidth,height=65mm]{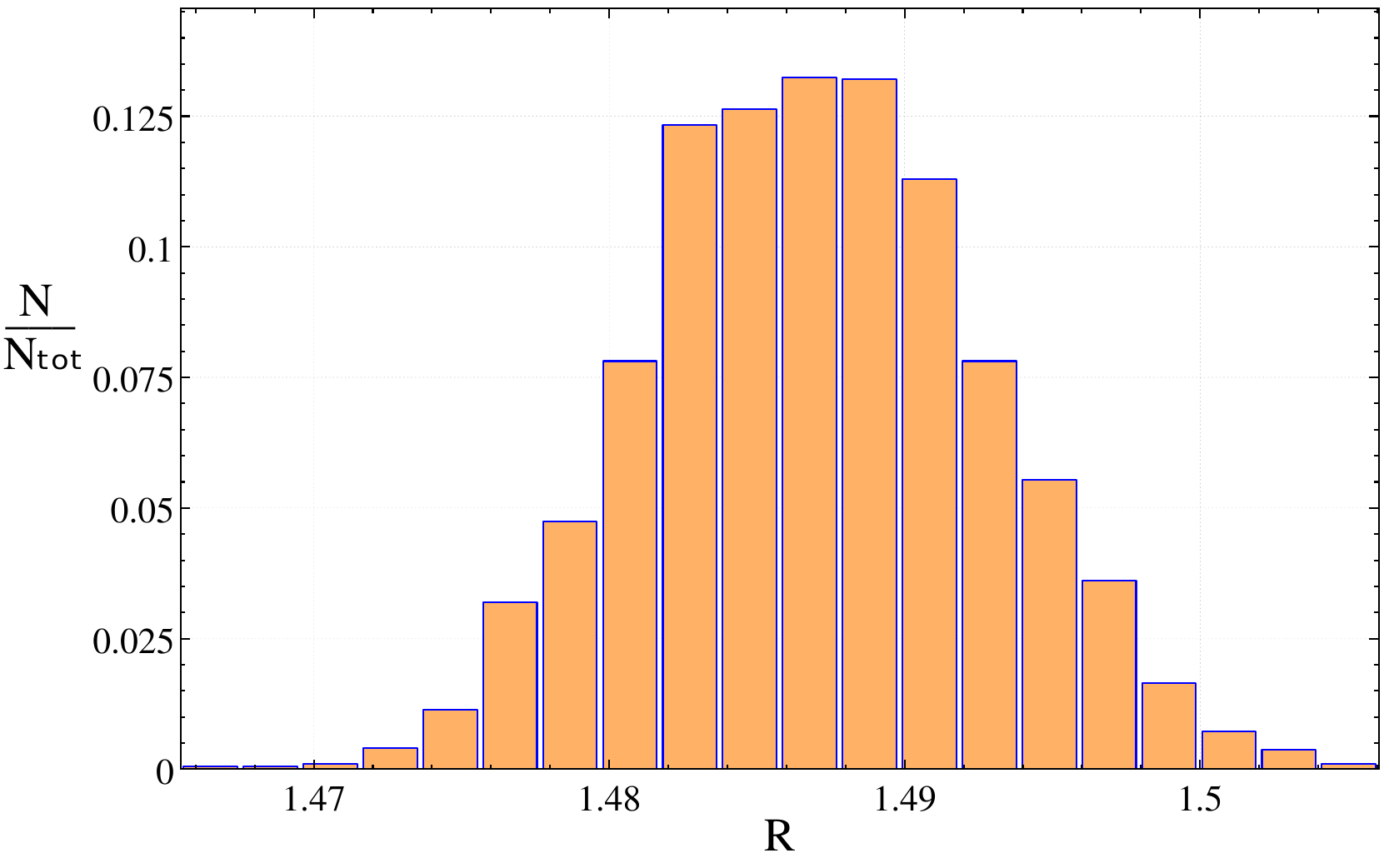}
	\includegraphics[clip,width=0.495\linewidth,height=65mm]{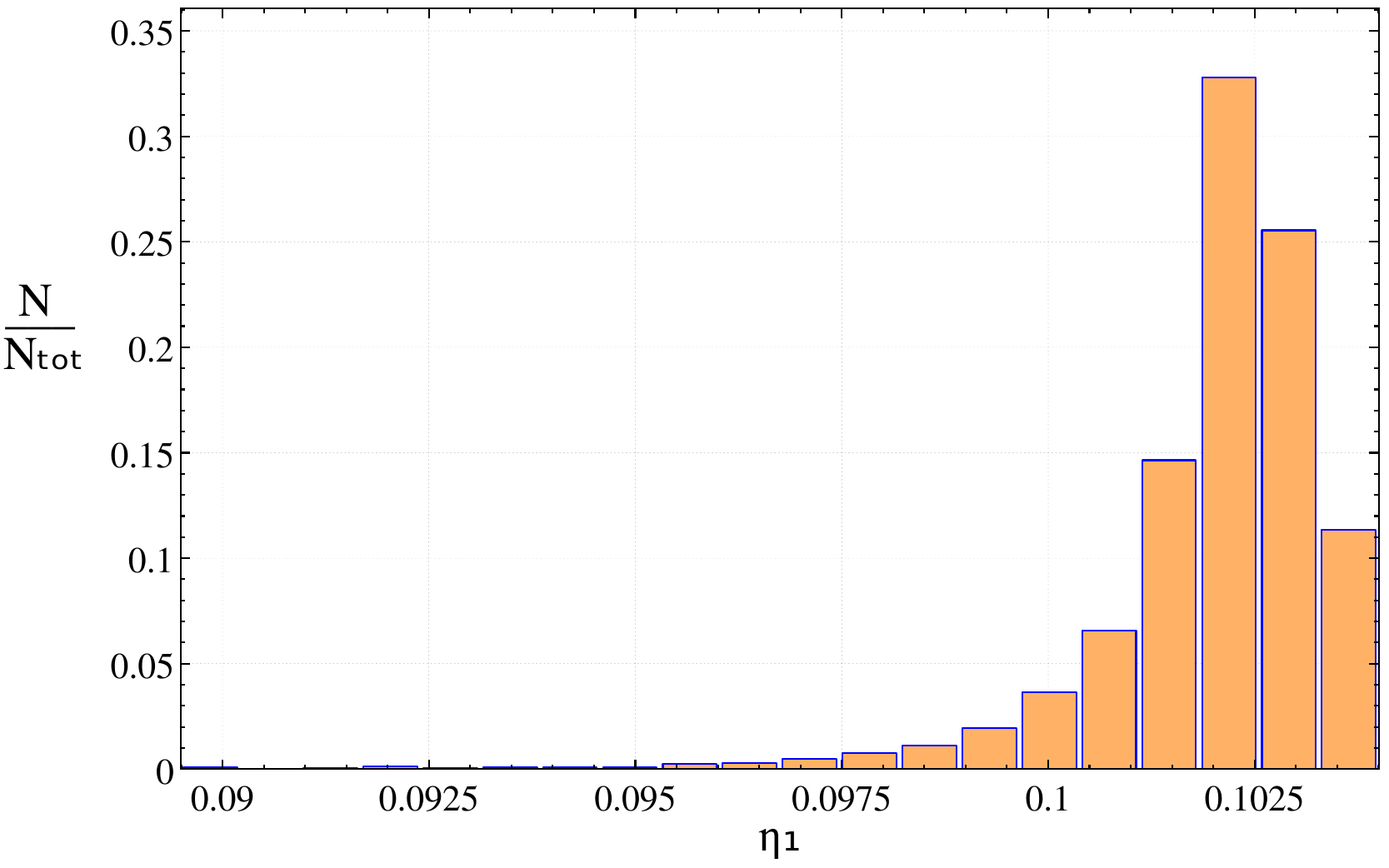}
	\caption{The same as in Fig.\,C1 but for NV326. Only seismic models with the rotational velocities $V_{\rm rot}>31\,\kms$ were included.}
	\label{histograms_V194}
\end{figure*}

\begin{figure*}
	\centering
	\includegraphics[clip,width=0.495\linewidth,height=53.5mm]{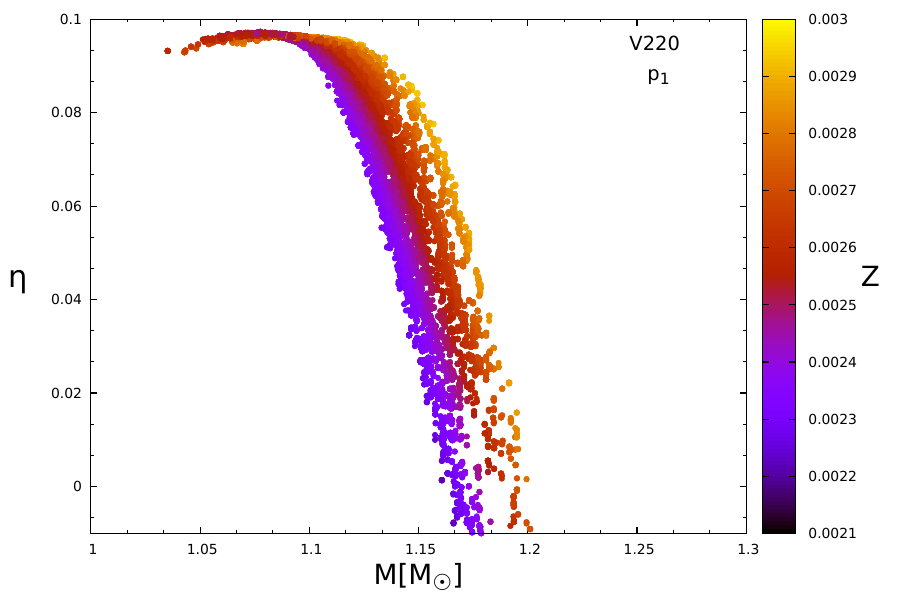}
	\includegraphics[clip,width=0.495\linewidth,height=53.5mm]{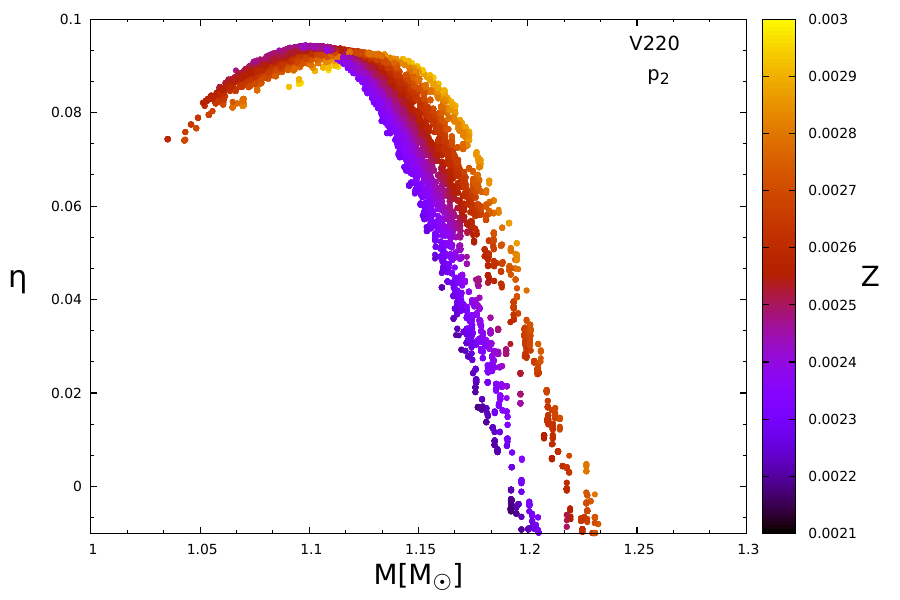}
	\includegraphics[clip,width=0.495\linewidth,height=53.5mm]{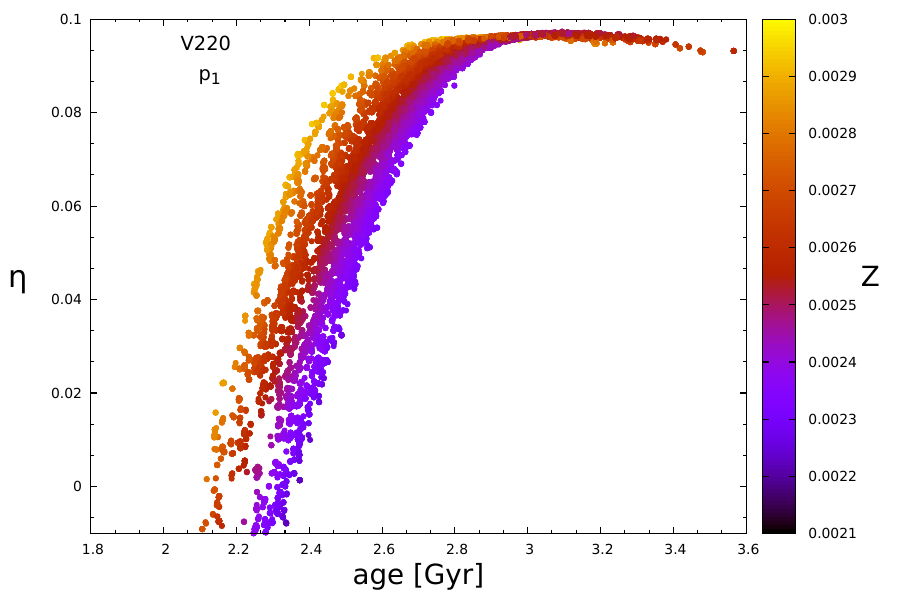}
	\includegraphics[clip,width=0.495\linewidth,height=53.5mm]{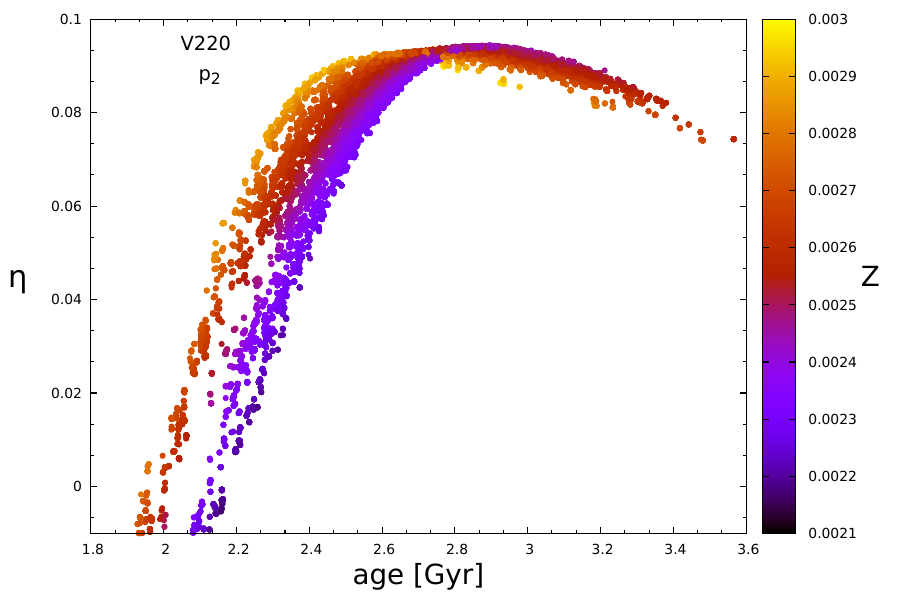}
	\caption{The instability parameter $\eta$ of the seismic models of V220 as a function of mass (upper) panels) and age (lower panels) 
		for the radial fundamental modes (left panels) and radial first overtone modes (right panels).}
	\label{eta_M_age_V220}
\end{figure*}

\begin{figure*}
	\centering
	\includegraphics[clip,width=0.495\linewidth,height=53.5mm]{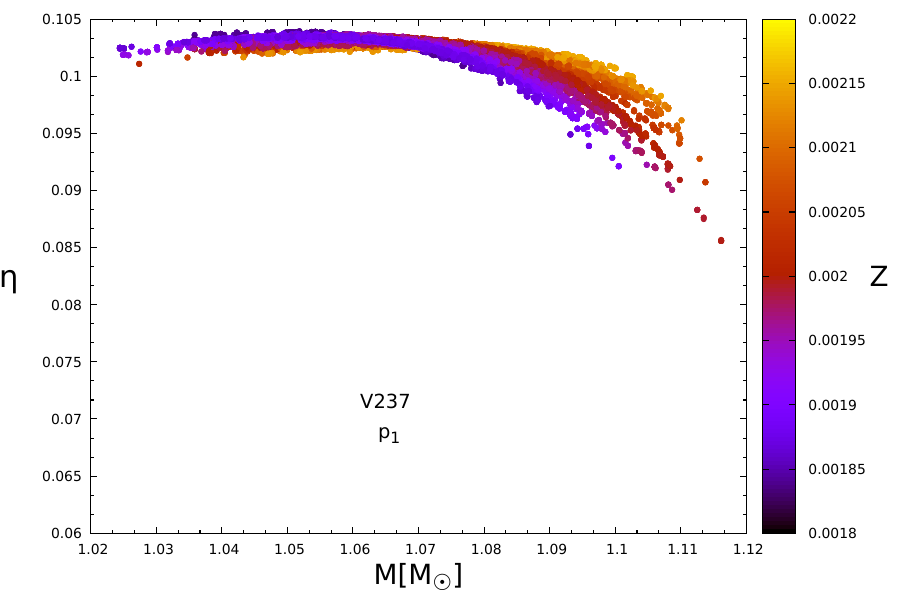}
	\includegraphics[clip,width=0.495\linewidth,height=53.5mm]{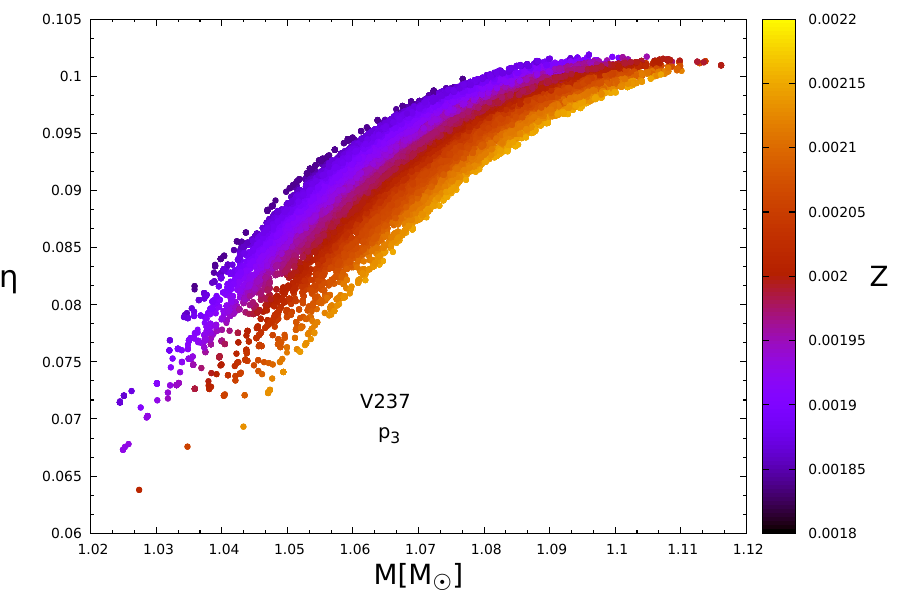}
	\includegraphics[clip,width=0.495\linewidth,height=53.5mm]{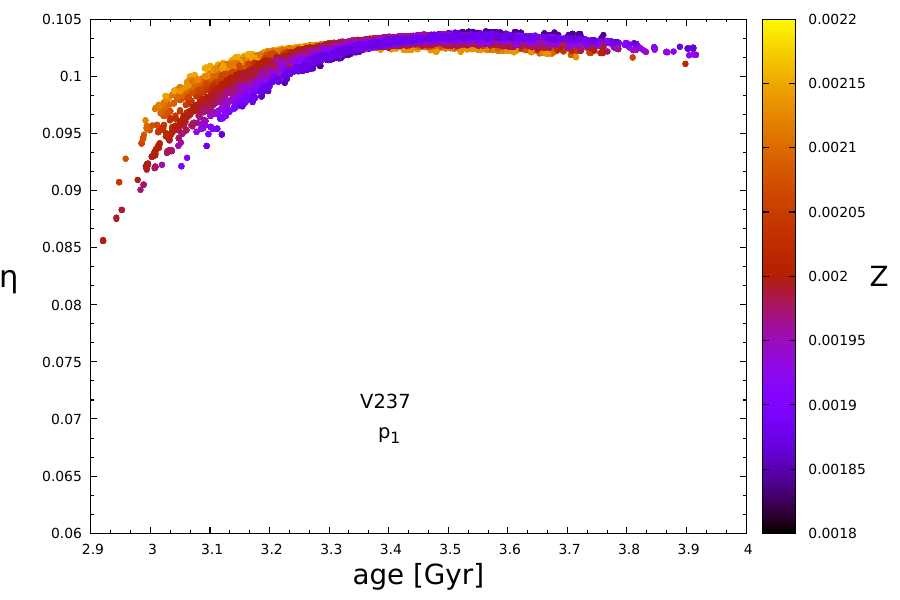}
	\includegraphics[clip,width=0.495\linewidth,height=53.5mm]{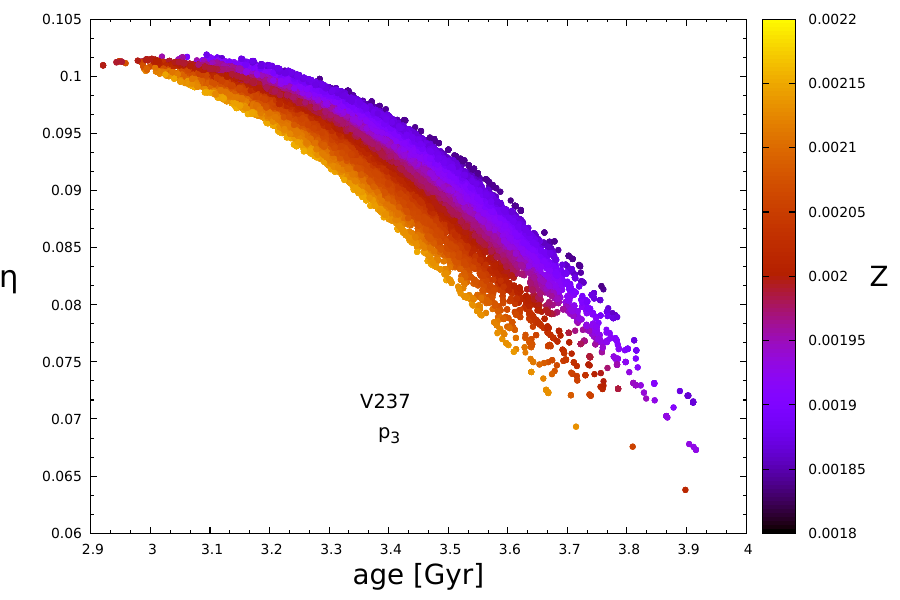}
	\caption{The same as in Fig.\,C6 but for the seismic models of V237 and with radial second overtone modes in the right panels.}
	\label{eta_M_age_V237}
\end{figure*}

\begin{figure*}
	\centering
	\includegraphics[clip,width=0.65\linewidth,height=8.6cm]{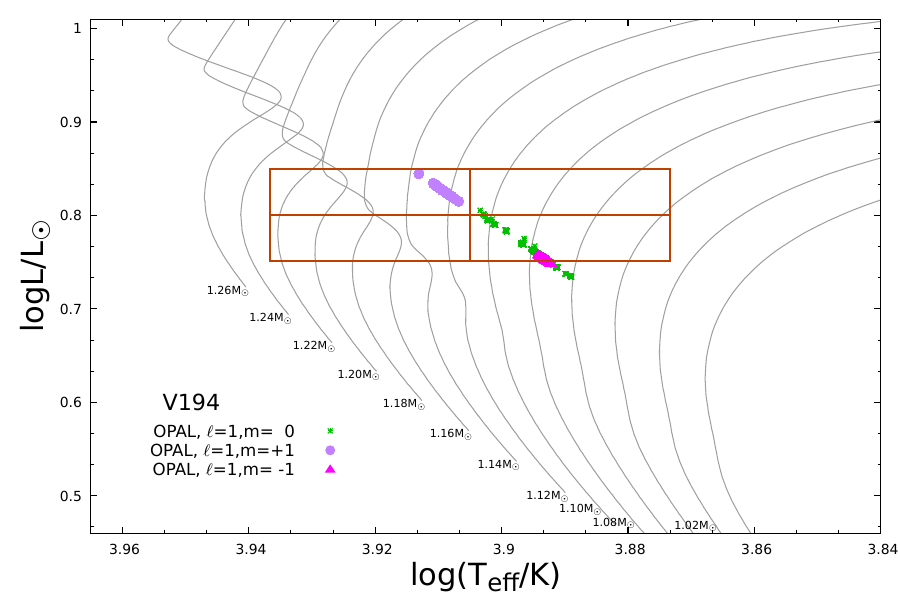}
	\caption{The HR diagram with the position of the SX Phe star V194 and its OPAL seismic models which reproduce the frequencies of two radial modes and one dipole mode with various azimuthal order $m$.}
	\label{V194_HR_l1allm}
\end{figure*}

\begin{figure*}
	\centering
	\includegraphics[clip,width=0.495\linewidth,height=64mm]{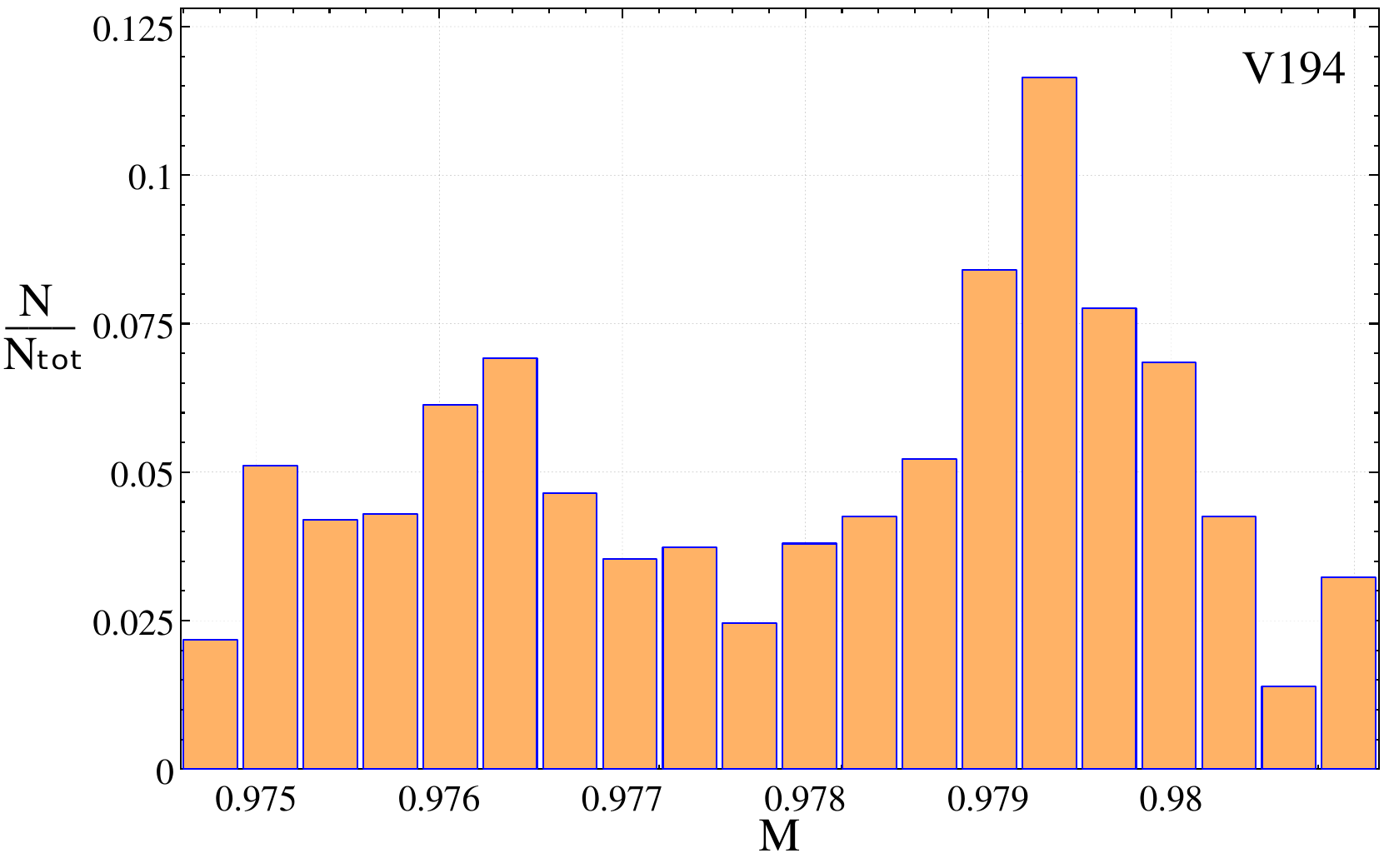}
	\includegraphics[clip,width=0.495\linewidth,height=64mm]{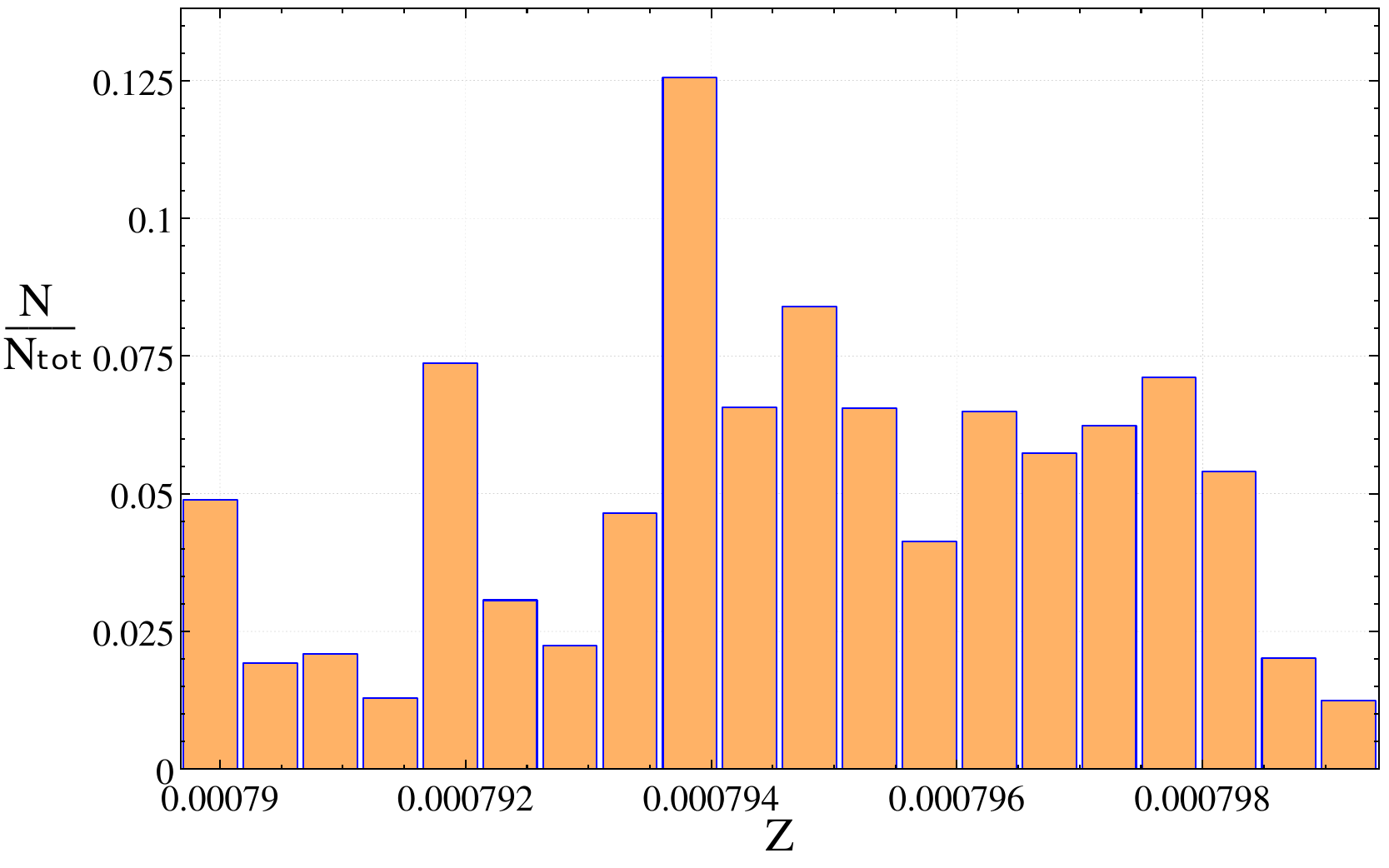}
	\includegraphics[clip,width=0.495\linewidth,height=64mm]{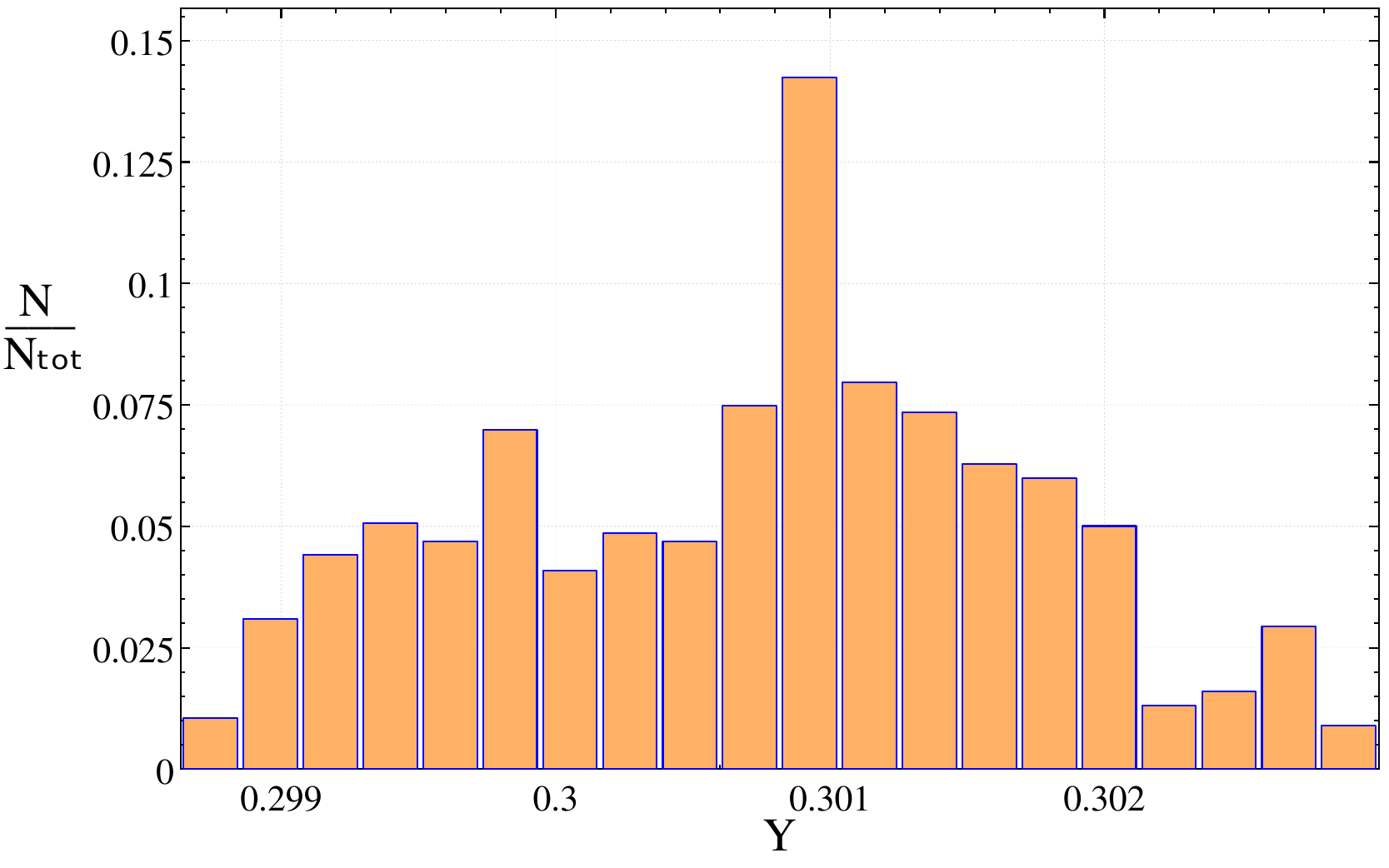}
	\includegraphics[clip,width=0.495\linewidth,height=64mm]{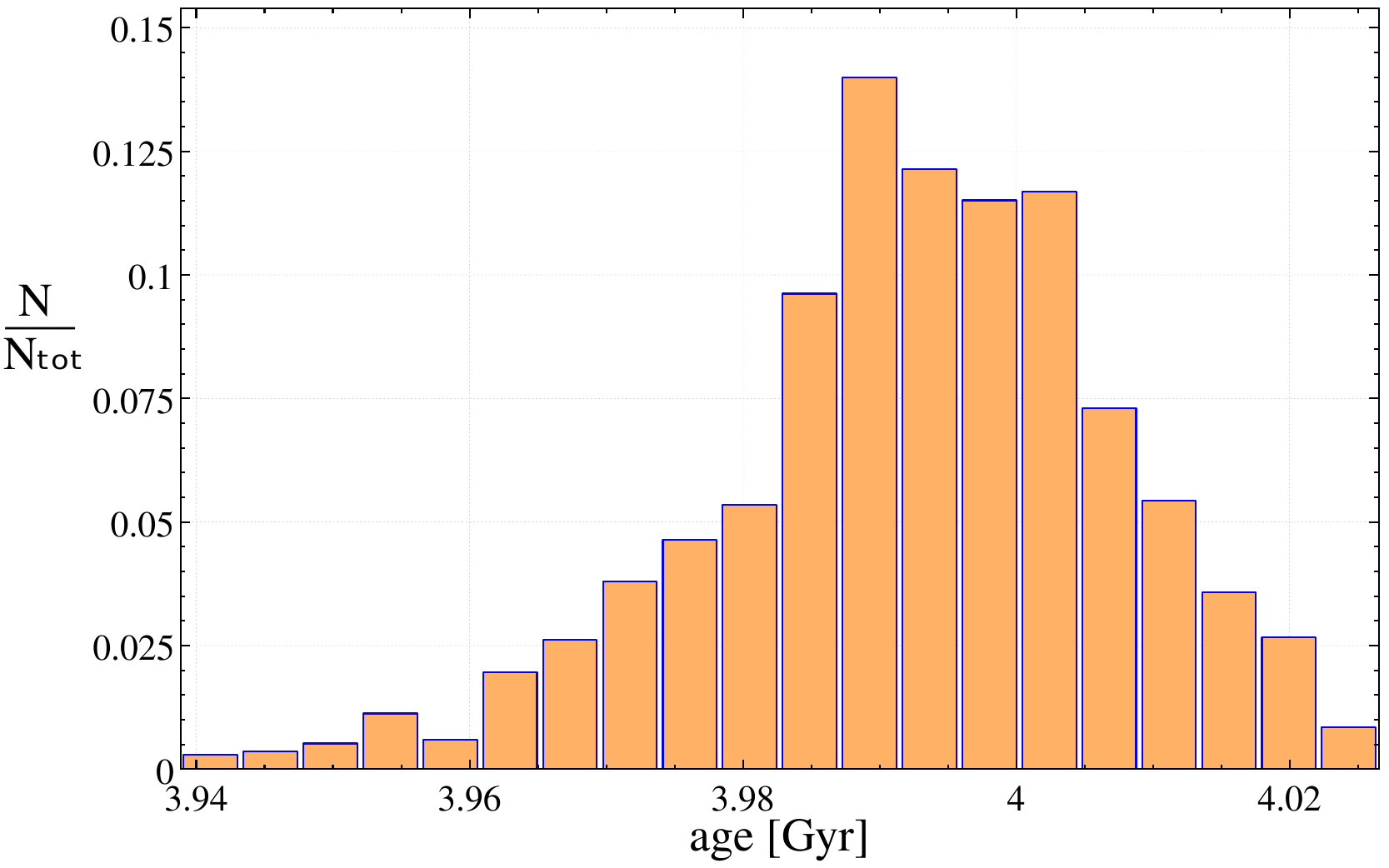}
	\caption{The normalized histograms for $M,~Z,~X_0$ and age of OPAL seismic models of V194 fitting additionally the frequency of a dipole retrograde mode.}
	\label{histograms_V194}
\end{figure*}

% Don't change these lines
\bsp	% typesetting comment
\label{lastpage}
\end{document}